\documentclass[12pt,english]{article}
\usepackage{mathptmx}
\usepackage{geometry}
\usepackage{xcolor}
\usepackage{array}
\usepackage{bbding}
\usepackage{multirow}
\usepackage[fleqn]{amsmath}
\usepackage{amssymb}
\usepackage{amsthm}
\usepackage{graphicx}
\usepackage[sort]{natbib}
\usepackage{lscape}
\usepackage[hyphens]{url}
\usepackage[hidelinks]{hyperref}
\usepackage{comment}
\usepackage{bm}
\usepackage[title]{appendix}
\usepackage{longtable}
\usepackage{mathtools}
\usepackage{chngcntr}
\usepackage{authblk}
\usepackage{babel}
\usepackage{dsfont}
\usepackage{subcaption}
\usepackage{fancyhdr}
\usepackage{cleveref}
\usepackage{abstract}

\makeatletter
\allowdisplaybreaks
\newcommand{\diff}{\,\mathrm{d}}

\date{}

\fancypagestyle{plain}{%
	\fancyhf{} 
	
	\lhead{\footnotesize \includegraphics[height=0.3cm]{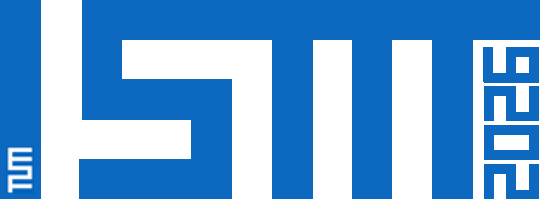}\ \ \textcolor{gray}{\textbf{}}}
	\chead{\footnotesize \textcolor{gray}{\textit{\shortauthors -- \runningtitle}}}
	\rhead{\footnotesize \textcolor{gray}{\thepage}}
}
\makeatletter
\def\blfootnote{\xdef\@thefnmark{}\@footnotetext}
\makeatother

\makeatletter
\def\titlepageext{
	\begin{center}	
	{\parindent0pt
		\rule{0.9\textwidth}{1pt}
		\begin{minipage}[t]{0.25\textwidth}
			\small {\it Keywords:}\\
			\keyword
		\end{minipage}%
		\hspace{3mm}
		\begin{minipage}[t]{0.6\textwidth}
			\small \abstract
		\end{minipage}%
		
		\rule{0.9\textwidth}{2pt}
	}
	\end{center}

	\blfootnote{* Corresponding author. E-mail address: \href{mailto:\corresemail}{\corresemail}.}
}
\makeatother

\usepackage[mathlines, pagewise]{lineno}
\usepackage{etoolbox} 

\newcommand*\linenomathpatchAMS[1]{%
	\expandafter\pretocmd\csname #1\endcsname {\linenomathAMS}{}{}%
	\expandafter\pretocmd\csname #1*\endcsname{\linenomathAMS}{}{}%
	\expandafter\apptocmd\csname end#1\endcsname {\endlinenomath}{}{}%
	\expandafter\apptocmd\csname end#1*\endcsname{\endlinenomath}{}{}%
}

\expandafter\ifx\linenomath\linenomathWithnumbers
\let\linenomathAMS\linenomathWithnumbers
\patchcmd\linenomathAMS{\advance\postdisplaypenalty\linenopenalty}{}{}{}
\else
\let\linenomathAMS\linenomathNonumbers
\fi

\linenomathpatchAMS{gather}
\linenomathpatchAMS{multline}
\linenomathpatchAMS{align}
\linenomathpatchAMS{alignat}
\linenomathpatchAMS{flalign}

\nolinenumbers

\usepackage{xcolor}
\theoremstyle{definition}
\newtheorem{assumption}{\normalfont\bfseries Assumption}
\newtheorem{remark}{\normalfont\bfseries Remark}

\title{From Micro to Macro Flow Modeling: Characterizing Heterogeneity of Mixed-Autonomy Traffic}

\def\shortauthors{Zhao and Yu}
\def\runningtitle{From micro to macro of traffic heterogeneity}

\author[a]{Chenguang Zhao}
\author[a,$\ast$]{Huan Yu}
\affil[a]{Thrust of Intelligent Transportation, The Hong Kong University of Science and Technology (Guangzhou)}
\def\corresemail{huanyu@ust.hk}

\def\abstract{Abstract. Most autonomous-vehicles (AVs) driving strategies are designed and analyzed at the vehicle level, yet their aggregate impact on macroscopic traffic flow is still not understood, particularly the flow heterogeneity that emerges when AVs interact with human-driven vehicles (HVs). Existing validation techniques for macroscopic flow models rely on high-resolution spatiotemporal data spanning entire road segments which are rarely available for mixed-autonomy traffic. AVs record detailed Lagrangian trajectories of the ego vehicle and surrounding traffic through onboard sensors. Leveraging these Lagrangian observations to validate mixed-autonomy flow models therefore remains an open research challenge. This paper closes the gap between microscopic Lagrangian data and macroscopic Euclidean traffic models by introducing a continuous traffic-heterogeneity attribute. We represent traffic flow with two coupled conservation laws with one for vehicle number and one for the traffic attribute. Reconstruction methods are designed to derive the traffic attribute from Lagrangian vehicle trajectories.
When abundant trajectory data are available, we characterize the traffic heterogeneity by extracting drivers’ desired speed and local behavioral uncertainty from trajectories. In data-scarce mixed traffic, we design an end-to-end mapping that infers the traffic heterogeneity solely from trajectories in the current spatiotemporal region. 
Experiments across multiple traffic datasets show that the proposed model effectively captures traffic heterogeneity by clustering the fundamental diagram scatter into attribute-based groups. The calibration errors of traffic flow dynamics are also reduce by 20~\% relative to the Aw-Rascle-Zhang model benchmark. 
Detailed analyses further show that the model generalizes well, maintaining nearly the same accuracy when evaluated under a variety of previously unseen traffic conditions.

}

\def\keyword{Mixed Autonomy\\Traffic Heterogeneity\\Autonomous Vehicle\\Traffic Flow Theory}

\begin{document}
\maketitle
\titlepageext

\section{Introduction}

Autonomous vehicles (AVs) have made significant advancements in real-world applications.  In mixed-autonomy traffic, where traditional human-driven vehicles (HVs) share the road with AVs, many research works have focused on the role of AVs in improving traffic systems at the microscopic scale. These studies often regulate the behavior of individual vehicles or inter-vehicle dynamics to optimize traffic characteristics,  such as capacity, stability~\citep{wang2021leading,shi2023deep,stern2018dissipation}, safety~\citep{zhao2023safety,zhao2024leveraging}, emissions~\citep{sun2020optimal,ercan2022autonomous}, and efficiency~\citep{li2022trade,barth2024co,jiang2024dynamic}. As the scale of the system grows from vehicle platoons to road-network traffic, microscopic modeling and design become computationally intensive, necessitating a shift to macroscopic approaches that describe the dynamics of aggregated traffic. Macroscopic traffic flow models are crucial for understanding the broader impact of AVs on mixed-autonomy systems.

\subsection{Flow Modeling for mixed-autonomy traffic}

Most research on mixed-autonomy modeling can be categorized into two approaches. The first approach  treats traffic systems as multi-agent systems and derives a macroscopic limit as the number of agents approaches infinity, by using kinematic limits or the mean-field theory. The second approach directly adopts flow models, such as LWR or ARZ, by adjusting model parameters or dynamic formulations to account for mixed traffic conditions.

The kinematic limit derives a continuum model, to which vehicle particle models converge as the number of vehicles approaches infinity, while maintaining a constant total mass~\citep{chiarello2020micro,dimarco2022kinetic}. The derived model provides rigorous theoretical results and helps identify the microscopic ingredients responsible for macroscopic terms. However, this approach has high analytical complexity, and a general mathematical theory remains unavailable. Consequently, only a few simple feedback-form AV controllers have been adopted to derive the macroscopic limit~\citep{chiarello2021statistical,holden2024continuum,tordeux2018traffic}.
The mean-field game theory assumes that each driver tries to optimize their own objective function decided by the traffic environment, and analyzes macroscopic traffic via the mean-field equilibrium, where no vehicle can reduce its driving cost by unilaterally changing its driving decisions~\citep{huang2020game,li2022equilibrium,mo2024game}. The equilibrium is obtained by solving a coupled backward Hamilton-Jacobi-Bellman (HJB) PDE, which describes the optimal actions of individual vehicles, and a forward Fokker-Planck-Kolmogorov (FPK) PDE, which describes the aggregated traffic flow dynamics. Deriving the equilibrium often involves strong assumptions that may not hold in real-world traffic conditions. For example, drivers are assumed to make decisions based on the position and speed of all other vehicles, whereas both human drivers and autonomous vehicles rely only on neighboring traffic information due to the limited vision and wireless communication ranges, respectively~\citep{li2025influence}. Moreover, solving the coupled forward-backward PDEs imposes a high computational burden.

To avoid the analytical and computational burden, the second approach models mixed-autonomy traffic using existing flow models, by adjusting critical model parameters for mixed traffic, such as   fundamental diagram, based on microscopic AV-controller design such as  AV penetration rate~\citep{qin2021lighthill,vander2020modeling} or controller gains~\citep{martinez2020stochastic}. 
The formulated model retains the structure of classical flow models, enabling the application of established analytical properties and numerical schemes. 
Related studies have adopted the first-order LWR model~\citep{vander2020modeling,qin2021lighthill} and the second-order ARZ model~\citep{aw2000resurrection,zhang2002non,imran2024macroscopic,hui2024anisotropic} to analyze mixed traffic.  The mapping from microscopic to macroscopic levels is clear and straightforward. However, these models treat all vehicles as a homogeneous group and provide limited insight, and sometimes are inconsistent with, how AVs interact with surrounding traffic on microscopic levels. 

To explicitly model the behavioral differences and interactions between AVs and HVs, two-class models have been proposed~\citep{logghe2008multi,zhang2024mean}. Each class's density follows its own continuity equation, while its desired speed depends on the total density. In two-class models, all HVs are categorized to one class and all AVs to the other.   As an extension to incorporate heterogeneity within AVs or HVs, multi-class models have also been proposed~\citep{wong2002multi,ngoduy2007multiclass,levin2016multiclass,qian2017modeling}.  But the analytical complexity and computational burden increase as the number of classes grows. 
Multi-class models represent AVs and HVs by their average attributes and implicitly assume that all AVs are identical, while all HVs share a distinct but uniform behavior. However, by analyzing real trajectory data, we show that the heterogeneity within the AV and HV classes may be even greater than that between the two classes. Therefore, it is inappropriate to impose a strict classification of AVs and HVs.

To overcome the limitations of discrete classification in capturing within-class heterogeneity, we propose to model driver characteristics as continuous attributes.
In this paper, we propose a traffic attribute variable as a bridge between microscopic and macroscopic traffic. We formulate the traffic attribute variable through an advection equation, indicating that it remains invariant along vehicle trajectories. In Lagrangian coordinates, the traffic attribute reflects the driver’s inherent behavioral characteristics, whereas in Euclidean coordinates, it represents the average traffic attributes at a given spatiotemporal location. In previous research, the model has been known as the Generic Second-Order Model (GSOM)~\citep{lebacque2007generic}.   Several physical interpretations have also been proposed for the traffic attribute variable, such as the desired gap~\citep{zhang2009conserved} or the free-flow speed~\citep{fan2014comparative,mo2024game}. However, existing reconstruction methods still operate at the macroscopic level and require complete spatiotemporal data, leading to high computational costs and limited feasibility for real-time applications. 
Moreover, under a macroscopic framework, it remains unclear how microscopic vehicle motions affect macroscopic traffic attributes. In this paper, we design a method to reconstruct the traffic attribute variable  from vehicle trajectories with a low computational cost and high interpretability.

\subsection{Data collection and utilization in mixed-autonomy traffic}

Developing and validating any flow model requires traffic data.  Early-stage data collection methods, such as those used in the PeMS dataset~\citep{choe2002freeway}, rely on road-based radars or loop detectors to measure vehicle flow and occupancy at fixed locations. Despite their ease of implementation, the data provided by these methods lacks detailed information on vehicle motion within a road segment and is thus insufficient for studying traffic wave propagation.  
With advancements in visual sensing techniques, traffic data across an entire road segment over a time period  has been collected using surveillance cameras (e.g., the NGSIM~\citep{NGSIM} and I24 Motion~\citep{gloudemans202324} datasets) or aerial videos from helicopters (e.g., the HighD dataset~\citep{gloudemans202324}). 
These videos capture the position of each vehicle in every frame, which are then used to extract detailed motion trajectories. 
Extracting accurate vehicle trajectories from video images is a challenging and time-consuming task. Besides, potential occlusions caused by bridges or foggy weather can reduce measurement accuracy or even result in data unavailability~\citep{gloudemans202324}.

To overcome the limitations of video-based trajectory extraction, alternative data sources with higher accuracy and real-time availability are gaining increasing attention.
The above datasets provide traffic data in Euclidean coordinates, where sensors are fixed and the collected data covers a spatiotemporal domain across the entire road segment. 
The deployment of automated vehicles is expected to provide abundant traffic data in Lagrangian coordinates.  
AVs are typically equipped with various onboard sensing devices, such as radar and LiDAR sensors, cameras, GPS, and Inertial Measurement Units (IMUs), which provide accurate real-time measurements of of position, speed, acceleration of the ego vehicle and also surrounding vehicles~\citep{liu2024survey,masello2022traditional}. Some researchers and manufacturers have published trajectory data collected by AVs from real traffic environments~\citep{sun2020scalability,zhou2024unified,ammourah2025introduction,makridis2021openacc}.

Lagrangian data directly provides high-accuracy vehicle motion information, thereby significantly facilitating the analysis of microscopic traffic dynamics. 
Several studies have deployed designed controllers on experimental vehicles and tested their performance in terms of  stability~\citep{jin2018experimental,zhou2022significance,hayat2025traffic,wang2023implementation} and safety~\citep{alan2024integrating,batkovic2023experimental}. In addition, some studies utilize commercial autonomous vehicles to collect traffic data or directly analyze open datasets to investigate the impact of AVs on traffic systems. 
Research has analyzed the effects of AVs from various aspects, such as string stability~\citep{gunter2019model,gunter2020commercially,yu2025dynamic}, safety~\citep{zhang2025anticipatory},  mobility~\citep{zhou2024unified,beigi2025impact},  fuel consumption~\citep{mao2024internal}, human's reaction to autonomous driving~\citep{wen2022characterizing}, calibration of AV dynamics~\citep{hu2022processing,li2021car}, and comparison between driving behaviors of AVs and human drivers~\citep{hu2023autonomous,li2024automated}. 
However, since the detection range of onboard sensors is typically restricted to a few nearby vehicles, the collected data fail to cover a complete spatiotemporal domain, which is essential for studying macroscopic mixed-autonomy traffic flow dynamics, i.e., the aggregated impact of AVs on traffic. 

For mixed-traffic modeling, large-scale Euclidean datasets capturing mixed-autonomy traffic remain scarce.  
Although recent efforts have collected AV data, such as the TGSIM dataset~\citep{ammourah2025introduction} and the 100-CAV experiment~\citep{lee2025traffic,ameli2025design}, the impact of AVs remains poorly understood due to their currently low penetration rate. Moreover, even with a higher AV penetration rate in the future, obtaining Euclidean data for mixed-traffic research will remain challenging. It is often impossible to determine whether a vehicle is equipped with autonomous driving capabilities or whether those capabilities are actively in use. In addition, many autonomous vehicles operate in multiple driving modes, and it is nearly impossible to identify the specific mode in use, as detailed driving algorithms are highly confidential. 
Given these challenges,  existing validation methods relying on Euclidean data may not be feasible for mixed-autonomy flow model research.

\subsection{Contribution}

\begin{figure}[!t]
    \centering    \includegraphics[width=0.9\linewidth]{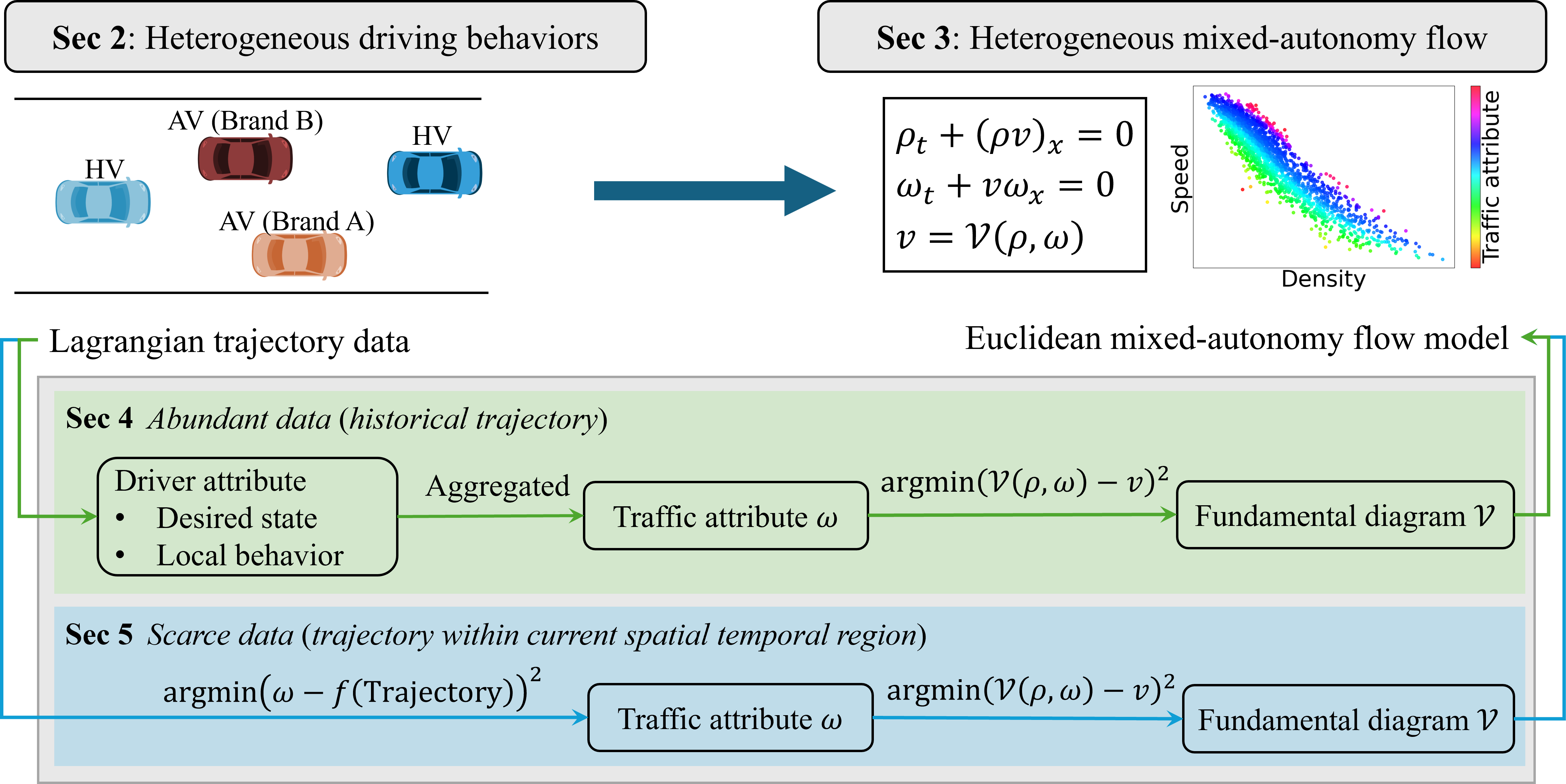}
    \caption{We use heterogeneity as a micro-macro bridge in mixed-autonomy traffic flow modeling. We analyze heterogeneous driving behaviors in~\Cref{sec:micro} and  introduce a traffic-heterogeneity attribute variable to explain how microscopic vehicle behaviors affect macroscopic flow dynamics in~\Cref{sec:macro}. We propose two reconstruction methods to obtain the traffic attribute from vehicle trajectories. When there is abundant data, the traffic attribute is decided from driver attributes that are extracted from trajectories as in~\Cref{sec:reconstruct model}. When there is scare data, we propose an end-to-end mapping to get the traffic attribute directly from trajectories in~\Cref{sec:data driven}. }
    \label{fig:intro}
\end{figure}

{To summarize, the main challenge in mixed-autonomy flow modeling is the gap between Lagrangian data and Euclidean models.}   While traditional flow model validation requires Euclidean data covering a long spatiotemporal period, mixed-traffic systems typically provide Lagrangian data limited to a small spatial range along the vehicle trajectory.  Therefore, a new modeling formulation is required that can obtain macroscopic Euclidean traffic variables directly from microscopic Lagrangian trajectories. In this paper, we bridge the micro–macro gap by characterizing heterogeneity in mixed traffic through a traffic heterogeneity attribute as shown in~\Cref{fig:intro}. To enable real-time application of the proposed model,  we further formulate a mapping from vehicle trajectories to the traffic heterogeneity attribute.

The main contribution of this work lies in bridging the gap between the scarcity of mixed-autonomy Euclidean traffic data and the challenge of validating corresponding flow models. We propose a mathematical formulation for deriving macroscopic traffic heterogeneity attributes from Lagrangian vehicle trajectories. 
The proposed formulation serves as a reconstruction method for generating Euclidean traffic variables from Lagrangian data, providing a foundation for both developing and validating mixed-autonomy macroscopic traffic flow models. 
Specifically, we cover  microscopic-level analysis and macroscopic-level validation:  
\begin{itemize}
    \item Section~\ref{sec:micro}: Analyzing heterogeneous attributes of AVs and HVs at the microscopic level.
    We analyze real AV and HV trajectories and identify two types of driver attributes that remain approximately constant during the driving process: the driver's expected traffic state and local behavioral uncertainty. We find that heterogeneity exists not only between AVs and HVs, but also within each of these classes. Moreover, the heterogeneity of two individual vehicles may exceed the difference between the AV-class average  and HV-class average. 
    \item Section~\ref{sec:macro}: Modeling mix-autonomy traffic via a continuous traffic-attribute variable. The heterogeneity within AV and HV classes implies that a clear-cut classification between them is improper, which motivates us to use  a continuous traffic heterogeneity attribute variable  to model mixed-autonomy dynamics. We describe dynamics of  the traffic attribute variable via an advection equation, which gives the conservation law of density-weighted attribute. Therefore, the mixed-autonomy flow model is formulated by two conservation laws: vehicle number and driver attribute. 
    \item Section~\ref{sec:reconstruct model}: Reconstruction of traffic attributes from vehicle trajectories. We propose to obtain the macroscopic traffic attribute variable from the two microscopic-level terms:   drivers’ expected states and local behavioral patterns.     The proposed formulation establishes a link from Lagrangian trajectory data to Euclidean traffic attributes, allowing us to analyze how vehicles’ microscopic behaviors influence macroscopic flow dynamics. Using real traffic datasets, we demonstrate that the proposed method yields lower modeling error in traffic flow prediction. 
    \item Section~\ref{sec:data driven}: Data-driven reconstruction with scarce trajectory data. 
    We consider more practical yet challenging scenarios where only limited or no historical trajectory data is available to calibrate drivers’ expected states.
    We first demonstrate that the driver’s local behavioral uncertainty, derived solely from current trajectories, remains approximately constant and thus serves as a driver attribute.
     We then employ a data-driven method to construct an end-to-end mapping from microscopic trajectories to macroscopic traffic attributes. Through experiments on multiple open-source traffic datasets, we show that the proposed method reduces calibration error from 20~\% to 2~\% and significantly closes the micro–macro gap.
     Further analysis validates that the proposed framework has a strong generalization ability and accurately captures traffic attributes under previously unseen  traffic conditions.  
\end{itemize}

\section{Microscopic Heterogeneity as  Driver Attribute}\label{sec:micro}

In this section, we extract  microscopic-level vehicle motion attributes from real AV and HV trajectory data. We identify two driver-related attributes: a driver’s desired speed, which reflects their preferred traffic state, and the variance in driving behavior, which captures their stochasticity.
 We show that there exists heterogeneity in the driving behaviors of the AV-class and HV-class. Besides, individual  AVs and HVs also exhibit heterogeneous driving behaviors.

\subsection{Driver attribute: desired speed}

The longitudinal driving decisions of human drivers are usually described by car-following models, which express the acceleration $\dot{v}$ as a function of speed $v$, gap $s$, and leading vehicle speed $v_{\mathrm{leader}}$. 
In this paper, we adopt the simple yet widely used  optimal velocity model (OVM): 
\begin{align}\label{eq:OVM}
	\dot{v}  = \frac{V_{\mathrm{opt}}(s) - v}{\tau} + \beta (v_{\mathrm{leader}} - v),
\end{align}
where $V_{\mathrm{opt}}$ is the driver's desired speed dependent on its gap with the leader vehicle, $\tau$ is the adaption time to the desired speed, and $\beta$ is the sensitivity parameter to the speed difference. We take the desired speed as:
\begin{align}\label{eq:micro Vopt}
	V_{\mathrm{opt}}(s) = \max \left\{ 0, \min \left\{ V^{\mathrm{f}}, \frac{s-s^{0}}{T} \right\} \right\},
\end{align}
with  $V^{\mathrm{f}}$ being the free-flow speed, $T$ being the time gap, and $s^{0}$ being the minimum gap. The OVM clearly describes how a driver reacts to the three traffic states: its own speed $v$, the leader speed $v_{\mathrm{leader}}$, and the gap $s$. The first term reflects how the driver adapts its acceleration based on gap, and the second term reflects how the driver adapts its acceleration based on the speed difference. Unlike human drivers, whose driving decisions are hard to be accurately captured, the motion of autonomous vehicles is governed by well-designed controllers.  However, the detailed controller algorithms in commercial AVs are often highly confidential. Therefore, we also calibrate  an OVM car-following model as in~\Cref{eq:OVM} for autonomous vehicles. 

Since we focus on the longitudinal dynamics in this section, we select the following datasets collected in varied car-following scenarios: Ring by~\citep{zheng2021experimental}, CATS collected by~\citep{shi2021empirical}, OpenACC from~\citep{makridis2021openacc}, and CentralOhio in~\citep{xia2023automated}. The first dataset is collected from an experiment involving 40 human drivers on an 800-meter ring road. There is only one lane on the road, which eliminates the influence of lane-changing. The other three datasets are collected from AVs equipped with adaptive cruise control (ACC) operating in real traffic scenarios. We use the processed data provided in~\citep{zhou2024unified}, which have extracted the trajectories of cruising behaviors.   In Appendix A, we give the detailed introduction of these datasets and also the distribution of collected speed and gap information from the datasets. These datasets cover  a wide range of traffic scenarios from free to congested, with the speed ranging from zero (fully stop) to 50 m/s (free flow) , and the gap ranging from to 5 m to 100 m.

\begin{figure}[!t]
	\centering
	\subcaptionbox{$\tau$}{\includegraphics[width=0.3\linewidth]{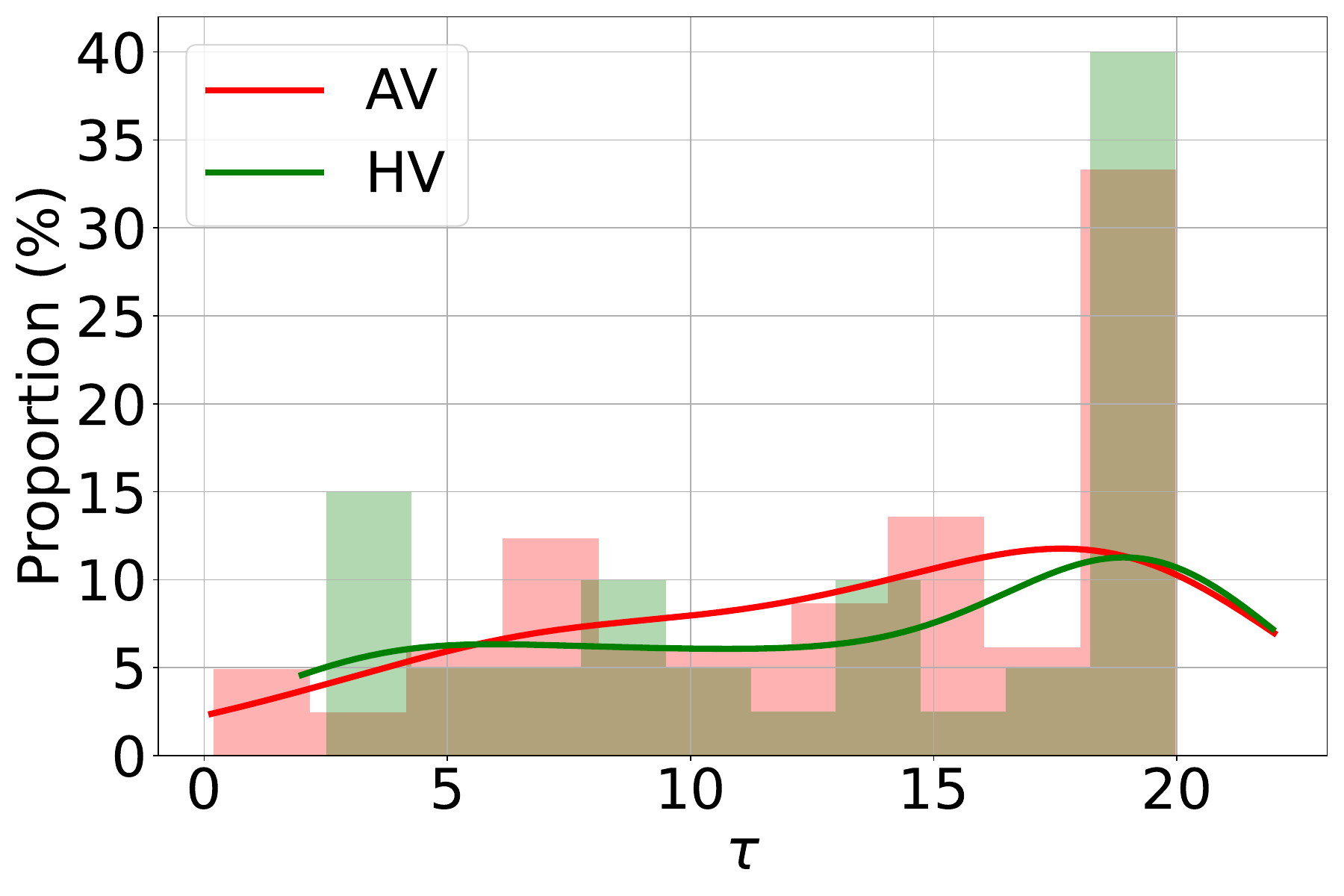}}
	\subcaptionbox{$\beta$}{\includegraphics[width=0.3\linewidth]{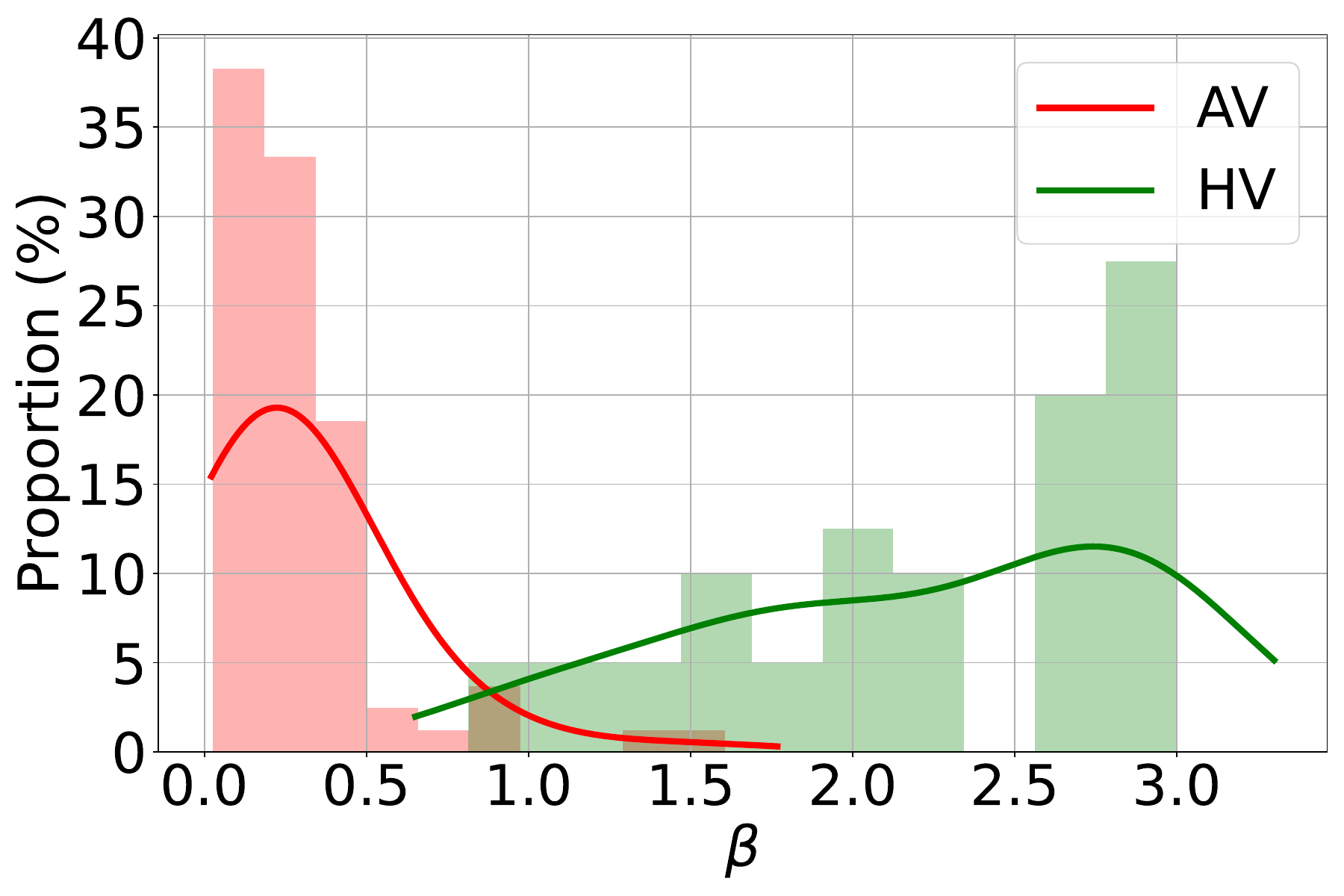}}\\
	\subcaptionbox{$s^0$}{\includegraphics[width=0.3\linewidth]{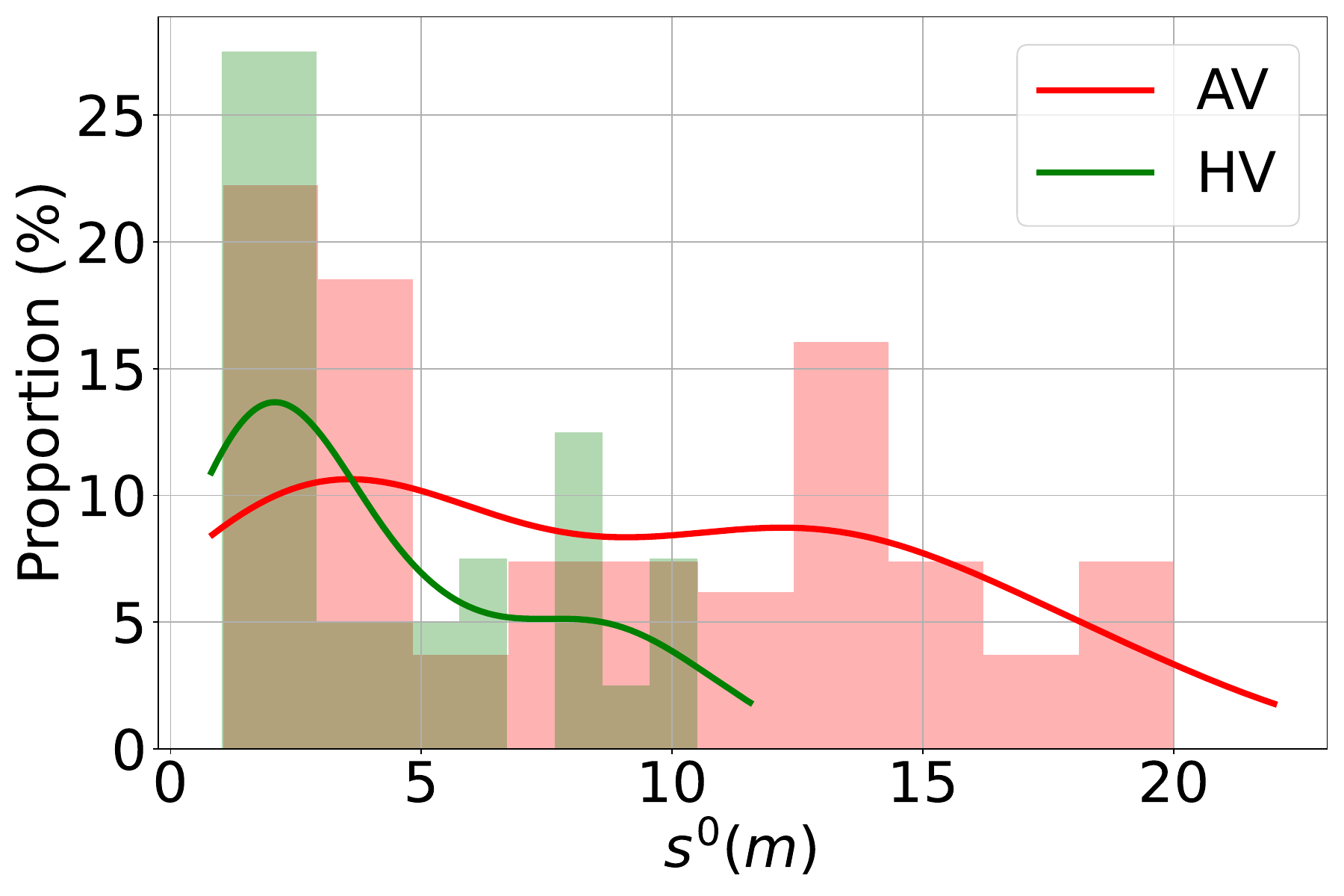}}
	\subcaptionbox{$T$}{\includegraphics[width=0.3\linewidth]{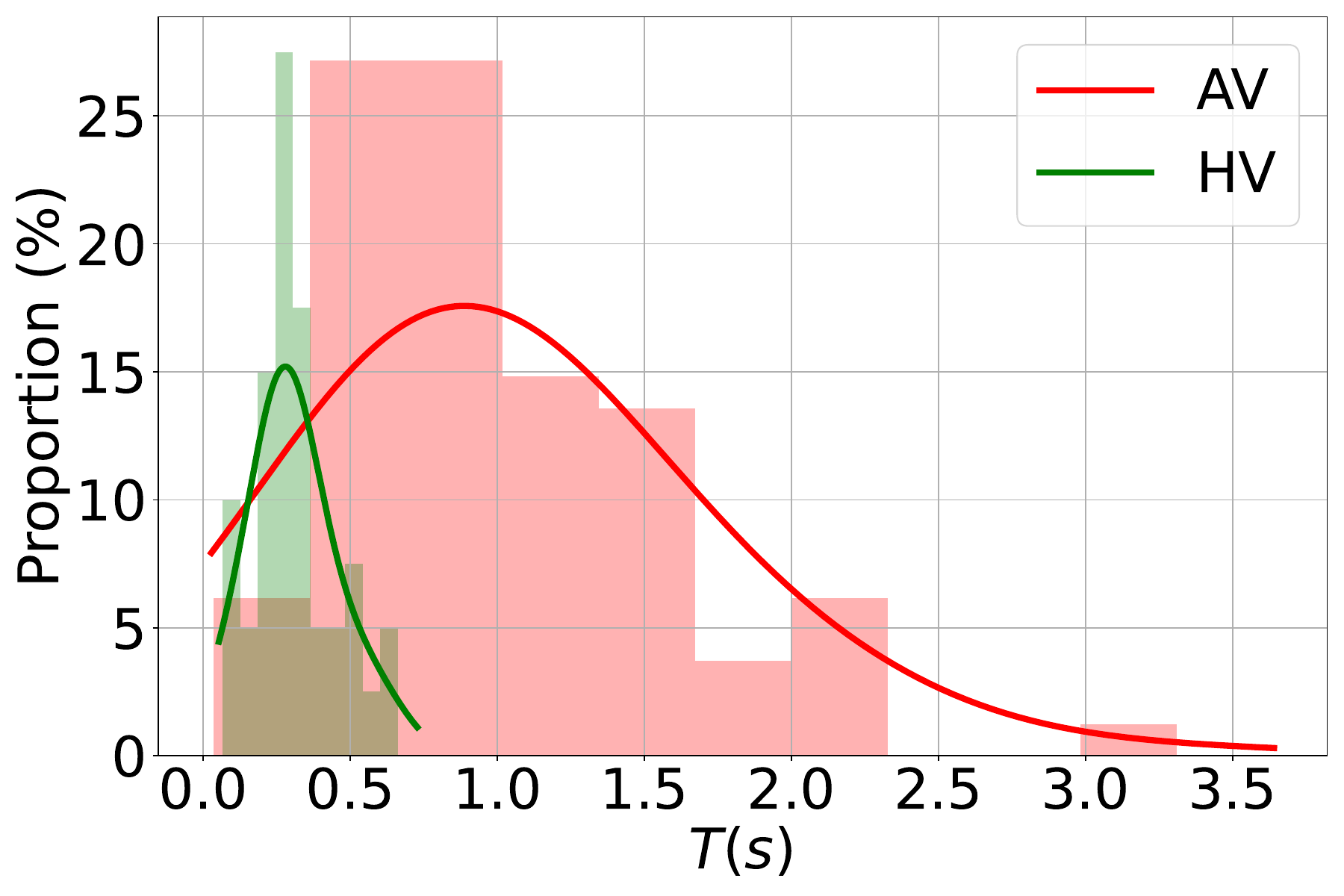}}
	\subcaptionbox{$v^{\mathrm{f}}$}{\includegraphics[width=0.3\linewidth]{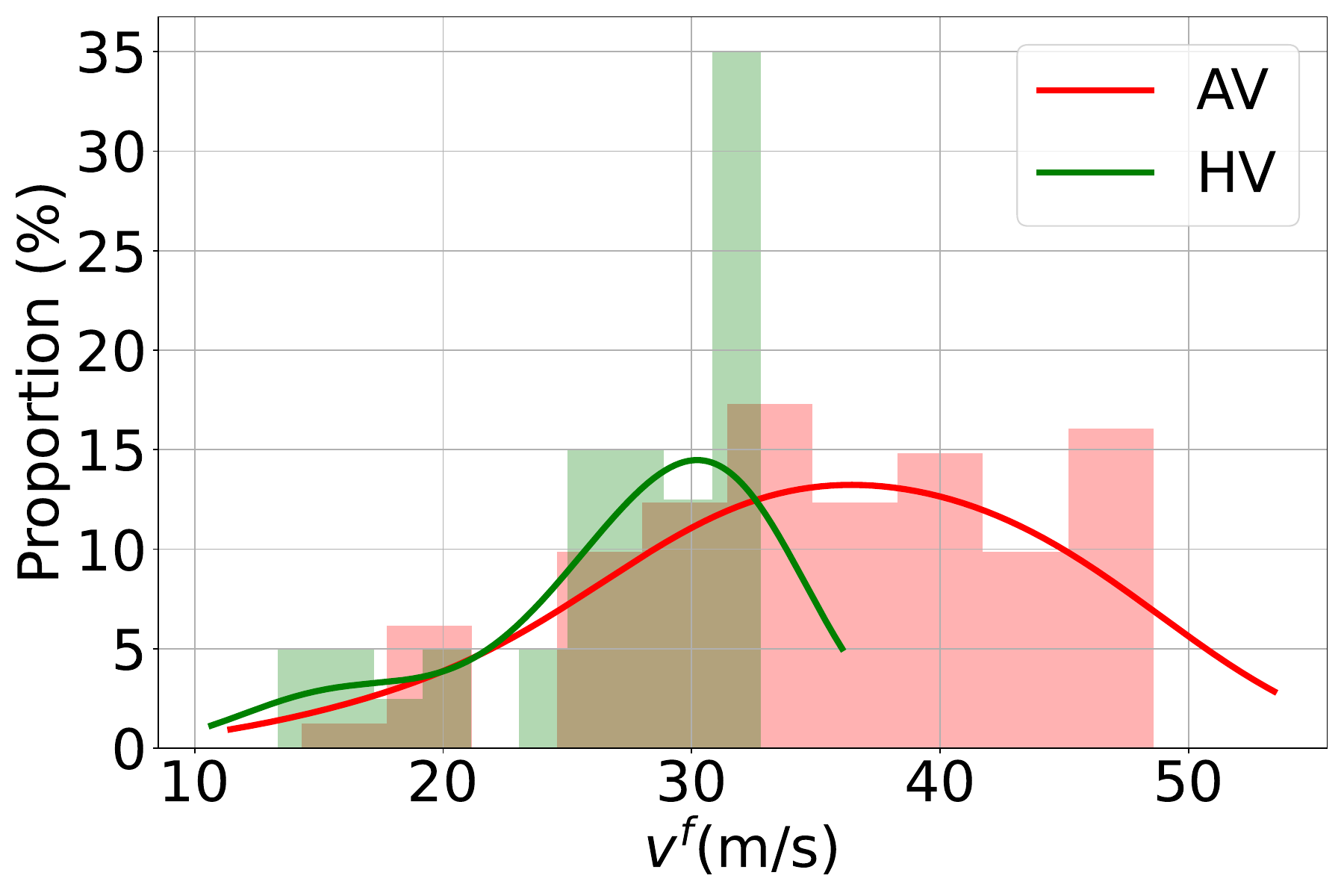}}
	\caption{Distribution of calibrated car-following model parameters for HV and AV. There exists heterogeneity not only in the AV-class and HV-class, but also within both classes.} \label{fig:micro car following parameters}
\end{figure}

We calibrate five parameters: $\tau$, $\beta$, $V^{\mathrm{f}}$, $T$, and $s^{0}$, from collected AV and HV trajectories. 
The detailed calibration algorithms are given in Appendix B. We give the distribution of calibrated parameters in~\Cref{fig:micro car following parameters}. For AVs, we give the distribution of calibrated parameters for all AVs in the three AV datasets. From the parameter distribution, we conclude the following two findings:
\begin{itemize}
    \item When HVs and AVs are considered as two distinct classes, significant differences can be observed between them across all five parameters. This observation is consistent with prior findings on the heterogeneity between AVs and HVs~\citep{zhou2024unified}. 
    For example, in the distribution of the speed sensitivity parameter $\beta$ shown in~\Cref{fig:micro car following parameters}(b), the majority of AVs have $\beta < 1$, whereas most human drivers have $\beta > 1$. 
    Similarly, for the stopping distance $s^0$ in~\Cref{fig:micro car following parameters}(c), some AVs adopt a conservative driving policy with $s^0$ values around 15–20 meters, while most human drivers have $s^0$ values below 5 meters.  
    \item In addition to the heterogeneity between the HV class and the AV class, we also observe heterogeneity within each class. It is well understood that human drivers exhibit considerable variation in model parameters. Take the same example  in~\Cref{fig:micro car following parameters}(b),  human drivers have speed sensitivity parameters ($\beta$) ranging from 1.0 to 3.0. Similarly, commercial AVs also exhibit noticeable variation in microscopic parameters. 
    For example, the distribution of the gap-sensitivity parameter $\tau$ in~\Cref{fig:micro car following parameters}(a) shows that both AVs and HVs have values ranging from 5 to 20. 
\end{itemize}    
The heterogeneity of each AV and each HV implies that it is improper to describe  mixed-autonomy traffic via 
multi-class models that takes AV as a class and HV as another class, and motivates us to propose continuous traffic-heterogeneity attribute variable in this paper.

\subsection{Driver attribute: stochasticity}

\begin{figure}[!t]
	\centering
	One example AV\\
	\subcaptionbox{Scatter plot of $C^{\mathrm{absave}}$ and $C^{\mathrm{std}}$}{\includegraphics[width=0.4\linewidth]{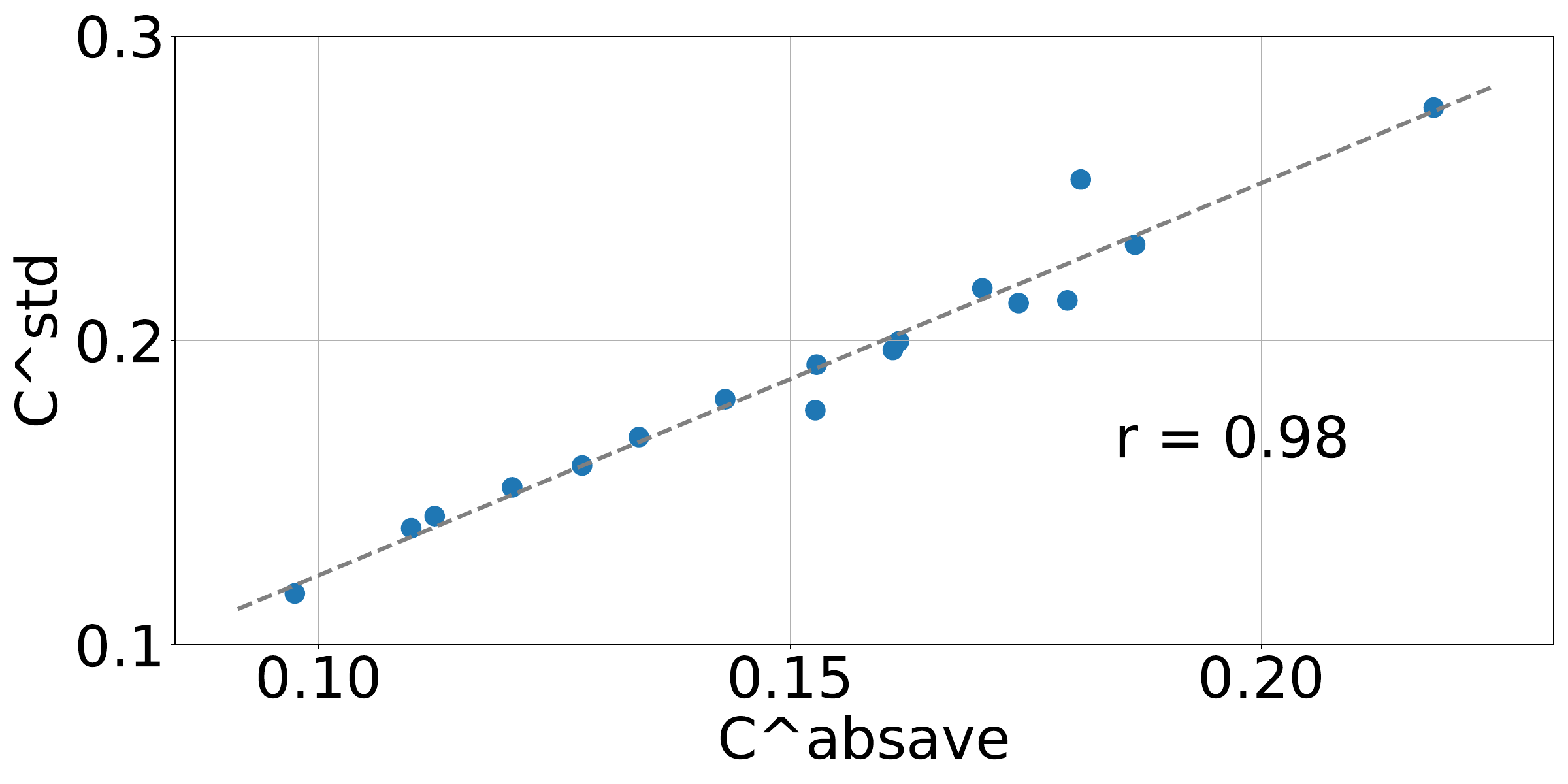}}
	\subcaptionbox{$\omega_k^{\mathrm{L}}$}{\includegraphics[width=0.4\linewidth]{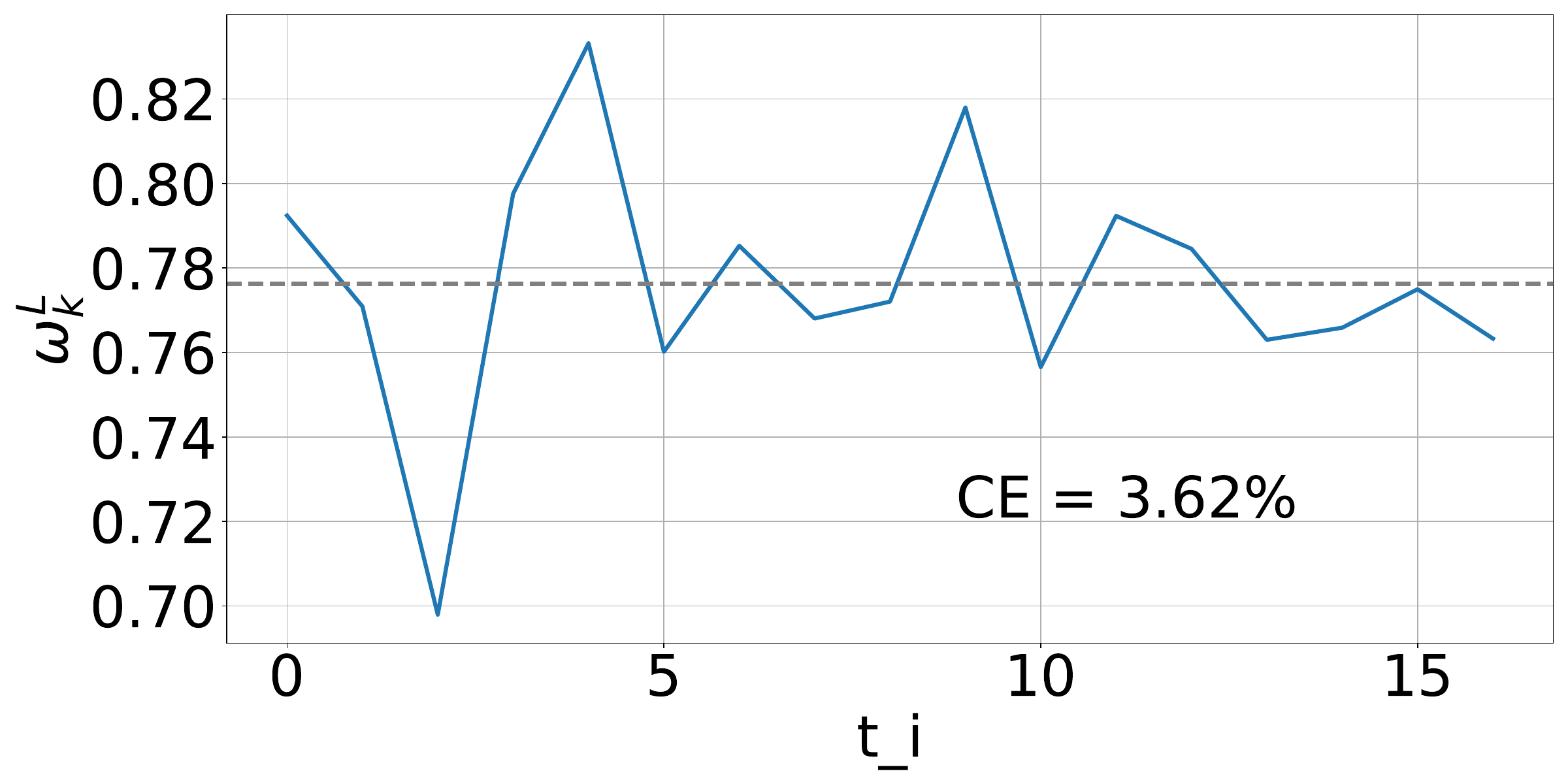}}
	\\
	One example HV \\
	\subcaptionbox{Scatter plot of $C^{\mathrm{absave}}$ and $C^{\mathrm{std}}$}{\includegraphics[width=0.4\linewidth]{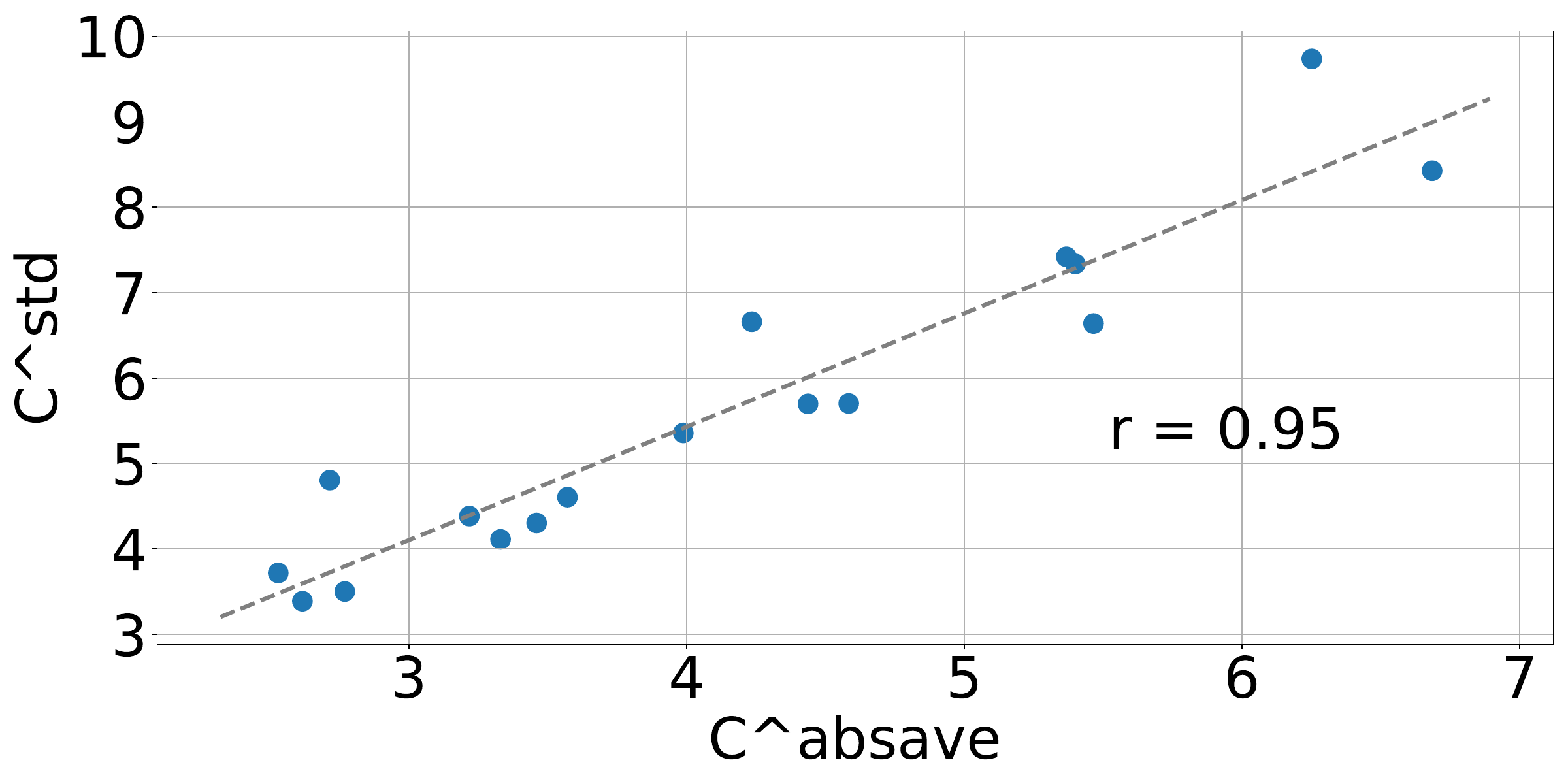}}
	\subcaptionbox{$\omega_k^{\mathrm{L}}$}{\includegraphics[width=0.4\linewidth]{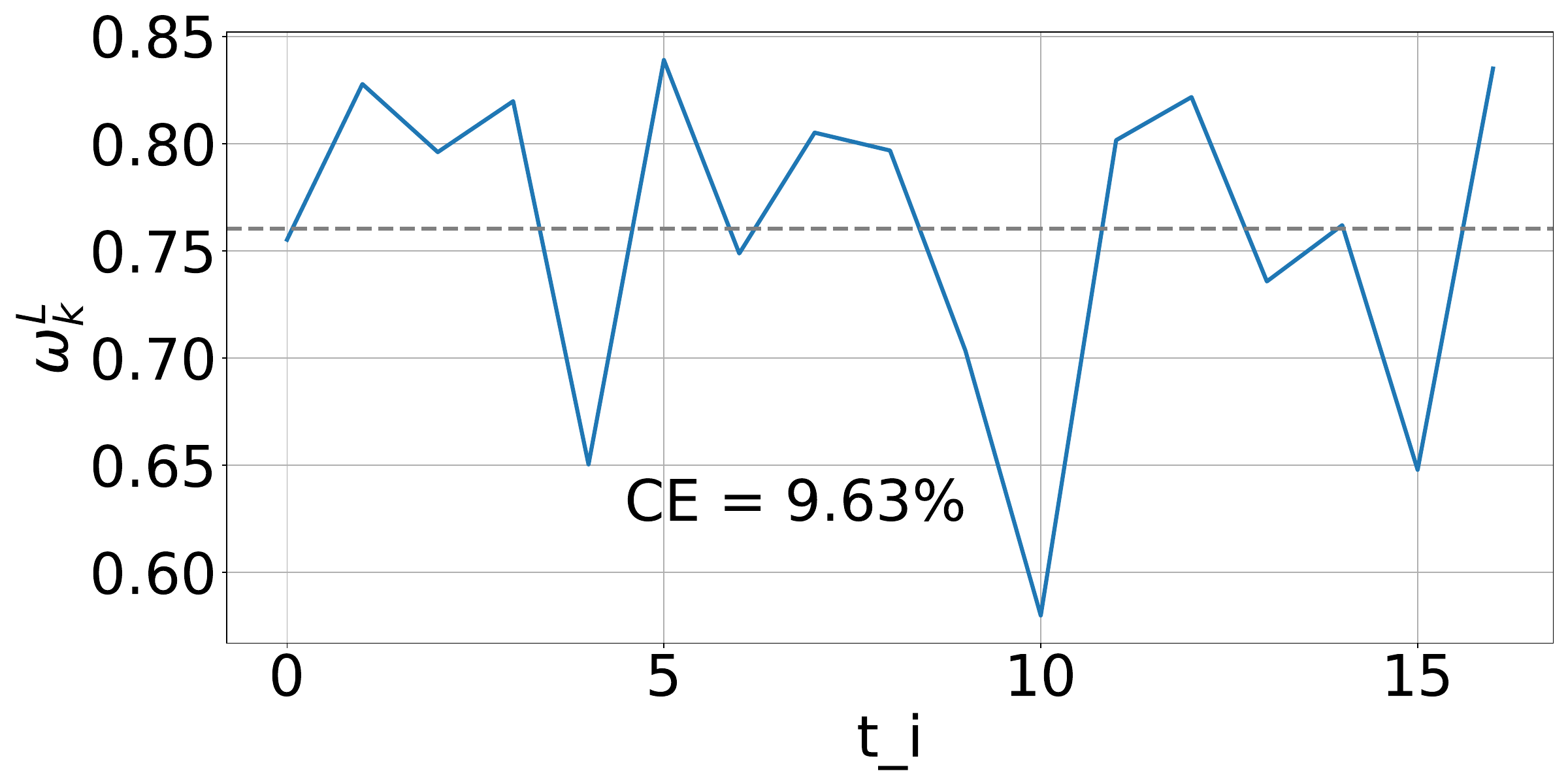}}
	\caption{The stochasticity-related vehicle attribute constructed from jerk profiles. There exists a linear relationship between the $C^{\mathrm{absave}}$ and $C^{\mathrm{std}}$ as shown in (a)(c). And we calculate a vehicle-attribute value $\omega^{\mathrm{L}}$, which has a low constancy error and remains almost constant during the driving process.  We give the result for one example AV and example HV in the CentralOhio dataset here. We further show in~\Cref{fig:jerk} that the proposed $\omega^{\mathrm{L}}$ also remains a low constancy error for all vehicles over other datasets.}\label{fig:jerk example}
\end{figure}

While drivers have a desired speed, they also exhibit uncertainty in practice. We propose a jerk-based metric to represent the stochastic characteristics of drivers and demonstrate that it remains approximately constant during the driving process.

For  a vehicle $k$, we divide it trajectory into $n$ segments $\mathcal{T}_{1},\mathcal{T}_{2},\cdots,\mathcal{T}_{n}$. We consider the trajectory during the time segment $i$,  and calculate the average absolute jerk:
\begin{align}
	 C_{k}^{\mathrm{absave}}(i)  = \frac{1}{T_{i}} \sum_{t\in \mathcal{T}_{i}} |j_{k}(t)|,
\end{align}
and the standard deviation of its jerk:
\begin{align}
	 C_{k}^{\mathrm{std}}(i) = \sqrt{\frac{1}{T_i} \sum_{t\in \mathcal{T}_i}  (j_{k}(t)-C_{k}^{a}(i))^2},
\end{align}
where $T_i$ is the number of sample data in the time segment, $j_k(t)$ is vehicle $k$'s jerk at time $t$, and $C_{k}^{a} (i) = \frac{1}{T_{i}} \sum_{t\in \mathcal{T}_{i}} j_{k}(t)$. We find that for both AVs and HVs, there is a linear relationship between $C_{k}^{\mathrm{absave}}$ and $C_{k}^{\mathrm{std}}$, as shown in~\Cref{fig:jerk example}(a) and in~\Cref{fig:jerk example}(c) respectively.

\begin{figure}[!t]
    \centering
    OpenACC \\
    \subcaptionbox{Constancy error}{\includegraphics[width=0.4\linewidth]{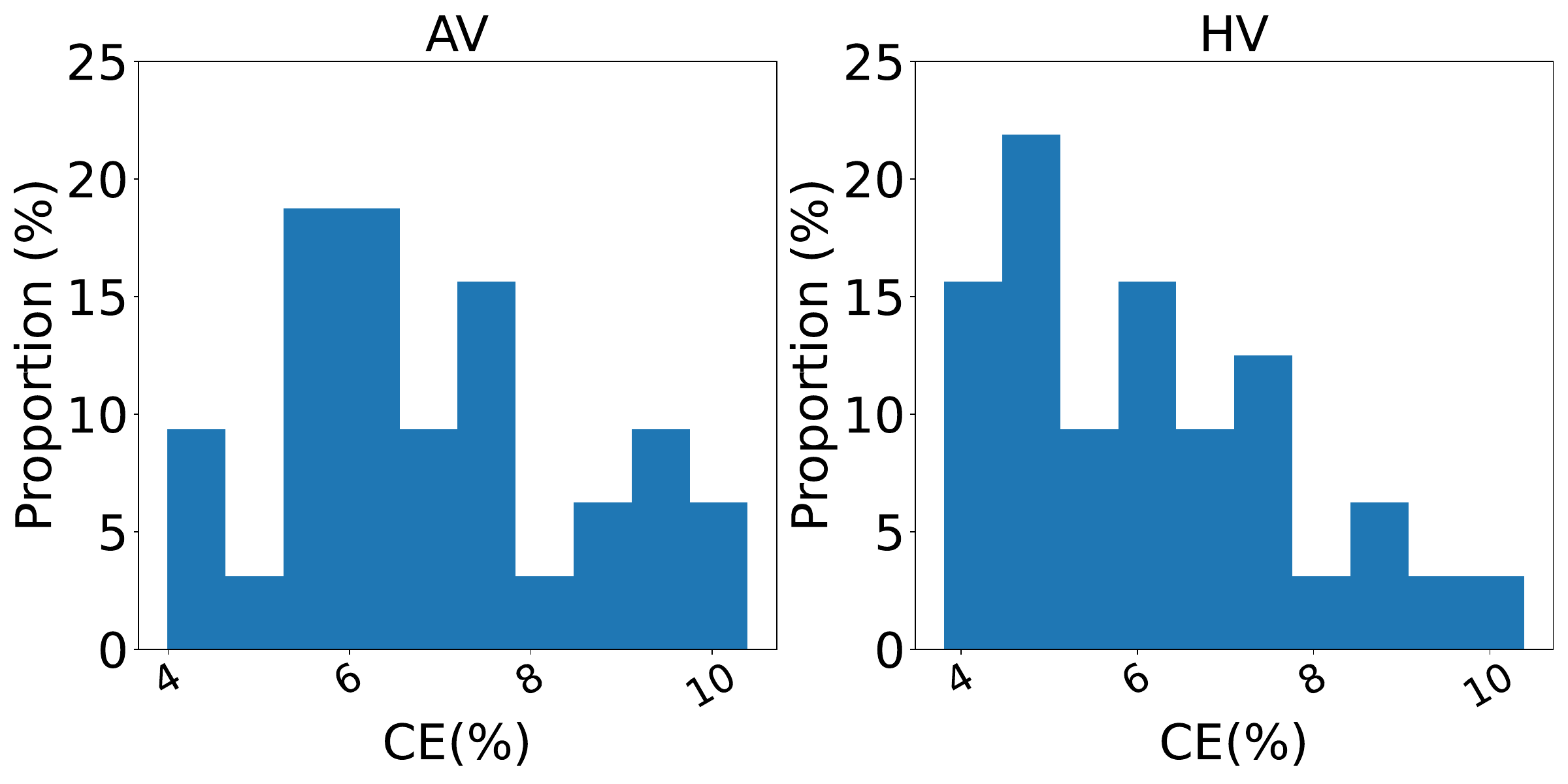}}
    \subcaptionbox{Driver attribute}{\includegraphics[width=0.4\linewidth]{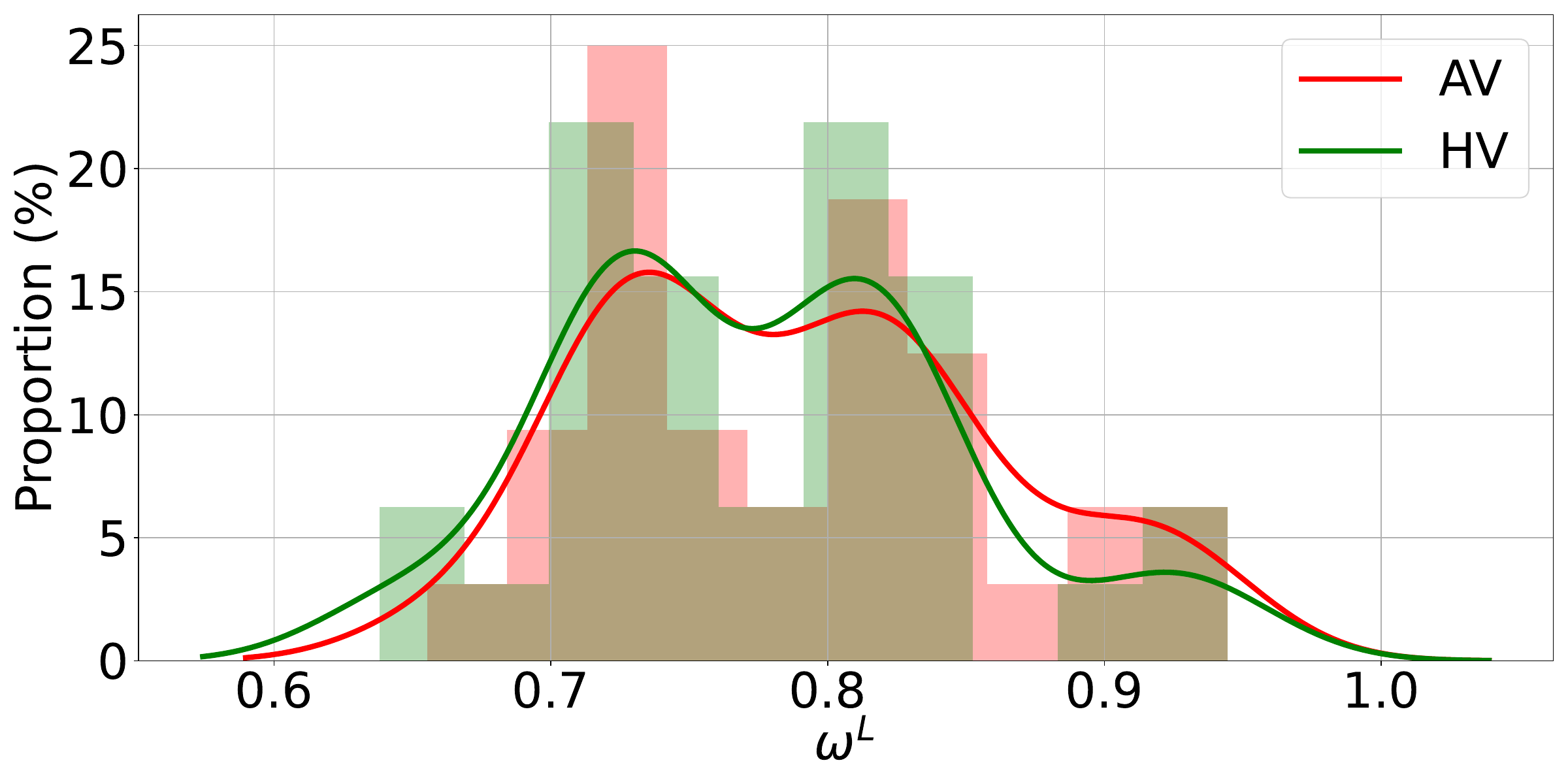}}
     \\
    CATS \\
    \subcaptionbox{Constancy error}{\includegraphics[width=0.4\linewidth]{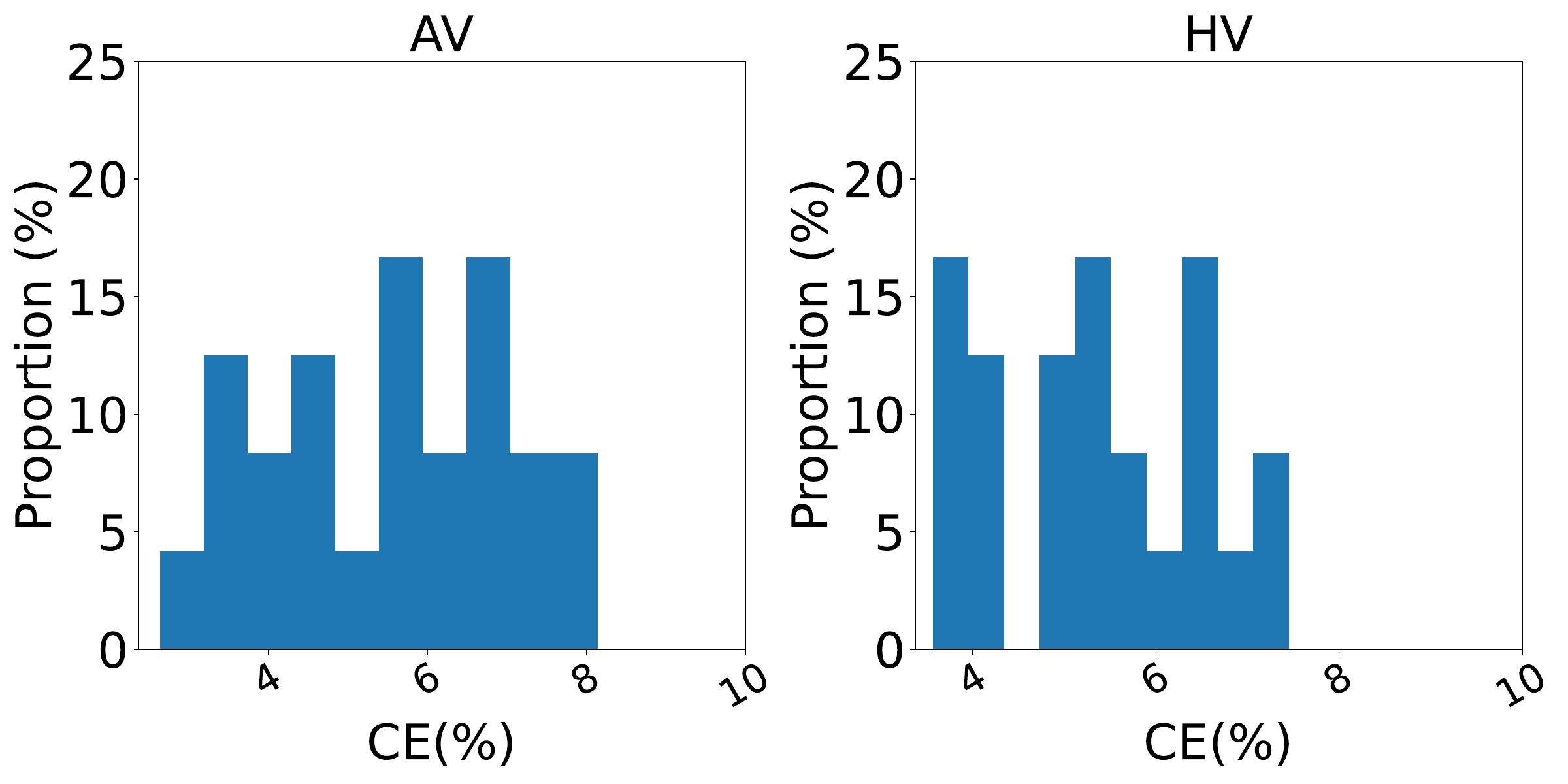}}
    \subcaptionbox{Driver attribute}{\includegraphics[width=0.4\linewidth]{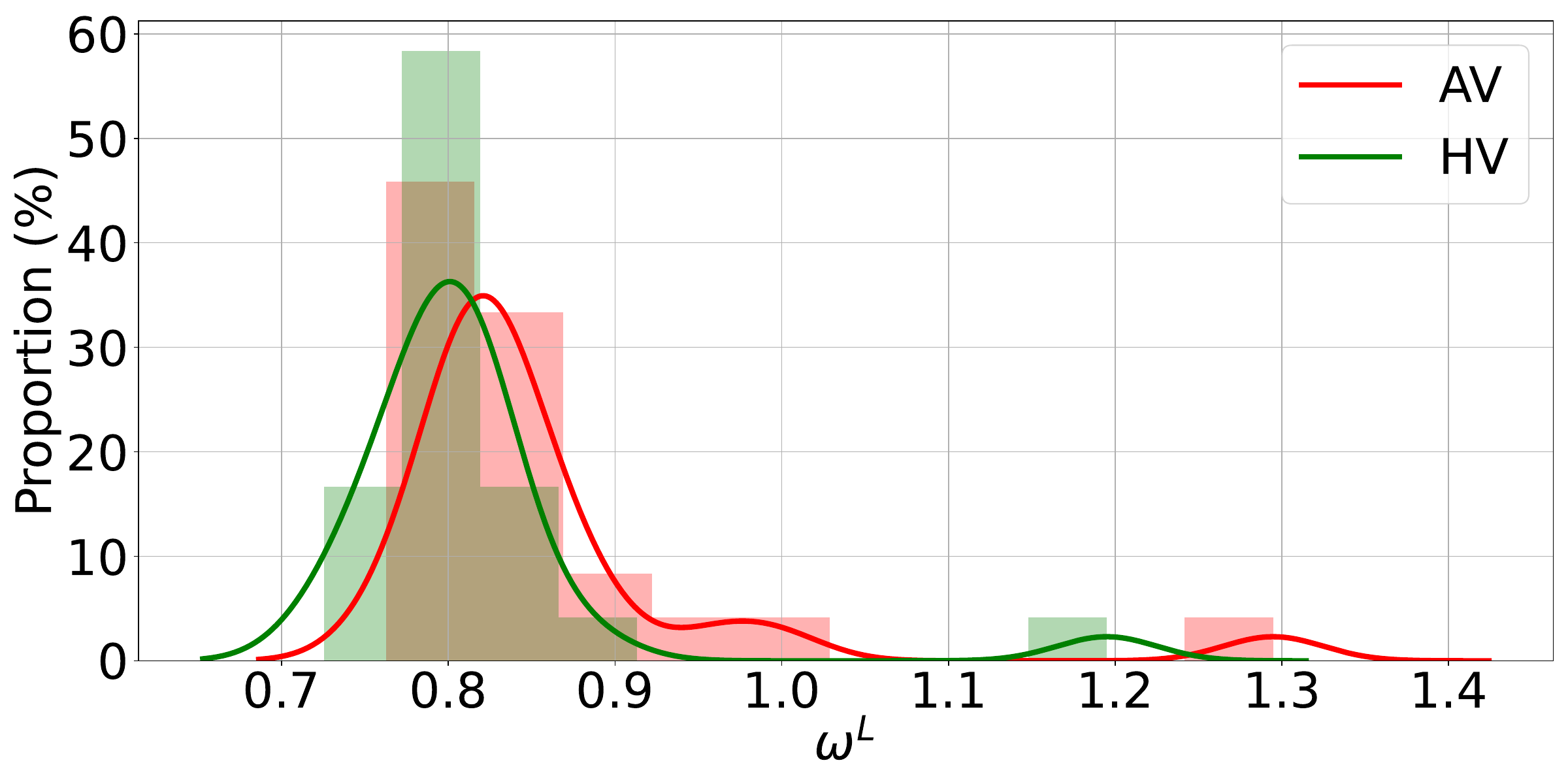}}
     \\
    CentralOhio \\
    \subcaptionbox{Constancy error}{\includegraphics[width=0.4\linewidth]{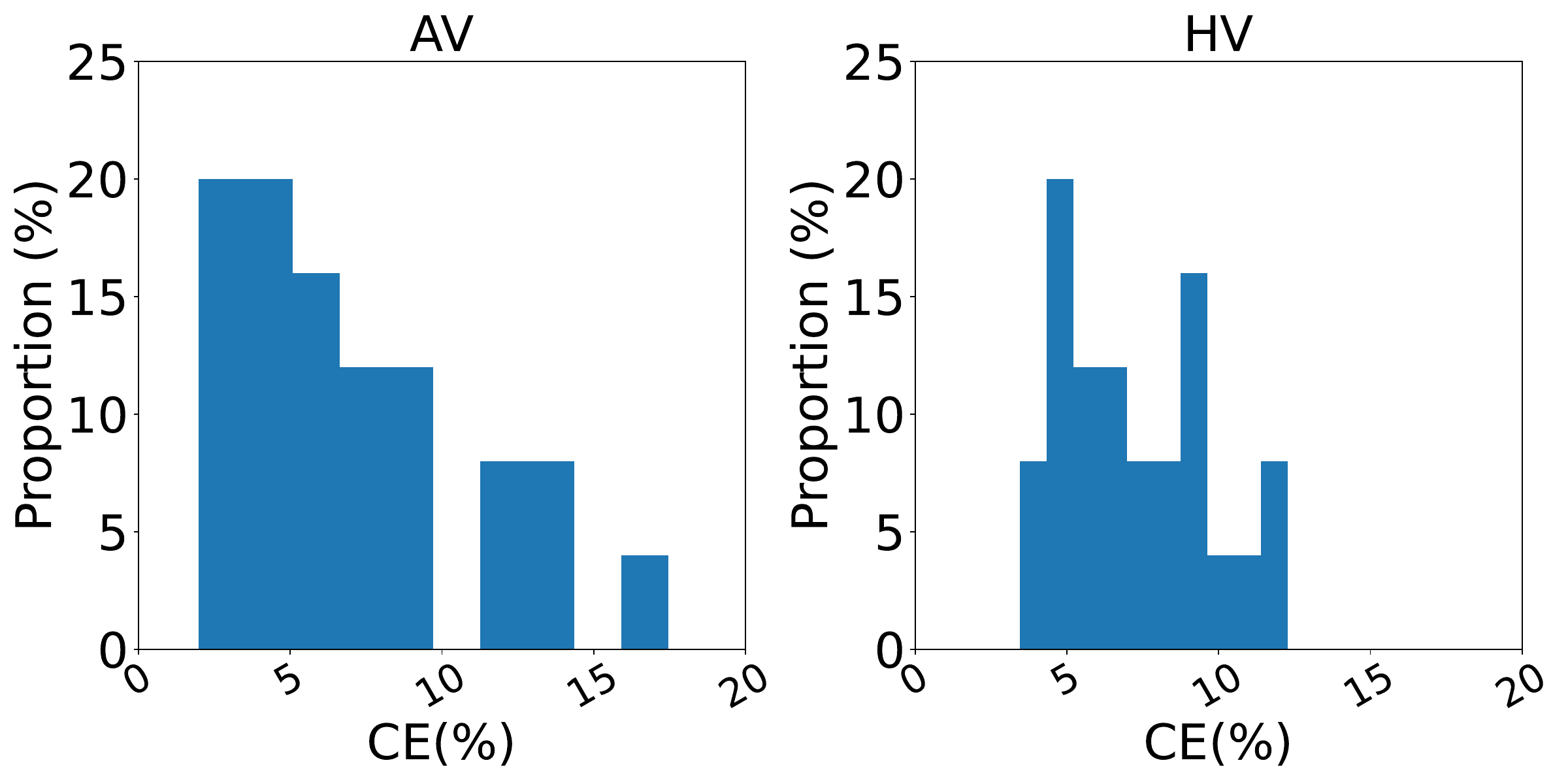}} \subcaptionbox{Driver attribute}{\includegraphics[width=0.4\linewidth]{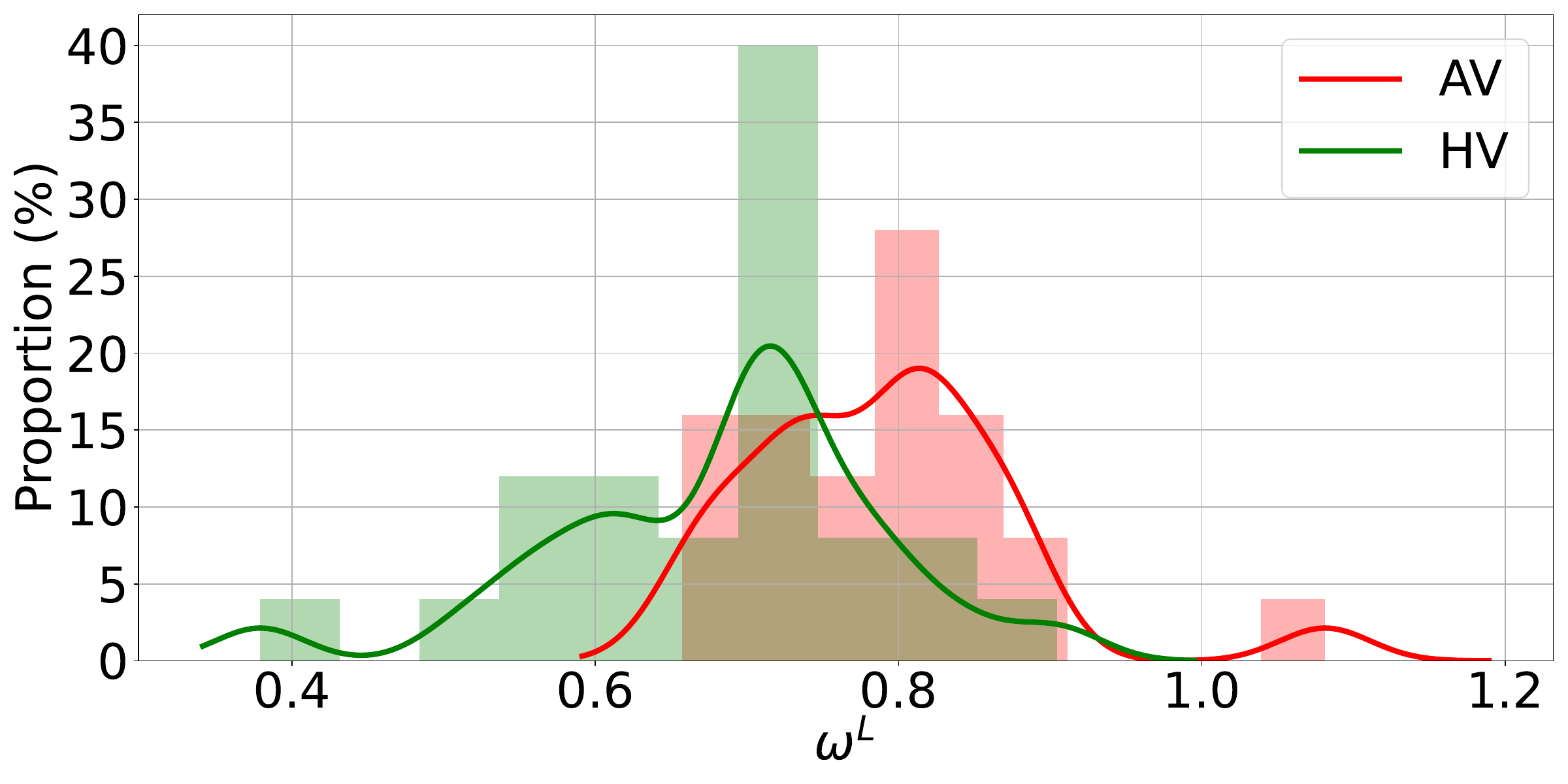}}
    
	\caption{We extract the driving stochasticity-related attribute from the jerk profile for the three datasets. The first column gives the distribution of constancy error (CE) of the proposed vehicle-attribute $\omega^{\mathrm{L}}$. Most vehicles have CE less than 10~\%, which indicates that the $\omega^{\mathrm{L}}$ remains almost constant and is  a metric of vehicle attributes. The right column compares  the distribution of $\omega^{\mathrm{L}}$ for HVs and HVs. As the figures show, there is both inter-class heterogeneity (when we treat AVs as a class and HVs as another class)  and intra-class heterogeneity (when we analyze AVs or HVs individually). }\label{fig:jerk}
\end{figure}

We solve a linear regression equation by the least-squares method as:
\begin{align}
	C^{\mathrm{std}}_{k} (i) = a_k+ C^{\mathrm{absave}}_{k } (i)  b_k.
\end{align}
For the  example AV and HV  given in~\Cref{fig:jerk example}, the correlation coefficients are 0.98 and 0.95 respectively, which indicates a strong linear correlation between the two jerk-related values. Based on the solved regression coefficient $a_k$, we further define  vehicle-$k$'s attribute at  each time segment $i$ as:
\begin{align}\label{eq:omega L jerk}
	\omega^{\mathrm{L}}_{k}(i)= \frac{C^{\mathrm{std}}_k (i) - a_k}{C^{\mathrm{absave}}_i(i) }.
\end{align}
We find that the value of $\omega^{\mathrm{L}}_k $ remains approximately constant during driving, as shown in~\Cref{fig:jerk example}(b) and~\Cref{fig:jerk example}(d). To quantitatively evaluate how much the $\omega^{\mathrm{L}}_k $ varies during the driving process, we define a \textit{constancy error} as:
\begin{align}
	CE_{k} = \frac{1}{n} \sum_i \left(\frac{\omega^{\mathrm{L}}_k (i) -  \omega^{\mathrm{L}}_k} { \omega^{\mathrm{L}}_k} \right)^2 \times 100\%,
\end{align}
with $\omega^{\mathrm{L}}_k = \frac{1}{n}\sum_{i}\omega^{\mathrm{L}}_k (i)$  being vehicle-$k$'s attribute value averaged over the whole driving process. 
The $CE$ for the example AV and HV are $3.62\%$ and $9.63 \%$ respectively. To demonstrate that the proposed jerk-based value $\omega^{\mathrm{L}}$ is a vehicle attribute for all vehicles, we give in~\Cref{fig:jerk}(a)(c)(e)  the  constancy  error $CE_k$ for all AVs and  HVs in the three AV datasets. We see that the  constancy error is distributed between $2\%$ and $20\%$, which indicates that for both AVs and HVs, the extracted $\omega^{\mathrm{L}}_k$ by~\Cref{eq:omega L jerk} has only a small variation during driving, and thus can be regarded as a driver attribute.

We now analyze the differences in driving behavior between AVs and HVs. 
\Cref{fig:jerk}(b)(d)(e) present the distribution of $\omega_k^{\mathrm{L}}$ for all AVs and HVs. The results reveal both inter-class heterogeneity (between AVs and HVs) and intra-class heterogeneity (among AVs or HVs individually).
\begin{itemize}
    \item When treating AVs and HVs as two distinct groups, they exhibit different distribution patterns. For example, in the CentralOhio dataset shown in~\Cref{fig:jerk}(f), most human drivers have $\omega_k^{\mathrm{L}} \le 0.7$, whereas a majority of AVs have $\omega_k^{\mathrm{L}} \ge 0.8$. Similarly, in the CATS dataset~\Cref{fig:jerk}(d), HVs exhibit values around 0.8, while the AV group centers from  0.8 to 0.9. 
    \item At the individual level, the behavioral differences between two AVs or two HVs may be even larger the average difference between the two classes.     For instance, in the OpenACC dataset shown in~\Cref{fig:jerk}(b), the class averages for AVs and HVs are both around 0.7. However, given the full distribution range from 0.6 to 1.0, the difference between two individual vehicles can be greater than the class-level average difference. These findings suggest that it is inappropriate to simply classify AVs and HVs into discrete groups. Consequently, traditional multi-class traffic flow models are insufficient for capturing such within-class heterogeneity.
\end{itemize}
In conclusion, the analysis on the microscopic level highlights that the heterogeneity between individual vehicles, regardless of AVs or HVs, can exceed the average inter-class heterogeneity, which motivates the use of a continuous traffic-heterogeneity attribute over discrete multi-class models. With access to more diverse and large-scale trajectory data in the future, the conclusion can be further validated and potentially refined.

\section{Macroscopic Heterogeneity as Traffic Attribute }\label{sec:macro}
We have demonstrated that AVs and HVs present heterogeneous microscopic behaviors. To  formulate the mixed-autonomy macroscopic traffic flow, we introduce a traffic-heterogeneity variable and use a generic second-order flow model to describe the macroscopic-level heterogeneity. To  verify the formulation, we run simulations to show that the microscopic-level heterogeneity leads to scatter in the fundamental diagram, i.e., macroscopic-level heterogeneity. And thus the heterogeneity is a link between micro to macro traffic.

In macroscopic traffic flow modeling, the three basic variables are density $\rho$, flow $q$, and traffic speed $v=q/\rho$. The density and flow are defined as the number of vehicles per unit length and per unit time respectively. By the definition, these variable provides information on `how many' vehicles, but not `what type' of vehicles. To reflect the vehicle behaviors at a macroscopic level, we introduce a traffic-attribute variable  $\omega$, and formulate the traffic flow as:
\begin{align}
    \partial_t \rho  + \partial_x (\rho  v) &= 0, \label{eq:GSOM rho} \\
    \partial_t \omega   + v \partial_x \omega &= 0, \label{eq:GSOM omega} \\
    v &= \mathcal{V}(\rho,\omega), 
\end{align}
where traffic speed $v(t,x)$ is determined by the density $\rho(t,x)$ and traffic attribute $\omega(t,x)$ through the two-variable fundamental diagram $\mathcal{V}(\rho,\omega)$. This model is refereed to as the generic second-order model (GSOM) in  related work~\citep{lebacque2007generic}.   
To ensure well-posedness of the model, we require that the  traffic speed is non-negative, and the the flow $Q = \rho\mathcal{V}(\rho,\omega)$ is concave.
\begin{assumption}\label{assump:FD}
    The fundamental diagram $\mathcal{V}(\rho,\omega)$ satisfies:
    \begin{align}
        &\mathcal{V}(\rho,\omega) \ge 0,\\
        &2 \partial_{\rho} \mathcal{V} (\rho,\omega) + \rho \partial_{\rho\rho} \mathcal{V} (\rho,\omega)<0.
    \end{align}
\end{assumption}

If we set $\mathcal{V}(\rho,\omega) = \omega - p(\rho)$ by a suitable pressure function $p(\rho)$, the GSOM becomes the ARZ model. If we assume that $\omega(t,x)$ remains a constant value in the whole spatial-temporal domain, then the GSOM becomes the LWR model. In related research, several  interpretations and corresponding reconstruction methods for the $\omega$ value has been proposed~\citep{zhang2009conserved,fan2014comparative,mo2024game}, but they have a high computation cost that cannot be applied to real-time traffic application. In this paper, we propose a reconstruction method to get the $\omega$ value from trajectories with a low computational cost.

\subsection{Conservation of traffic attribute}
It is well known that the continuity equation in~\Cref{eq:GSOM rho} represents the conservation law of the number of vehicles. For a road segment of $x\in[0,L]$, denote the number of vehicles as 
\begin{align}
    N(t) = \int_0^L \rho(t,x) \diff x.
\end{align}
We have: 
\begin{align}\label{eq:conservation N}
    \dot{N}(t) = Q_{\mathrm{in}}(t)  - Q_{\mathrm{out}} (t),
\end{align}
with $Q_{\mathrm{in}}(t)=q(t,0)$ and $Q_{\mathrm{out}}(t)=q(t,L)$ being the flux entering and exiting the road segment.

The dynamics of $\omega$ in~\Cref{eq:GSOM omega} implies that the traffic attribute follows a second conservation law. To see this, we define $y = \rho\omega$, and~\Cref{eq:GSOM rho}-~\Cref{eq:GSOM omega}   gives the dynamics of $y$ as
\begin{align}
    \partial_t y + \partial_x (vy) = 0.
\end{align}
Therefore, the density-weighted traffic attribute is also conservative. Following the definition of $N$,  we define:
\begin{align}
    I(t) = \int_0^L \rho(t,x) \omega(t,x)  \diff x.
\end{align}
And we have:
\begin{align}\label{eq:conservation omega}
    \dot{I}(t) = A_{\mathrm{in}}(t) - A_{\mathrm{out}}(t),
\end{align}
with $A_{\mathrm{in}}(t) =v(t,0)\rho(t,0)\omega(t,0)$ and $  A_{\mathrm{out}}(t) = v(t,L)\rho(t,L)\omega(t,L)$ being the driver attributes entering and exiting the road segment.  

\begin{remark}[Physical interpretation of conservation law]
For a road segment without any ramps, variations in the number of vehicles are determined solely by the entering and exiting flux, i.e., `how many' vehicles enter and exit the road, as formulated in~\Cref{eq:conservation N}.
If we assume that driver attributes remain constant during the driving process, then the dynamics of the aggregated attribute $I(t)$ are determined by the entering and exiting attributes, i.e., the drivers `who' enter and exit the road. 
\end{remark}

\begin{figure}[t]
	\centering
	\subcaptionbox{Homogeneous HV}{\includegraphics[width=0.4\linewidth]{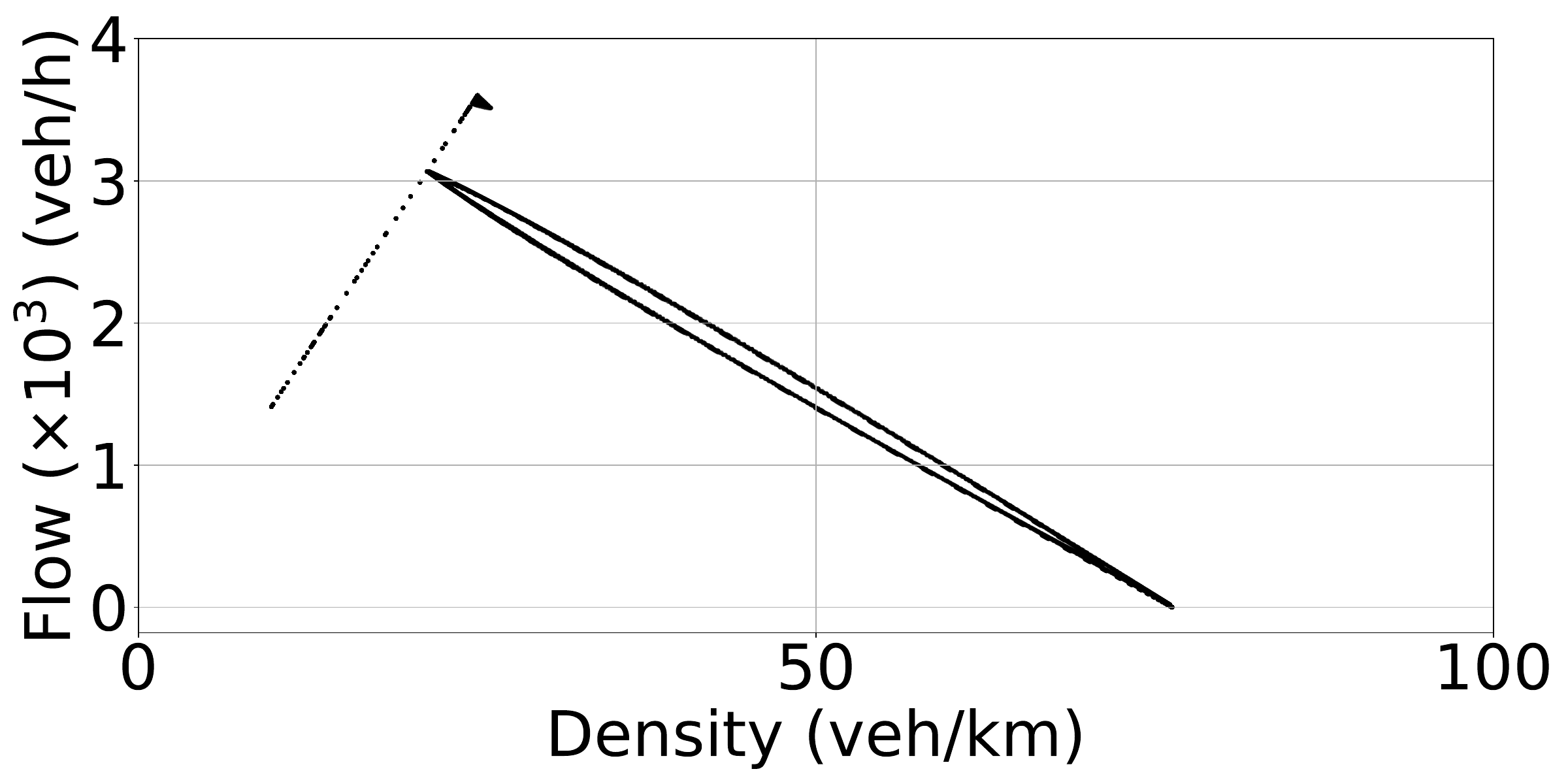}}
	\subcaptionbox{Homogeneous AV}{\includegraphics[width=0.4\linewidth]{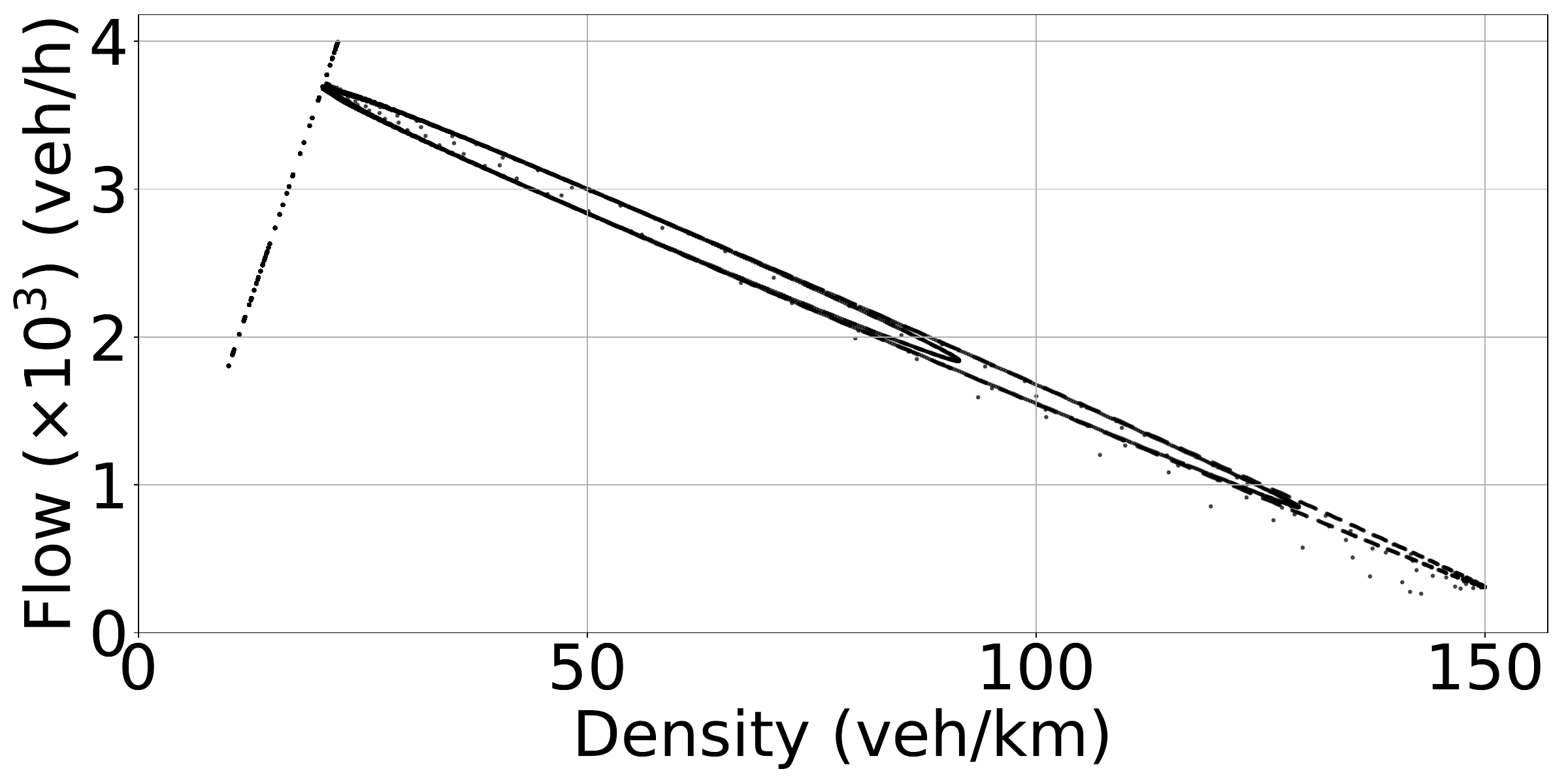}}\\
	\subcaptionbox{AV $p=50\%$, one type of AV}{\includegraphics[width=0.4\linewidth]{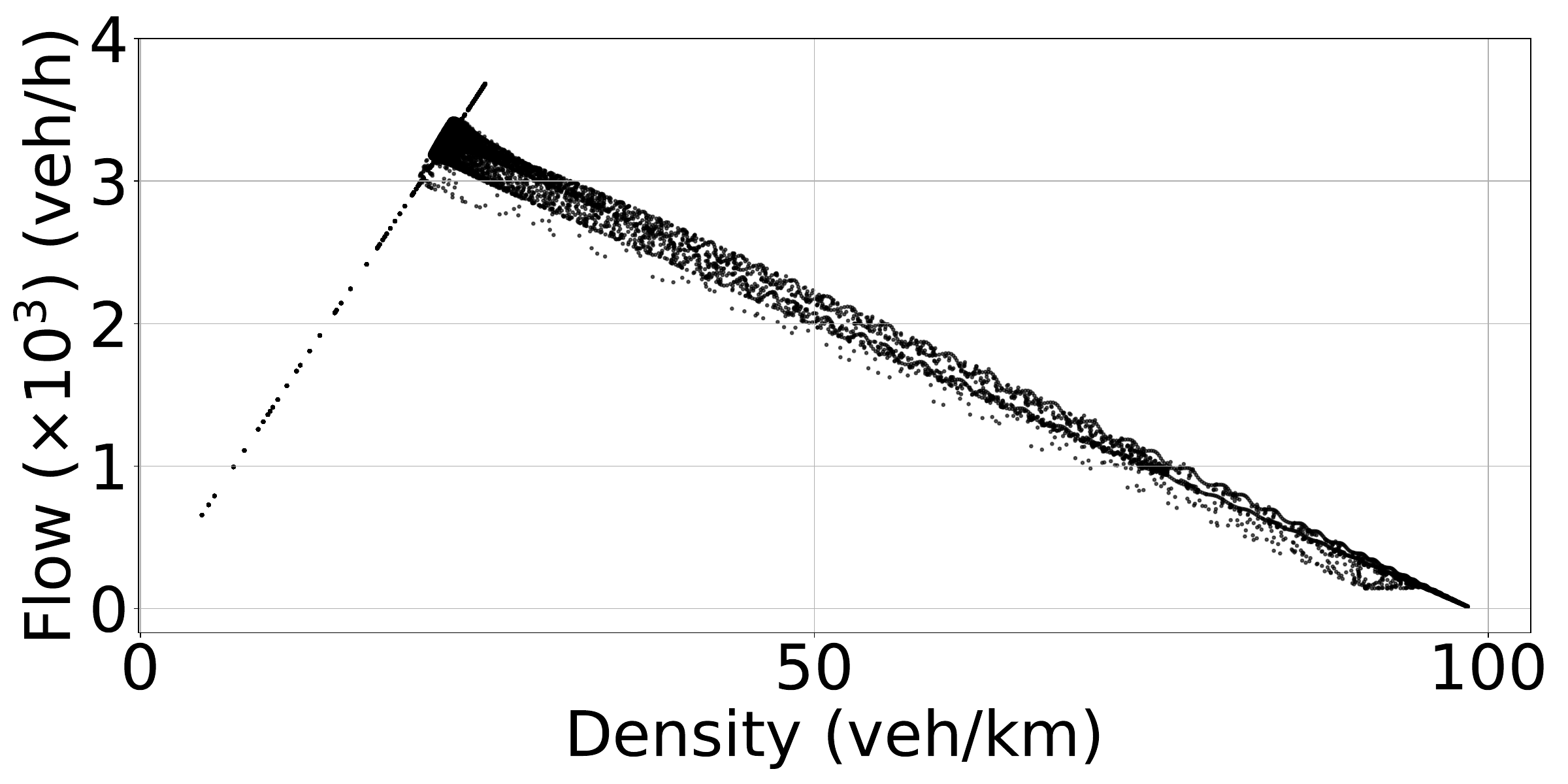}}
	\subcaptionbox{AV $p=50\%$, two types of AVs}{\includegraphics[width=0.4\linewidth]{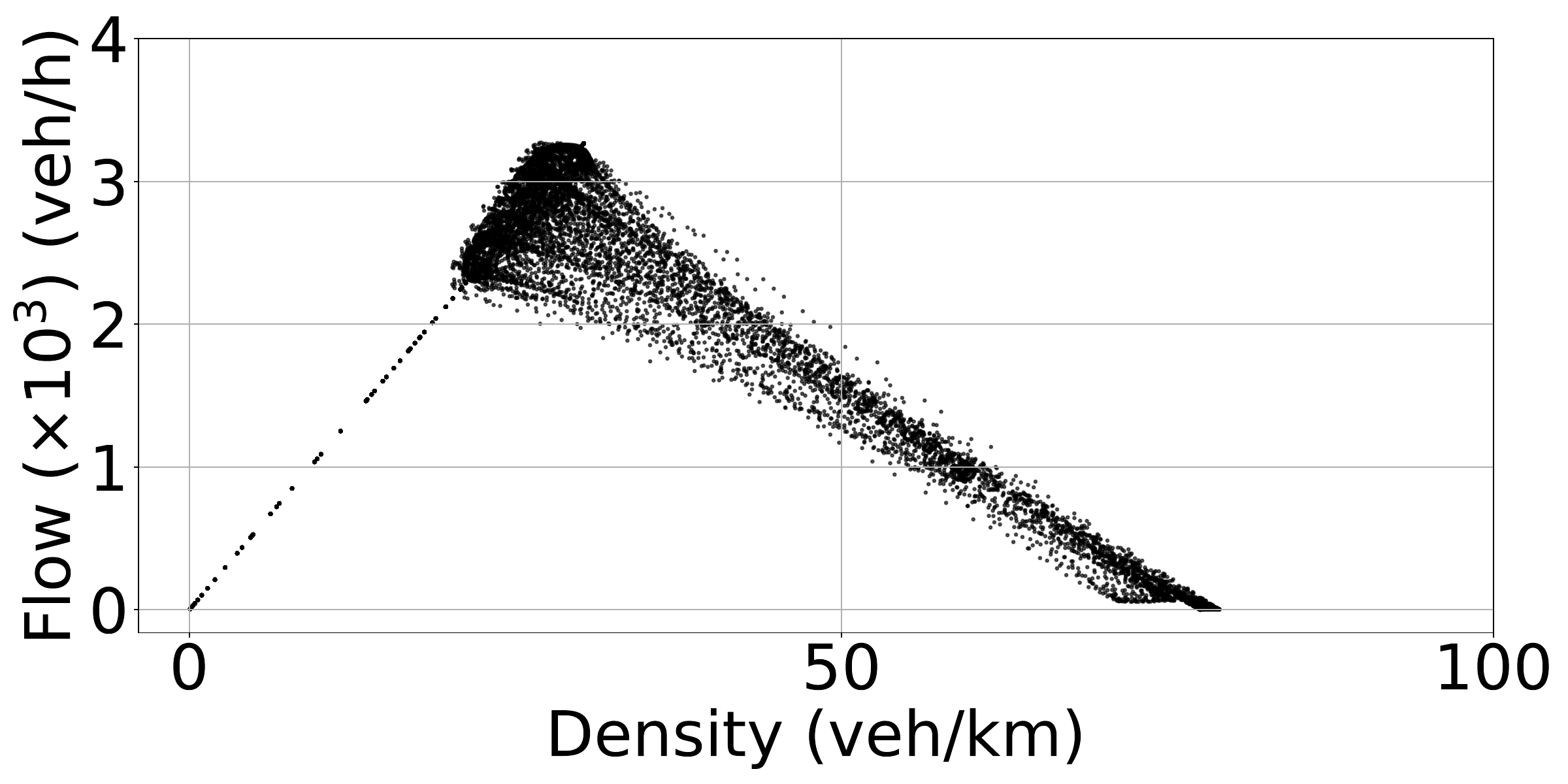}}
	\caption{We run simulations   to get vehicle's trajectories and  reconstruct the macroscopic density and flow. For homogeneous traffic, the $q-\rho$ basically follows the theoretical result of a triangular form. For heterogeneous mixed traffic, the fundamental diagram presents a spread, and the spread increases for more heterogeneous traffic.}  \label{fig:example FD}
\end{figure}

\subsection{Traffic heterogeneity as a micro-macro bridge}

In this paper, we interpret the traffic-attribute as a measure of driver heterogeneity, and use  heterogeneity as a bridge between microscopic and macroscopic traffic. To see this, we show in this part via simulation that an increase in heterogeneity on microscopic vehicle behaviors causes more heterogeneous macroscopic traffic dynamics.

We simulate the motion of a vehicle chain to generate trajectories and reconstruct the macroscopic traffic density $\rho$, speed $v$, and flow $q$. The detailed simulation settings are given in Appendix C. We consider one homogeneous HV traffic, one homogeneous AV traffic and two heterogeneous mixed-traffic.  \Cref{fig:example FD}  gives the collected $q$-$\rho$ scatter plot for the four settings. In~\Cref{fig:example FD}(a) and~\Cref{fig:example FD}(b), the system is homogeneous; i.e., all vehicles have identical car-following model parameters. We see that there is only a small variation on the  fundamental diagram.  Now consider mixed-autonomy traffic with an AV penetration rate of $50\%$, whose simulation result is given in Figures~\ref{fig:example FD}(c) and (d). In heterogeneous traffic,  the observed $q-\rho$ relationship does not follow a single line but lies within a region. The spread in the fundamental diagram means that the macroscopic traffic is heterogeneous, i.e., there is multiple flow corresponding to the same density. Comparing~\Cref{fig:example FD}(c) with one type of AV and~\Cref{fig:example FD}(d) with two types of AVs, we also find that, even with a fixed penetration rate, an increase in  AV heterogeneity  enlarges the spread range in the fundamental diagram. Therefore, with more heterogeneous  microscopic behaviors, the heterogeneity in macroscopic traffic also increases.

In this section, we introduce a continuous traffic-heterogeneity attribute to model mixed-autonomy traffic, motivated by the observation that heterogeneity exists not only between AVs and HVs but also within each class.
To make the model applicable in practice, it is essential to solve the two problems. First, we design a reconstructionist method to get the traffic attribute from vehicle trajectories, i.e.,
\begin{align}
    \omega = f(\mathrm{trajectory}).
\end{align}
Second, the fundamental diagram $\mathcal{V}(\rho,\omega)$ is calibrated by solving the constrained optimization problem: 
\begin{align}\begin{array}{ll}
     &\arg\min_{\theta}  \sum_i \left( \hat{\mathcal{V}}(\rho_i,\omega_i;\theta) - v_i \right)^2  \\
    \mathrm{s.t.} \quad &  \hat{\mathcal{V}}(\rho,\omega;\theta) \ge 0,\\
        &2 \partial_{\rho} \hat{\mathcal{V}} (\rho,\omega;\theta) + \rho \partial_{\rho\rho} \hat{\mathcal{V}} (\rho,\omega;\theta)<0, 
\end{array}\label{eq:FD optimization}
\end{align}
with $\theta$ being parameters in the fundamental diagram. The objective function is to minimize the calibration error, and the two constraints are derived from Assumption~\ref{assump:FD}  to ensure well-posedness of the learned fundamental diagram. Sections \ref{sec:reconstruct model}–\ref{sec:data driven} formulates the mapping and fundamental diagram  under data-rich and data-scarce scenarios respectively.

\section{Reconstruction of Traffic Heterogeneity with Abundant Data} \label{sec:reconstruct model}

In this section, we focus on scenarios with abundant traffic data and propose a formulation to get the traffic attribute from vehicle trajectories.   Due to the lack of real large-scale mixed-autonomy data, we validate the proposed mixed-autonomy model and reconstruction method on real heterogeneous HV data from~\citep{zheng2021experimental}. 
As analyzed in Section~\ref{sec:micro}, the heterogeneity within HVs may be even larger than that between HVs and AVs. So the validation still provides insights for future research if large-scale mixed-traffic data becomes available.

\subsection{Reconstruction of traffic heterogeneity  from driver heterogeneity}

We divide the whole temporal-spatial domain into $N_t\times N_x$ cells. We denote $\mathcal{N}_{i,j}$ as the set of vehicles in the cell $(i,j)$.  For the cell $(i,j)$, the traffic attribute $\omega_{i,j}$ is reconstructed as:
\begin{align}\label{eq:omega from tra}
    \omega_{i,j} &= \underbrace{ \frac{1}{N_{i,j}}\sum_{k\in \mathcal{N}_{i,j}}   V^{\mathrm{f}}_k}_{\text{Driver   attribute}} + \underbrace{
		\frac{1}{\sum_{k\in \mathcal{N}_{i,j}} e^{-\omega^{\mathrm{L}}_k}}\sum_{k\in \mathcal{N}_{i,j}} e^{-\omega^{\mathrm{L}}_k }\left( V_{i,j,k} - V_{i,j,k}^{\mathrm{desired}}\right)}_{\text{Driver's local behavior with interactions in traffic}}, 
\end{align}
where $V_k^{\mathrm{f}}$ is vehicle $k$'s free-flow speed calibrated from~\Cref{eq:micro Vopt},   $V_{i,j,k} = X_{i,j,k}/T_{i,j,k}$ is vehicle $k$'s local spatial-temporal average speed in the cell $(i,j)$ with $X_{i,j,k}$ and $T_{i,j,k}$ being the distance and time vehicle-$k$ travels in the cell, and $V_{i,j,k}^{\mathrm{desired}} = V^{\mathrm{opt}}_k \left(\frac{1}{\rho_{i,j}}-l_k\right) $ is  vehicle $k$'s gap-dependent desired speed given current traffic density $\rho_{i,j}$ with $l_k$ being its length. To explain the intuition and rationale behind these two terms, we begin by analyzing collected traffic data.

\begin{figure}[!t]
    \centering
    \subcaptionbox{The scatter speed-density points for four vehicle groups}{\includegraphics[width=0.45\linewidth]{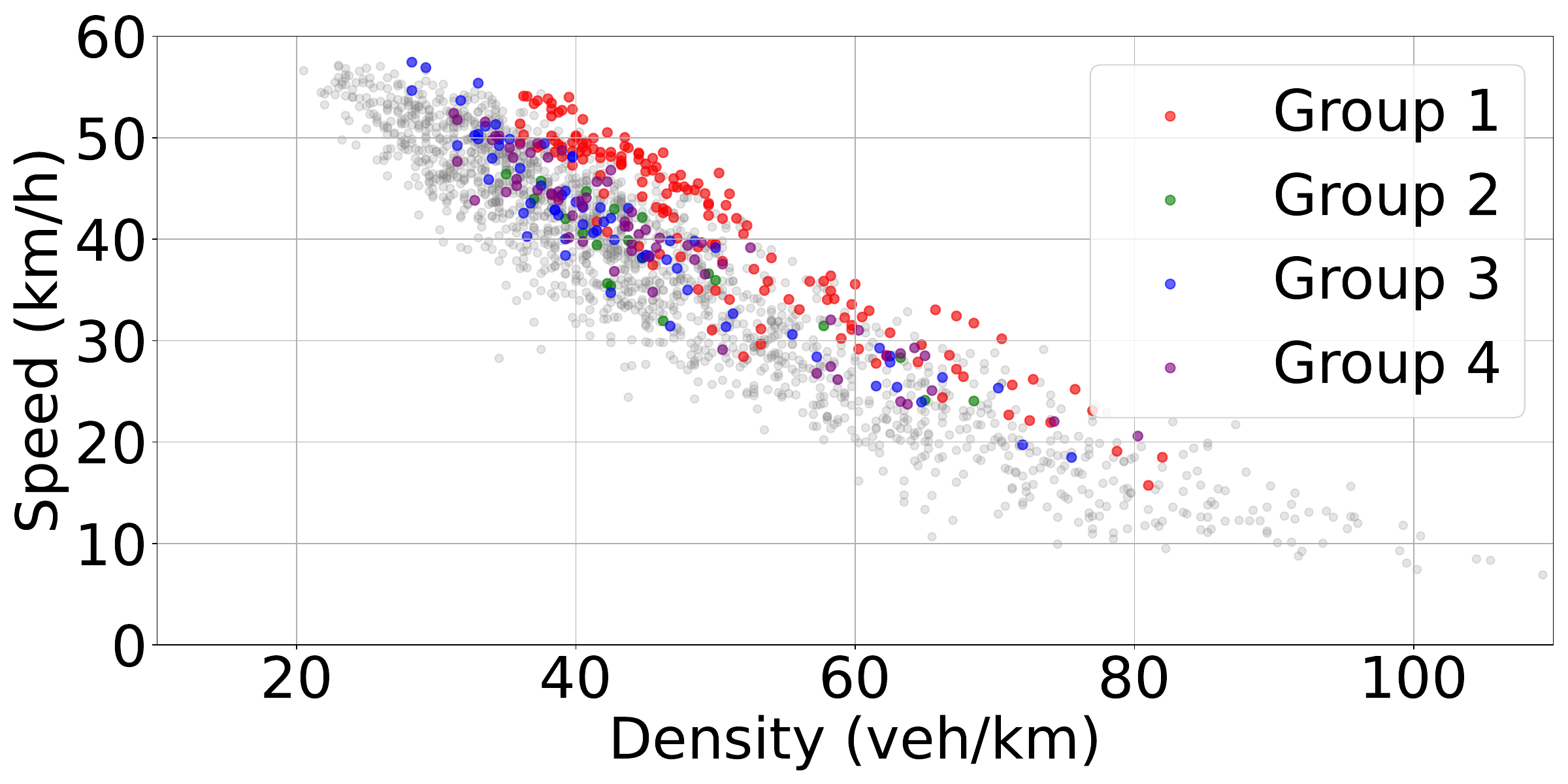}}
    \subcaptionbox{The speed-density  scatter plot with continuous traffic attribute}{\includegraphics[width=0.45\linewidth]{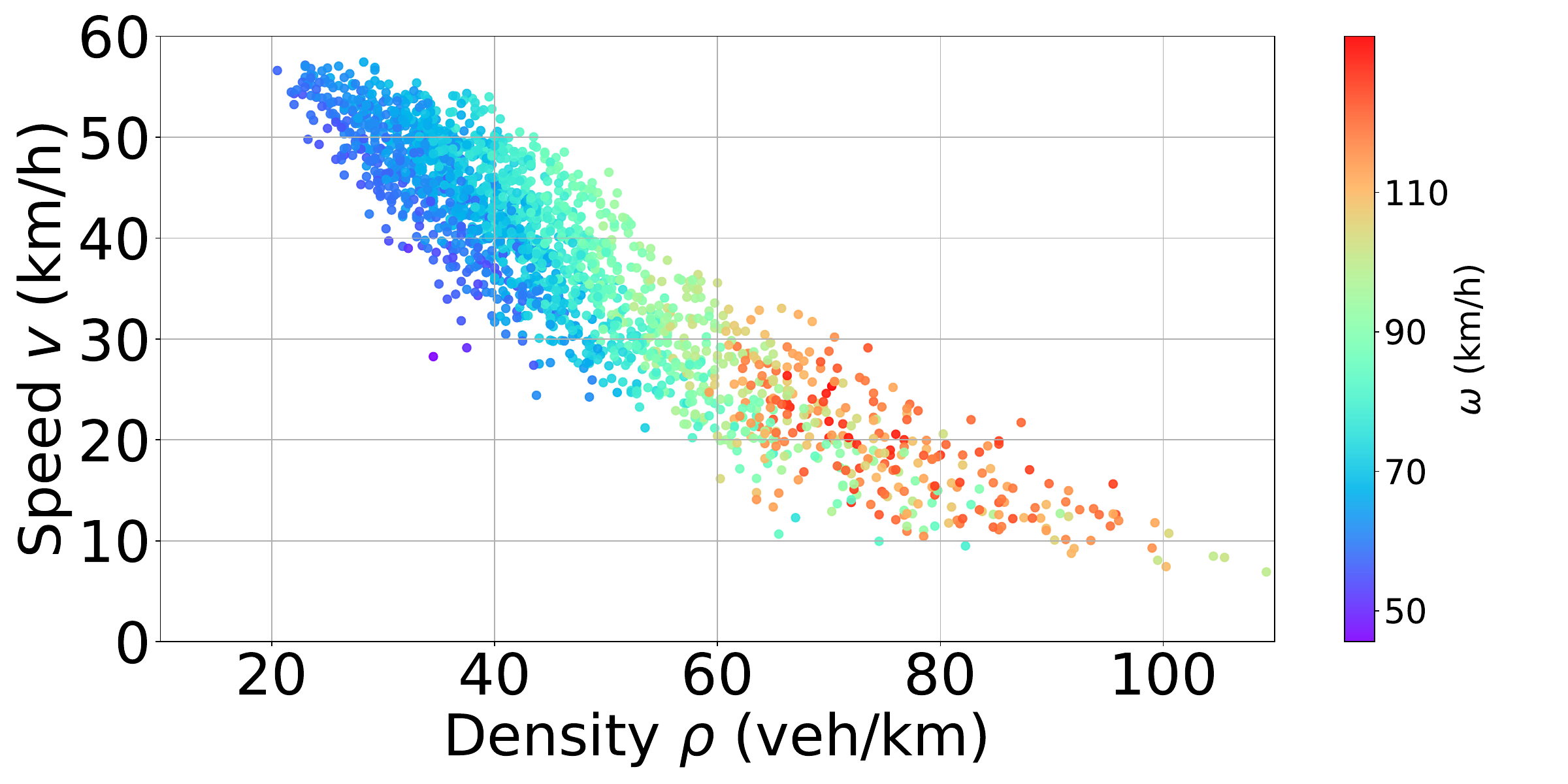}}
    \caption{Subfig(a): The scatter speed-density points for four vehicle groups. The grey points are the $v-\rho$ points for the whole spatial-temporal domain. Since we select cells with all vehicles in the corresponding group, there are cells unlabeled. Subfig(b): The speed-density ($v$-$\rho$)  scatter plot with {continuously-varying} $\omega$ value.  }
    \label{fig:FD scatter omega} \label{fig:omega} \label{fig:vehicle sep FD}
\end{figure}

We divide the 40 drivers into four groups: Group One (Drivers 1–10), Group Two (Drivers 11–20), Group Three (Drivers 21–30), and Group Four (Drivers 31–40). For Group One, we select cells containing all ten drivers and mark the corresponding density-speed scatter points in red in~\Cref{fig:vehicle sep FD}(a). For the other three groups, we follow the same procedure and mark their density-speed scatter points in~\Cref{fig:vehicle sep FD}(a). From~\Cref{fig:vehicle sep FD}(a), we have two key findings:
\begin{enumerate}
    \item For Group One, its macroscopic average speed is higher than that of the other groups at the same density. Moreover, Group one maintains a higher speed at almost the entire range of densities.
    \item For each Group, the speed also varies at the same density. Take Group One as an example, when the density is approximately 40 veh/km, the observed speed ranges from 45 km/h to 55 km/h.
\end{enumerate}
The two terms in the reconstruction of $\omega$ in~\Cref{eq:omega from tra} are designed based on the above two findings. The first finding indicates that some vehicles consistently exhibit higher speed regardless of density, which motivates the first term in~\Cref{eq:omega from tra} to capture such density-independent speed trends. We calculate the average free-flow speed $V^{\mathrm{f}}_k$ for Group One to be 30 m/s;  while for other groups, it is  around 27 m/s. For the second term in~\Cref{eq:omega from tra}, we note that $V_{i,j,k}$ is vehicle $k$'s \textit{actual} speed, while $V^{\mathrm{desired}}_{i,j,k}$ represents its \textit{desired} speed at density $\rho_{i,j}$. The difference $\Delta V_{i,j,k} = V_{i,j,k} - V^{\mathrm{desired}}_{i,j,k}$  reflects the \textit{aggressive speed} presented by driver $k$ in current traffic conditions. With a larger $\Delta V_{i,j,k}$, driver $k$ tends to drive faster, and thus the corresponding  $(\rho,v)$ point will move upward  in the fundamental diagram. 
We also note that for vehicle $k$, a higher $\omega_k^{\mathrm{L}}$ implies more uncertainty in its driving behavior. So even if a driver has a larger $\Delta V_{i,j,k}$, it may result from its more uncertain driving behavior, and thus it should contribute less to the cell value $\omega_{i,j}$.

We give the speed-density ($v$-$\rho$)  scatter plot in Figure~\ref{fig:FD scatter omega}(b).  For the same density $\rho$, the scattered speed  is approximately separated by the reconstructed $\omega$.  The observed separation shows the potential to calibrate a fundamental diagram $\mathcal{V}(\rho,\omega)$ using $\rho$ and  $\omega$ as two independent variables, which will give present a lower calibration error than the traditional one-dimensional $V(\rho)$. 

\subsection{Calibration of two-variable fundamental diagram}
In this subsection, we calibrate a two-variable fundamental diagram $\mathcal{V}(\rho,\omega)$ to capture the dependency of traffic speed on density $\rho$ and traffic attribute $\omega$ by solving the constrained optimization problem in~\Cref{eq:FD optimization}. 
{Traditional optimization-based fitting methods, such as polynomial regression or parametric curve fitting,  require a pre-determined functional form. In our case, the fundamental diagram $\mathcal{V}(\rho,\omega)$ involves interactions between density $\rho$ and heterogeneous driver attributes $\omega$, which may not be easily captured by fixed-form models. In contrast, neural networks (NNs) offer greater flexibility since there is no assumptions on the functional form of the fundamental diagram.  Therefore, we adopt a neural network  to learn the mapping from $(\rho, \omega)$ to $v$, allowing for accurate  calibration without relying on restrictive assumptions about the underlying functional structure.} The effectiveness of NN-based calibration for single-variable fundamental diagram has been shown in our previous work~\citep{zhao2024observer}. In this paper, we further extend to calibrate the two-variable fundamental diagram $\mathcal{V}(\rho,\omega)$.

We convert the constrained optimization problem in~\Cref{eq:FD optimization}  to an unconstrained problem using the penalty method and  train the neural network by minimizing the loss function:
\begin{align}
    L(\theta) =   L_d(\theta) + p L_p(\theta),
\end{align}
where the data loss $L_d$ and the penalty loss $L_p$  are calculated based on the objective function and constraints in~\Cref{eq:FD optimization}, and $p>0$ is a penalty coefficient. To evaluate the calibration accuracy, we randomly select $N_d$ points $(\rho_i,\omega_i,v_i)$ from all reconstructed $(\rho,\omega,v)$ in the whole spatial-temporal domain, and get  the data loss $L_d$ as the difference between the learned speed $\hat{\mathcal{V}}(\rho_i,\omega_i)$ and observed speed $v_i$:
\begin{align}
	L_d = \frac{1}{N_d} \sum_{i=1}^{N_d} (\hat{\mathcal{V}}(\rho_{i},\omega_{i};\theta) - v_{i})^2. 
\end{align}
To constrain the learned fundamental diagram, we   randomly select  $N_p$ points of $(\rho_j,\omega_j)$  with $\rho_i \in [0,\rho_{\max
}]$, $\omega_j\in [\omega_{\min},\omega_{\max}]$ and get the penalty  loss  as:
\begin{align}
	L_p = & \frac{1}{N_p} \sum_{j=1}^{N_p} \left(\min \left \{0,  \hat{\mathcal{V}}(\rho_j,\omega_j;\theta) \right\}\right)^2 + \frac{1}{N_p} \sum_{j=1}^{N_p} \left(\max \left\{ 0, 2\partial_{\rho} \hat{\mathcal{V}}(\rho_j,\omega_j;\theta) + \rho_j \partial_{\rho\rho} \hat{\mathcal{V}}(\rho_j,\omega_j;\theta) \right\}\right)^2. \label{eq:loss FD constraint}
\end{align}
The partial derivatives $\partial_{\rho} \mathcal{V}$ and $\partial_{\rho \rho} \mathcal{V}$ are calculated using the automatic differentiation provided by the TensorFlow.  The first term in $L_p$ constrains that the fundamental diagram always have non-negative speed, and the second term is designed so that the traffic flow dynamics is hyperbolic.  When evaluating the penalty loss, the choice of the points $(\rho_j,\omega_j)$ is independent of the points $(\rho_i,\omega_i)$ that are used to evaluate the data loss $L_d$.

We set the NN as a fully connected feedforward  neural network with 3 hidden layers and 50 neurons in each hidden layer. The input dimension of the NN is two, one for density and one for traffic attribute. The output of the NN is traffic speed. We run 50,000 epochs of the  ADAM optimizer to train the NN. To evaluate the data loss, we randomly select 60 \% points from all reconstructed $(\rho,v,\omega)$ data. To evaluate the penalty loss, we select 200 $(\rho,\omega)$ points uniformly distributed within the range  $[0,\rho_{\max}]\times [\omega_{\min},\omega_{\max}]$. We set the penalty coefficient $p$  as 1,000.

\begin{figure}[!t]
    \centering
    \subcaptionbox{}{\includegraphics[width=0.4\linewidth]{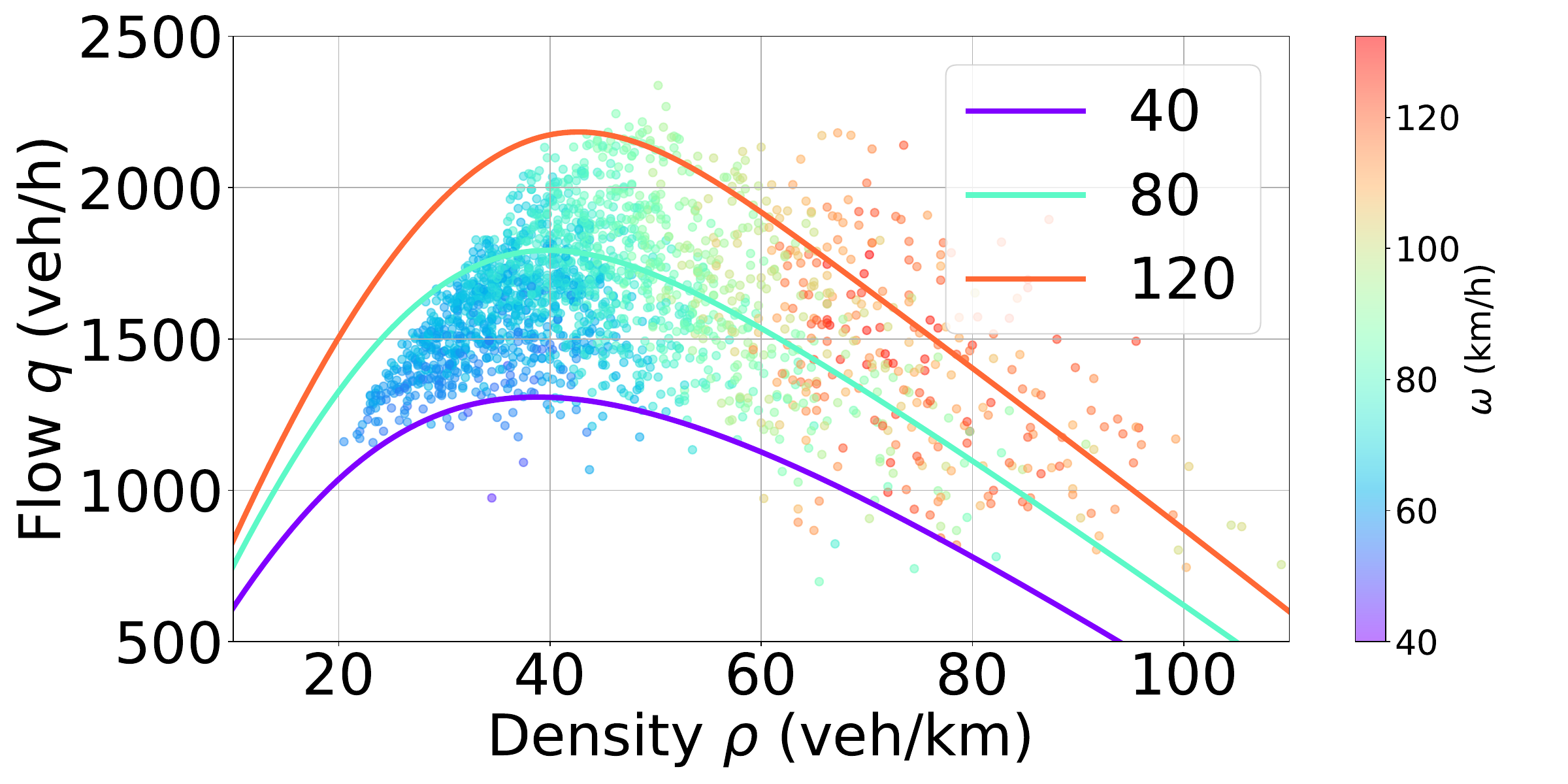}}
    \subcaptionbox{}{\includegraphics[width=0.4\linewidth]{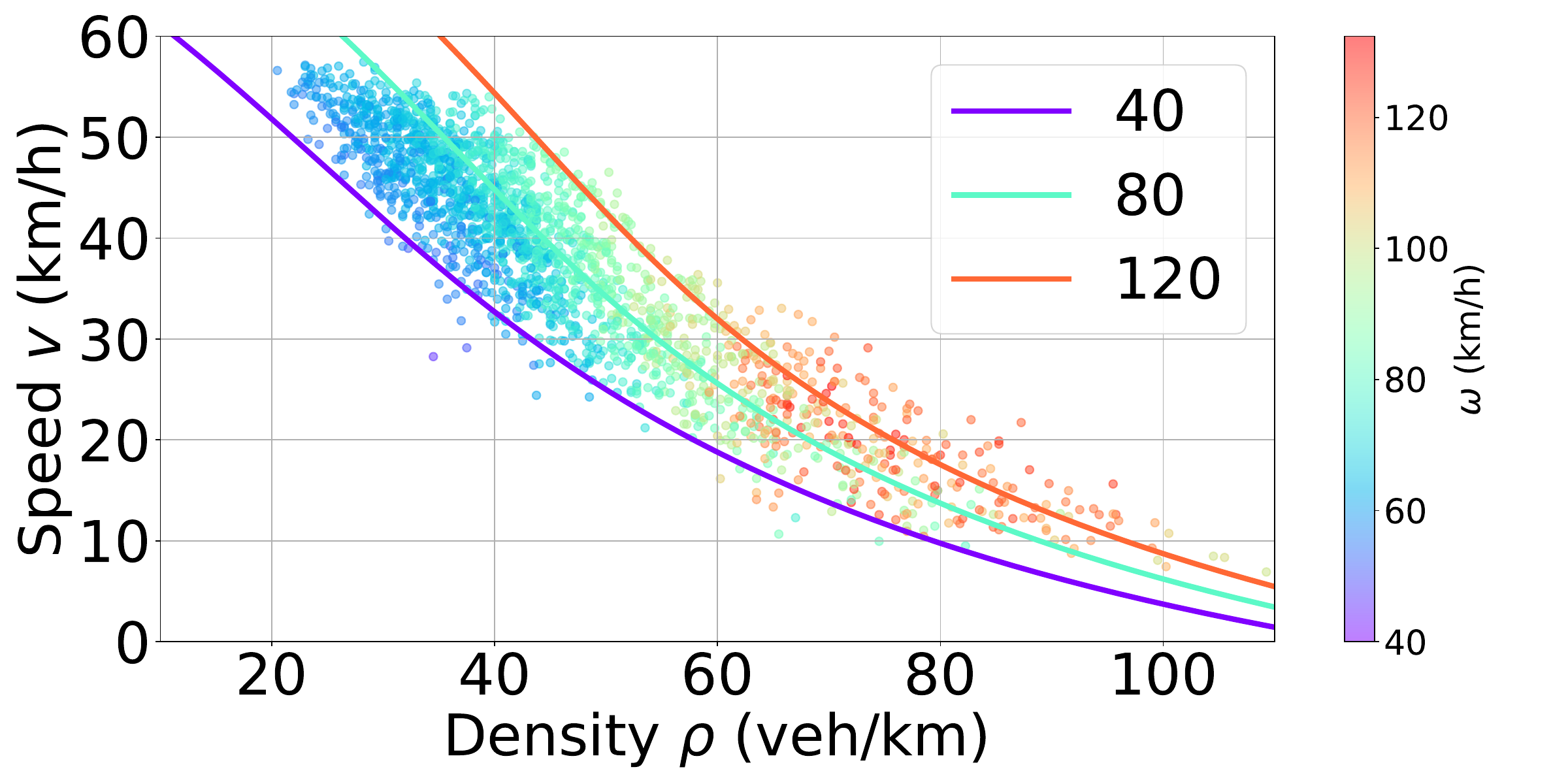}}
    \caption{The calibrated fundamental diagram.}
    \label{fig:FD by NN}
\end{figure}

To ensure consistency and robustness across datasets, we apply min-max normalization to the input variables of $\rho$ and $\omega$. The normalization ensures stable neural network training and facilitates generalization to datasets with different resolutions and scales~\citep{huang2023normalization}. For the fundamental diagram calibration in~\Cref{fig:FD by NN}, and also the results in~\Cref{fig:TGSIM:FD} for the TGSIM  dataset and \Cref{fig:NGSIM:FD} for the NGSIM dataset, we use the same  NN training settings, including the optimizer, the training epochs, and the penalty coefficient. The accuracy validates that the proposed NN-based calibration works across various traffic conditions.

\begin{figure}[!t]
    \centering
    Ground truth wave\\
    \subcaptionbox{Density (veh/km)}{\includegraphics[width=0.3\linewidth]{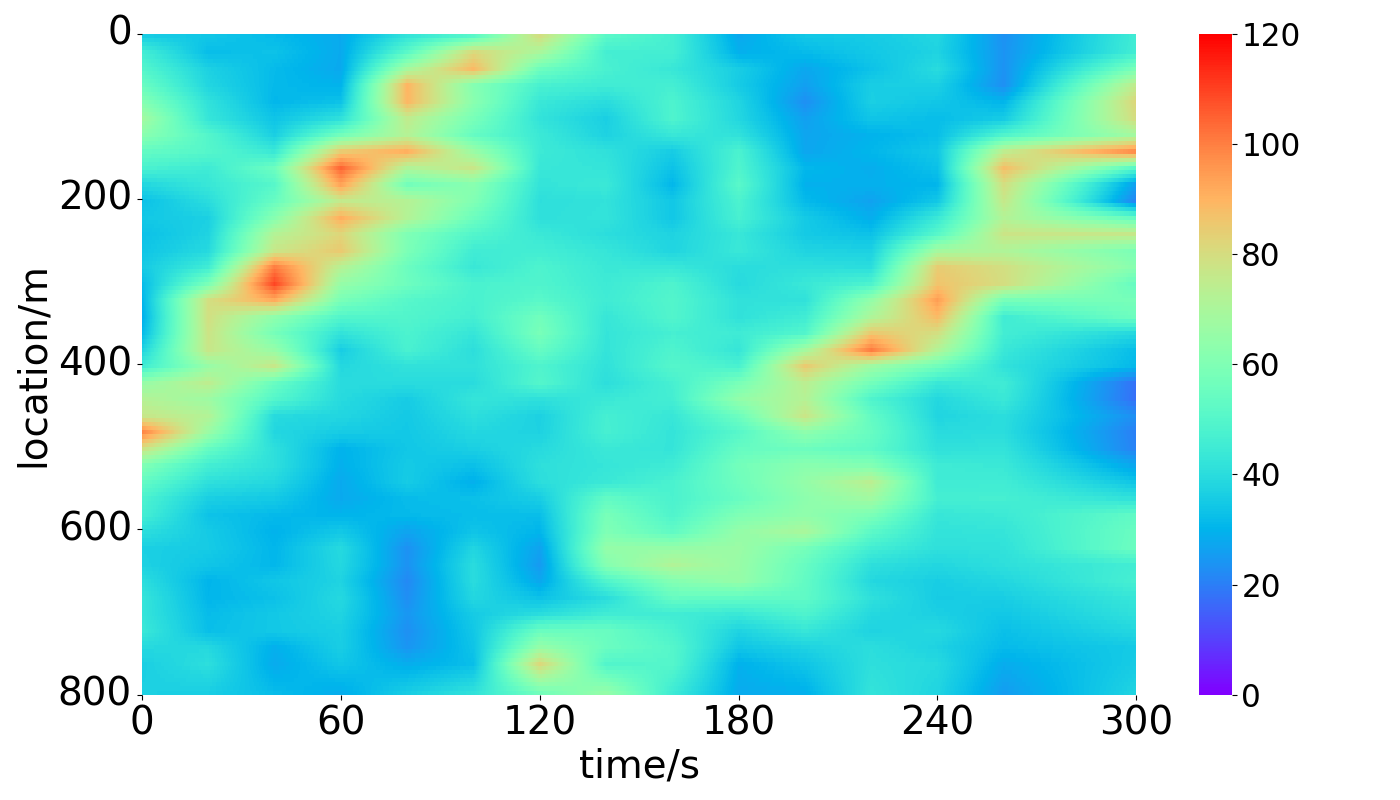}}
    \subcaptionbox{Speed (km/h)}{\includegraphics[width=0.3\linewidth]{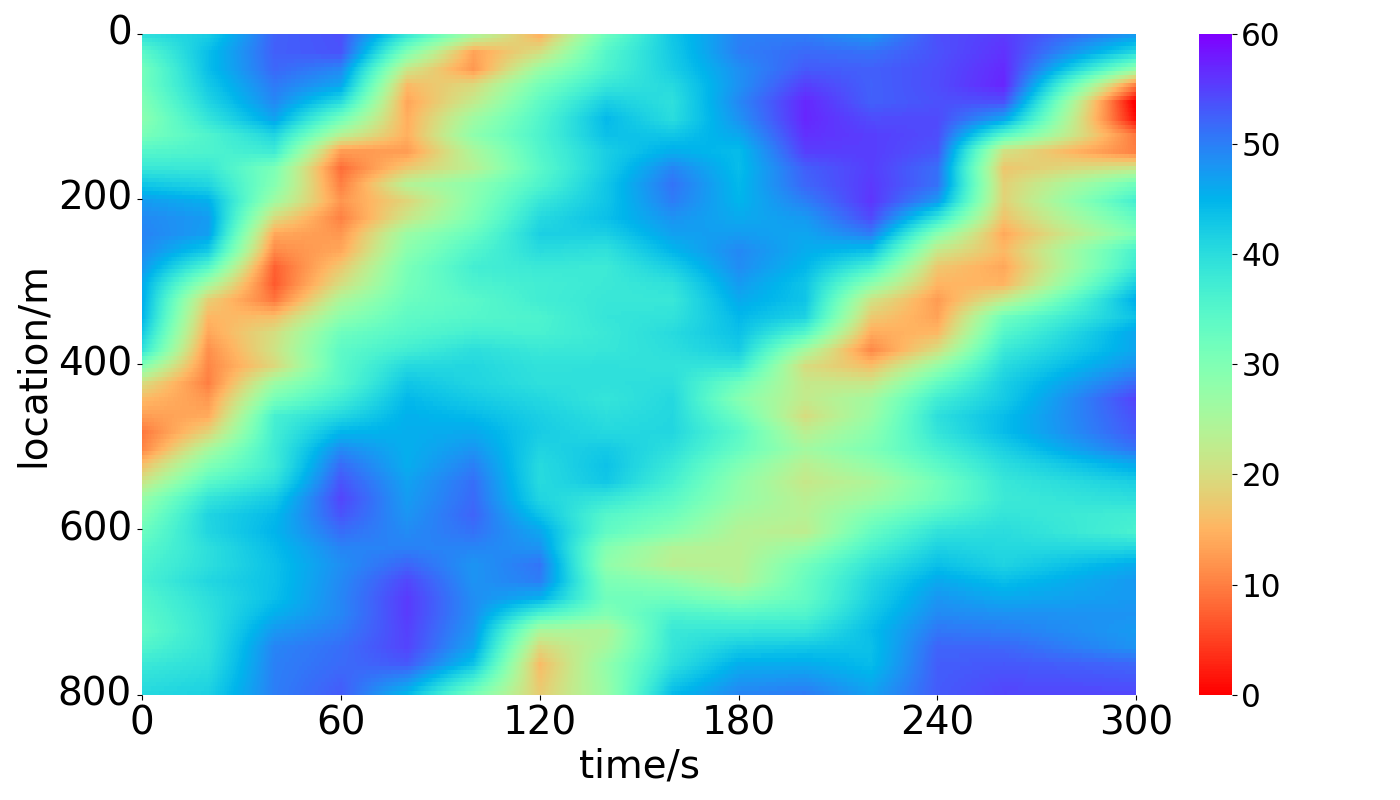}}
    \subcaptionbox{Flow (veh/h)}{\includegraphics[width=0.3\linewidth]{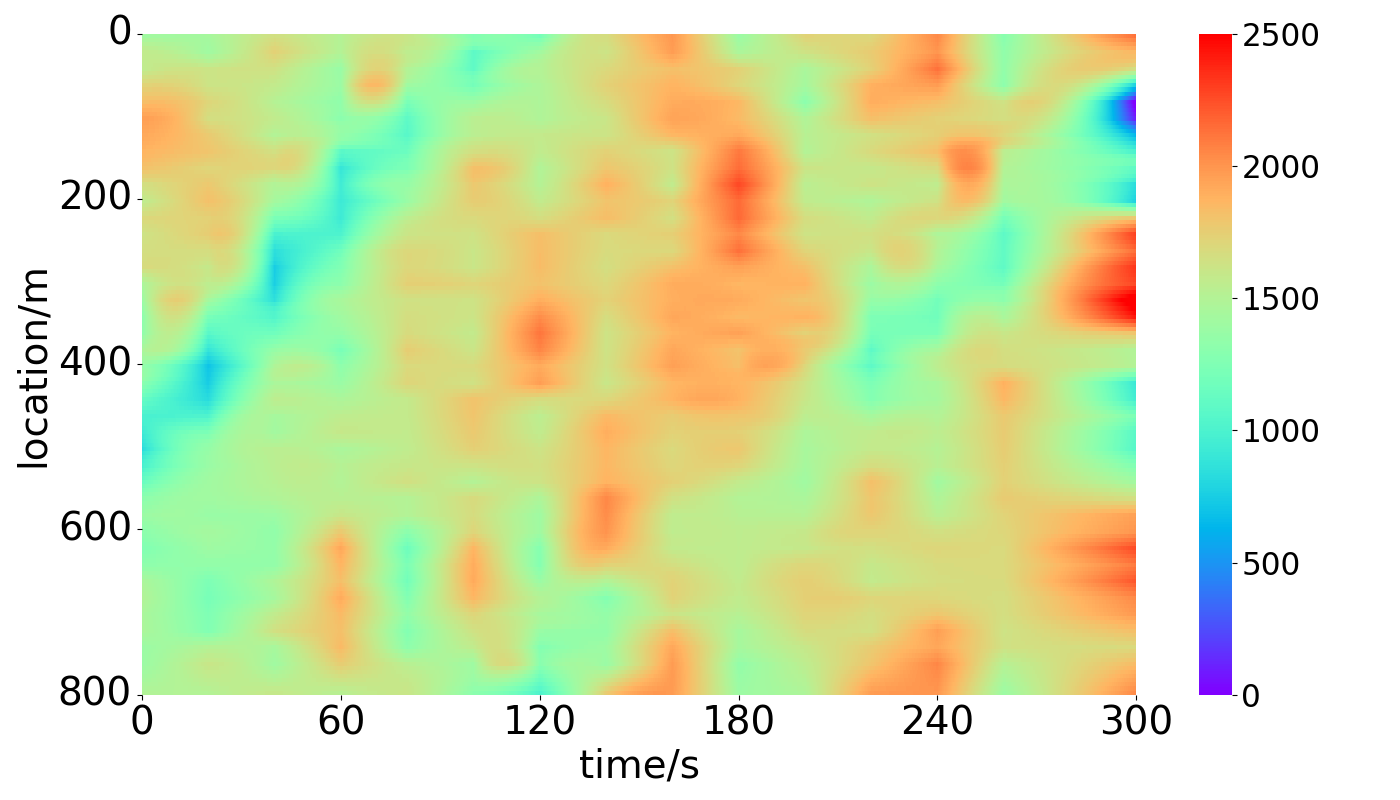}} \\
    Simulated wave\\
    \subcaptionbox{Density (veh/km)}{\includegraphics[width=0.3\linewidth]{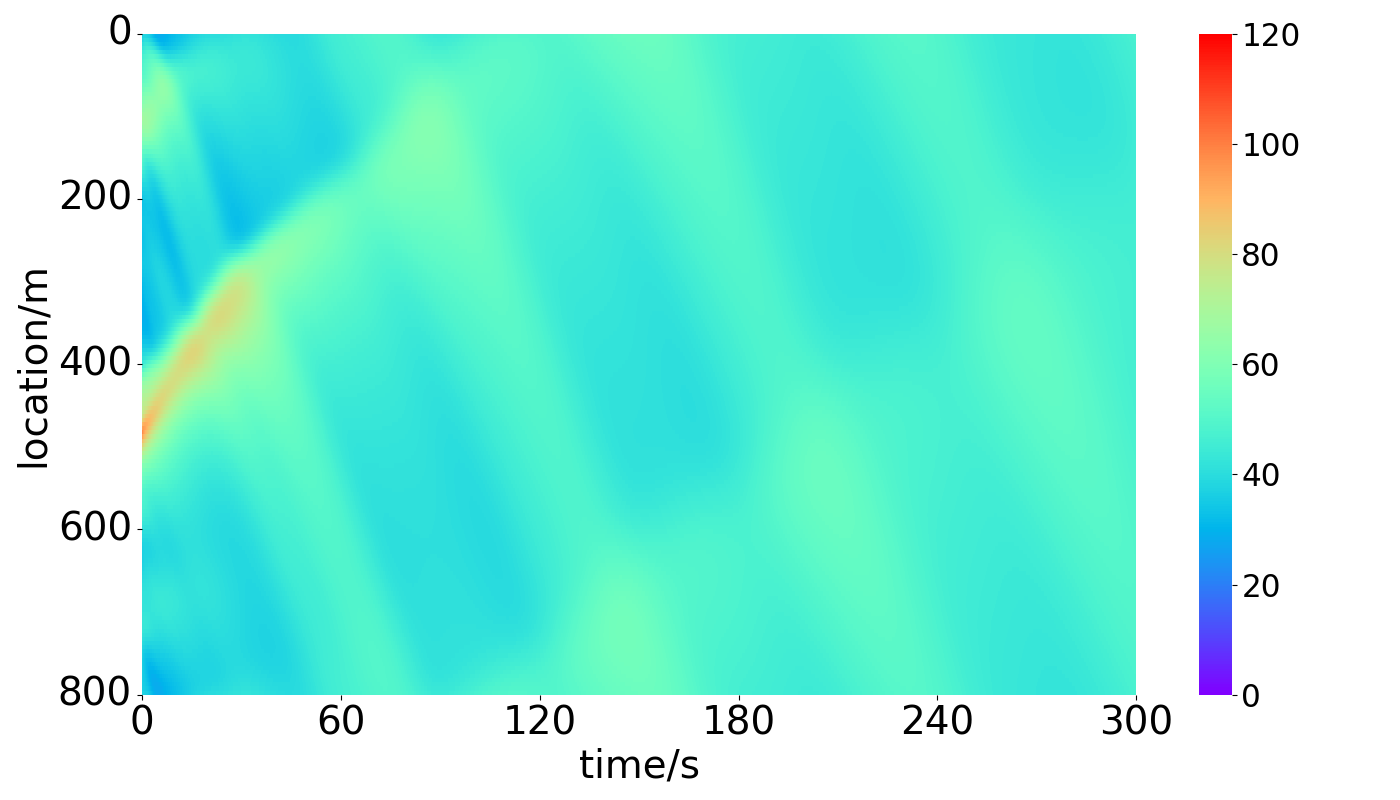}}
    \subcaptionbox{Speed (km/h)}{\includegraphics[width=0.3\linewidth]{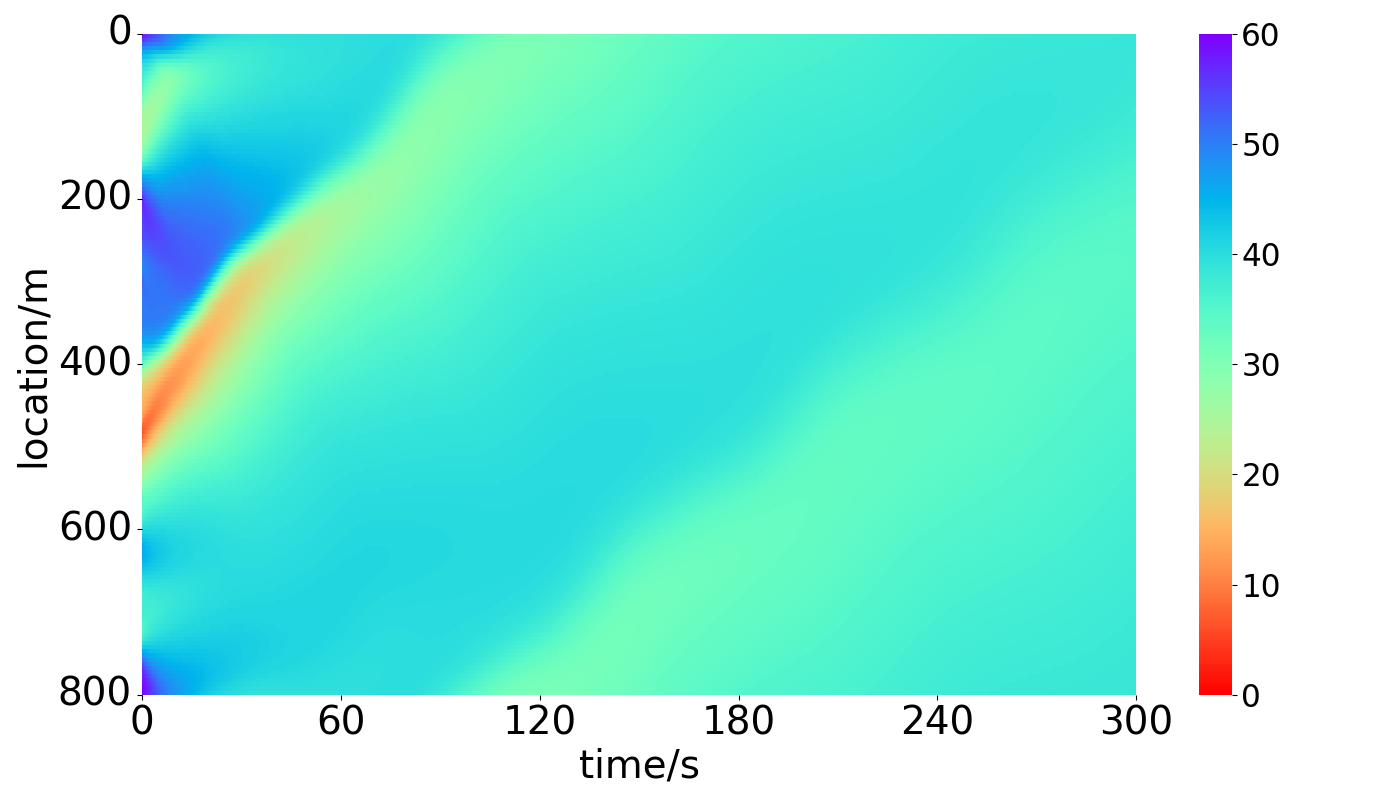}}
    \subcaptionbox{Flow (veh/h)}{\includegraphics[width=0.3\linewidth]{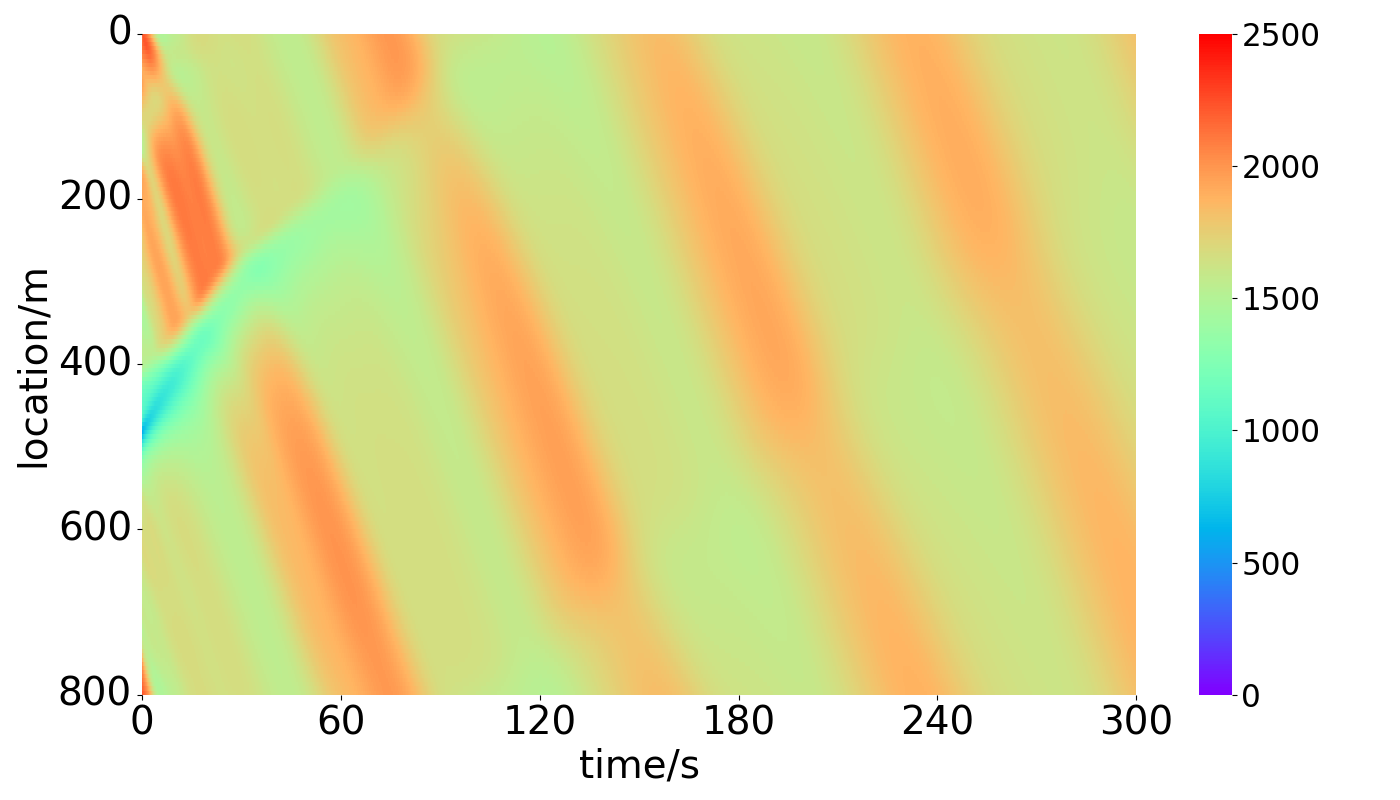}}
    \caption{The ground-truth traffic wave and simulated trajectories via the GSOM model. }
    \label{fig:wave}
\end{figure}

\begin{table}[!t]
    \centering
    \caption{The simulation error  for the wave propagation using the ARZ and GSOM model.}
    \label{tab:error ARZ GSOM}
    \begin{tabular}{c|c|c|c}
            & Density & Speed & Flow \\ \hline 
       ARZ  &  30.86 \%& 30.18  \% & 15.06  \%\\ \hline
       GSOM &  25.77 \% &  24.36 \%  & 14.09  \% \\ \hline
       Accuracy improvement & 17.49 \% & 19.28 \% & 6.44 \% \\ \hline 
    \end{tabular}
\end{table}

We evaluate the calibration error using the root-mean squared error (RMSE):
\begin{align}
	E_{FD} = \sqrt{ \frac{1}{N_x N_t} \sum_{i,j} \left(\frac{\hat{\mathcal{V}}(\rho_{i,j},\omega_{i,j};\theta) - v_{i,j}}{v_{i,j}}\right)^2}  \times 100\%.
\end{align}
We get the calibration error as $E_{FD} = 12.2\%$, which shows that the two-variable fundamental diagram provides an accurate calibration of traffic systems. We present the calibrated fundamental diagram in~\Cref{fig:FD by NN}. We also numerically solve the GSOM \Cref{eq:GSOM rho}-\Cref{eq:GSOM omega} with the learned fundamental diagram. We denote the simulated density, speed and flow as  $\hat\rho$, $\hat v$, $\hat q$ respectively. We set the initial condition $\hat \rho(0,x)$ and $ \hat \omega(0,x)$ the same as observed data $\rho(0,x)$ and $\omega(0,x)$. Since it is a ring road, we use periodic boundary conditions. We give in~\Cref{fig:wave} the ground truth and simulated traffic states. We see that the GSOM model captures the wave propagation.  We quantitatively measure the model error as 
\begin{align}
    & E_{Y}  =  \sqrt{ \frac{1}{N_x N_t} \sum_{i,j}\left(\frac{\hat Y_{i,j} - Y_{i,j}}{Y_{i,j}}\right)^2 } \times 100\%,
\end{align}
with $Y=\rho,v,q$.
We also run numerical simulations for the ARZ model and compare the model error for ARZ and GSOM in~\Cref{tab:error ARZ GSOM}. As the results show, by considering traffic heterogeneity, the GSOM model provides a more accurate estimation of all three macroscopic traffic values: density, speed and flow. For example, the speed simulation error using  the GSOM is reduced by around 20~\%.

\section{Reconstruction of Traffic Heterogeneity with Scarce Trajectory Data}\label{sec:data driven}

In the previous analysis, we focus on longitudinal traffic and adopt a micro-to-macro framework, where we first analyze vehicles’ driving behavior and then formulate a macroscopic traffic attribute from individual behavior patterns. To adopt the above methods to real-world traffic involves two main challenges: First, the previous analysis relies on trajectory data collected in  experimental settings, where sufficient data are available for detailed behavioral analysis. However, in real-world traffic, vehicle trajectories are often observed over only short time intervals, and there is little or even no historical data available to construct accurate behavioral models. Second, vehicles, whether HVs or AVs, exhibit not only longitudinal car-following behaviors but also lateral dynamics such as lane-changing and merging. These behaviors significantly complicate the analysis of driver attributes, particularly the calibration of desired speed.

In this section, we further extend the proposed micro-to-macro formulation to more complex traffic scenarios  with limited data availability. We first analyze vehicle trajectories from real highway datasets and show that the proposed stochasticity-related value remains approximately constant during the driving process, thus continuing to work as an indicator of driver attributes. Then, we propose an end-to-end learning algorithm that directly gets macroscopic traffic attribute values from vehicle trajectories.  We use the TGSIM and NGSIM datasets which are collected in real highways~\cite{talebpour2024third,NGSIM}. For the TGSIM dataset, we use three datasets: I90-I94 Stationary, I90-I94 Movings, and I294 L1. For the NGSIM dataset, we use the data collected in three time periods: 04:00-04:15, 05:00-05:15, and 05:15-05:30.

\subsection{Calibration of driver heterogeneity}

\begin{figure}[!t]
	\centering
	One example AV\\
	\subcaptionbox{Scatter plot of $C^{\mathrm{absave}}$ and $C^{\mathrm{std}}$}{\includegraphics[width=0.4\linewidth]{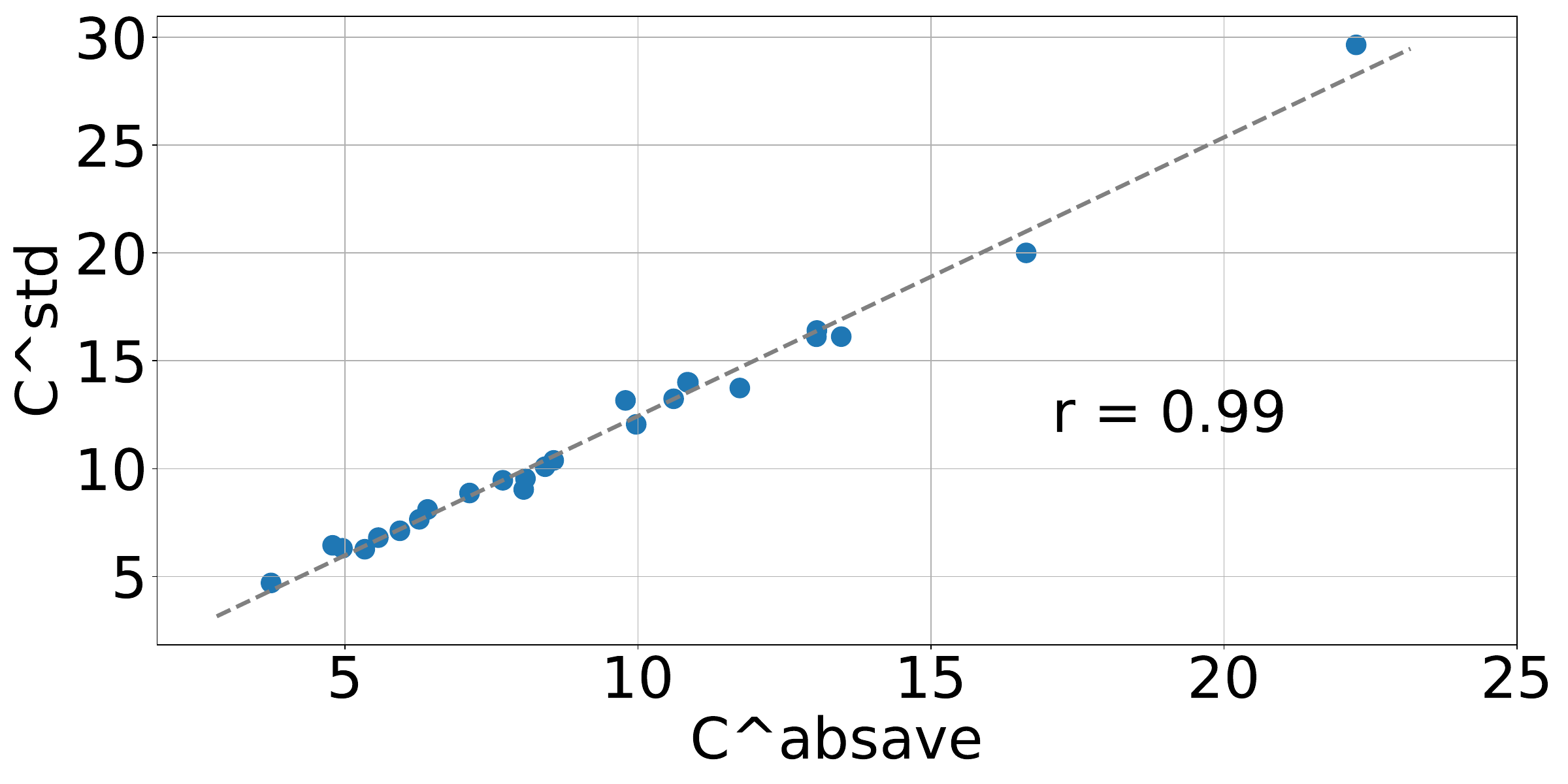}}
	\subcaptionbox{$\omega_k^{\mathrm{L}}$}{\includegraphics[width=0.4\linewidth]{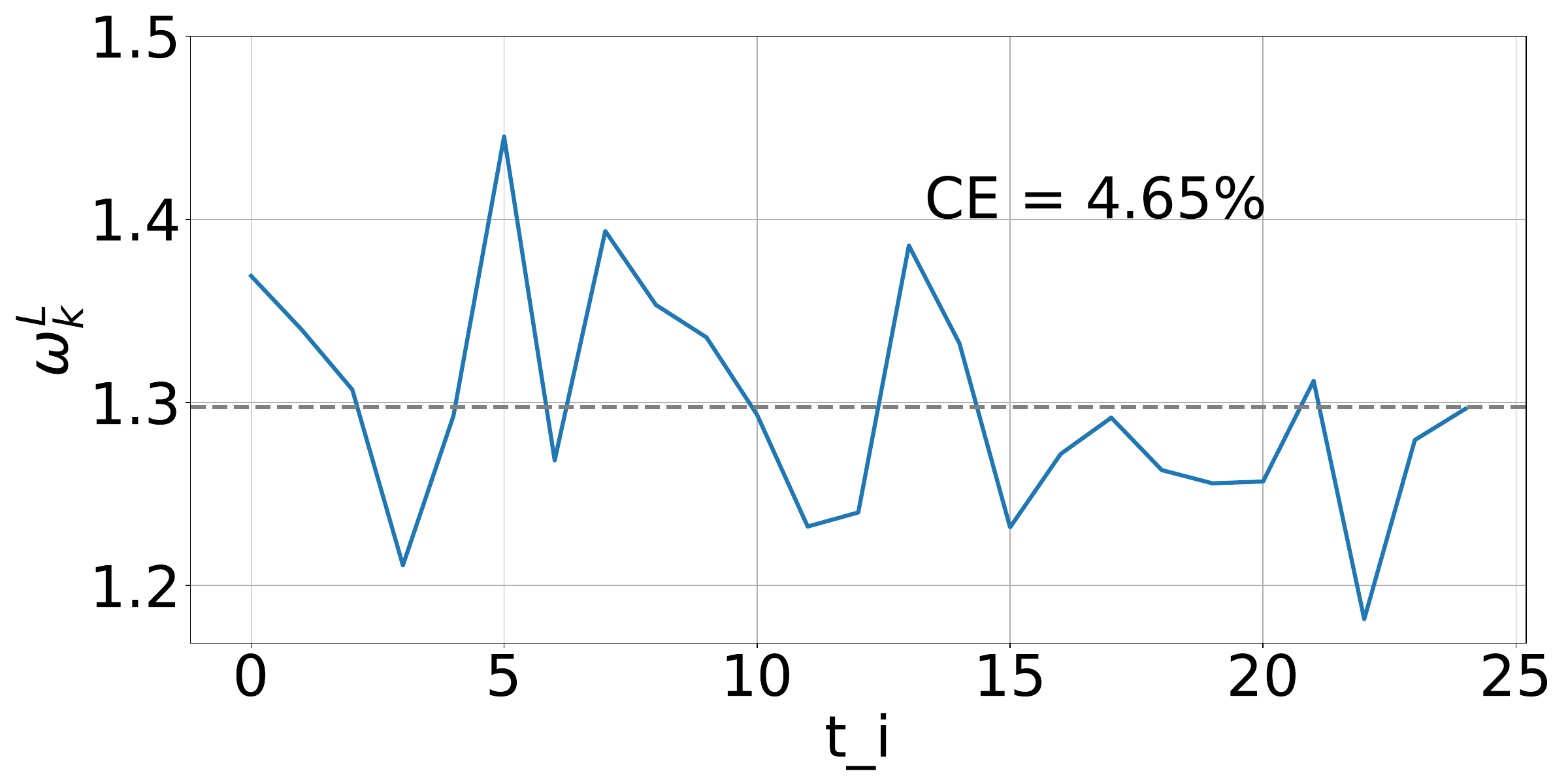}}
	\\
	One example HV \\
	\subcaptionbox{Scatter plot of $C^{\mathrm{absave}}$ and $C^{\mathrm{std}}$}{\includegraphics[width=0.4\linewidth]{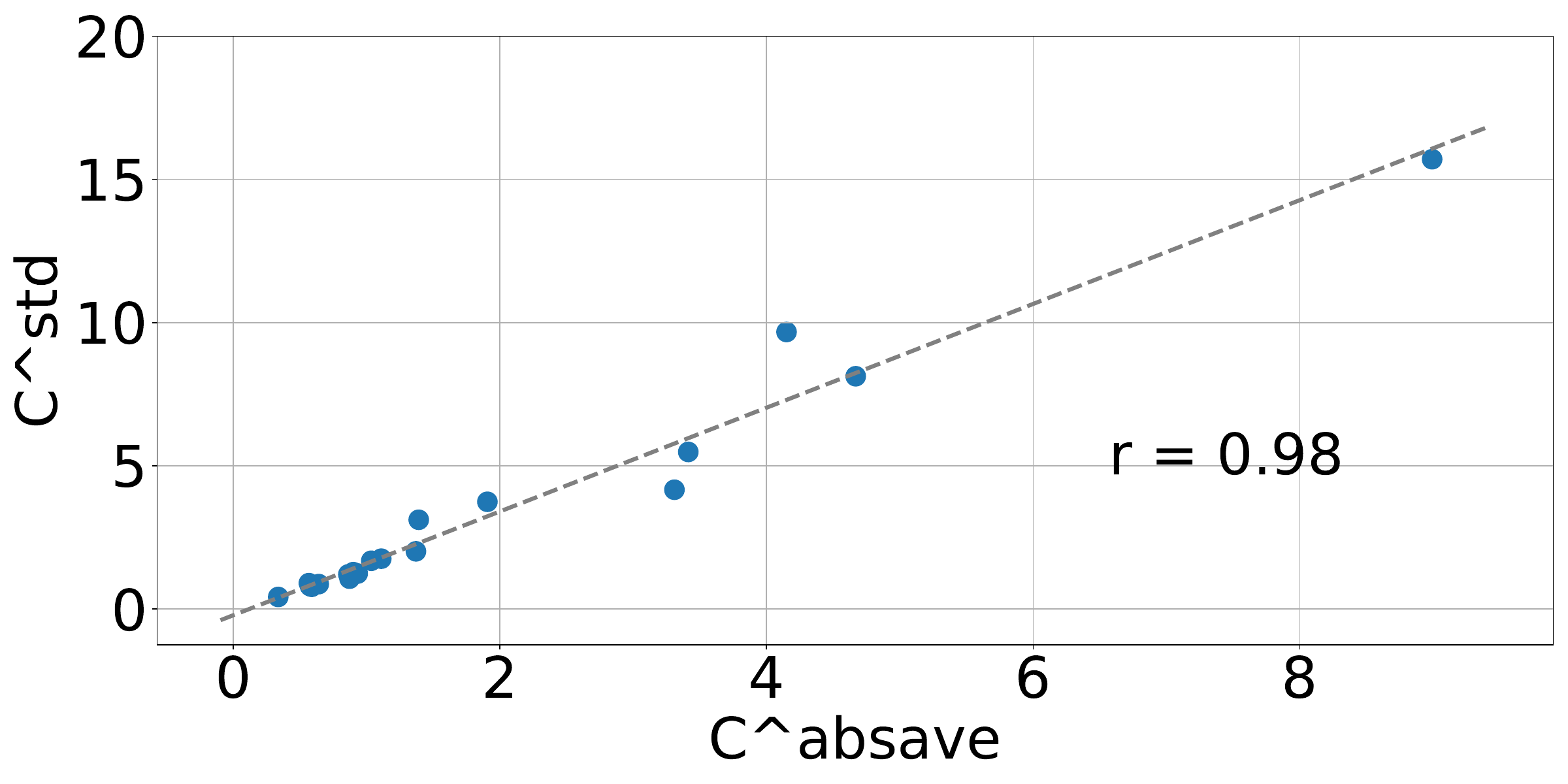}}
	\subcaptionbox{$\omega_k^{\mathrm{L}}$}{\includegraphics[width=0.4\linewidth]{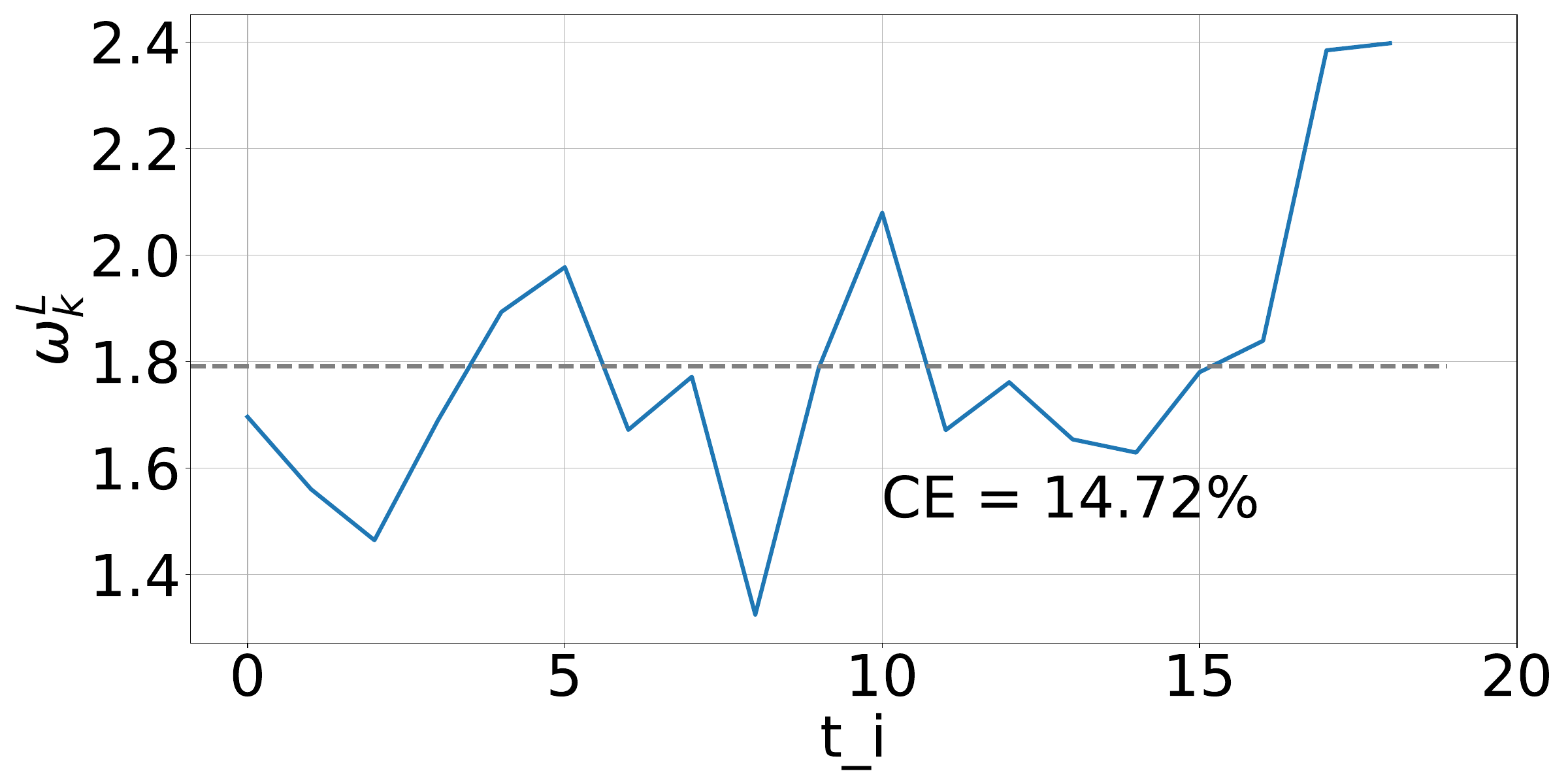}} 
	\caption{Vehicle attribute for an example AV and an example HV from the TGSIM dataset. In the TGSIM dataset, the vehicle has both longitudinal cruising and lateral lane-changing behaviors. In such a  more complicated traffic scenario,  there also exists a linear relationship between the $C^{\mathrm{absave}}$ and $C^{\mathrm{std}}$, which is used to extract vehicle attribute in this work.}\label{fig:TGSIM:jerk example}
\end{figure}

\begin{figure}[!t]
	\centering
	I-90 I-94 Stationary \\
	\subcaptionbox{Constancy error}{\includegraphics[width=0.4\linewidth]{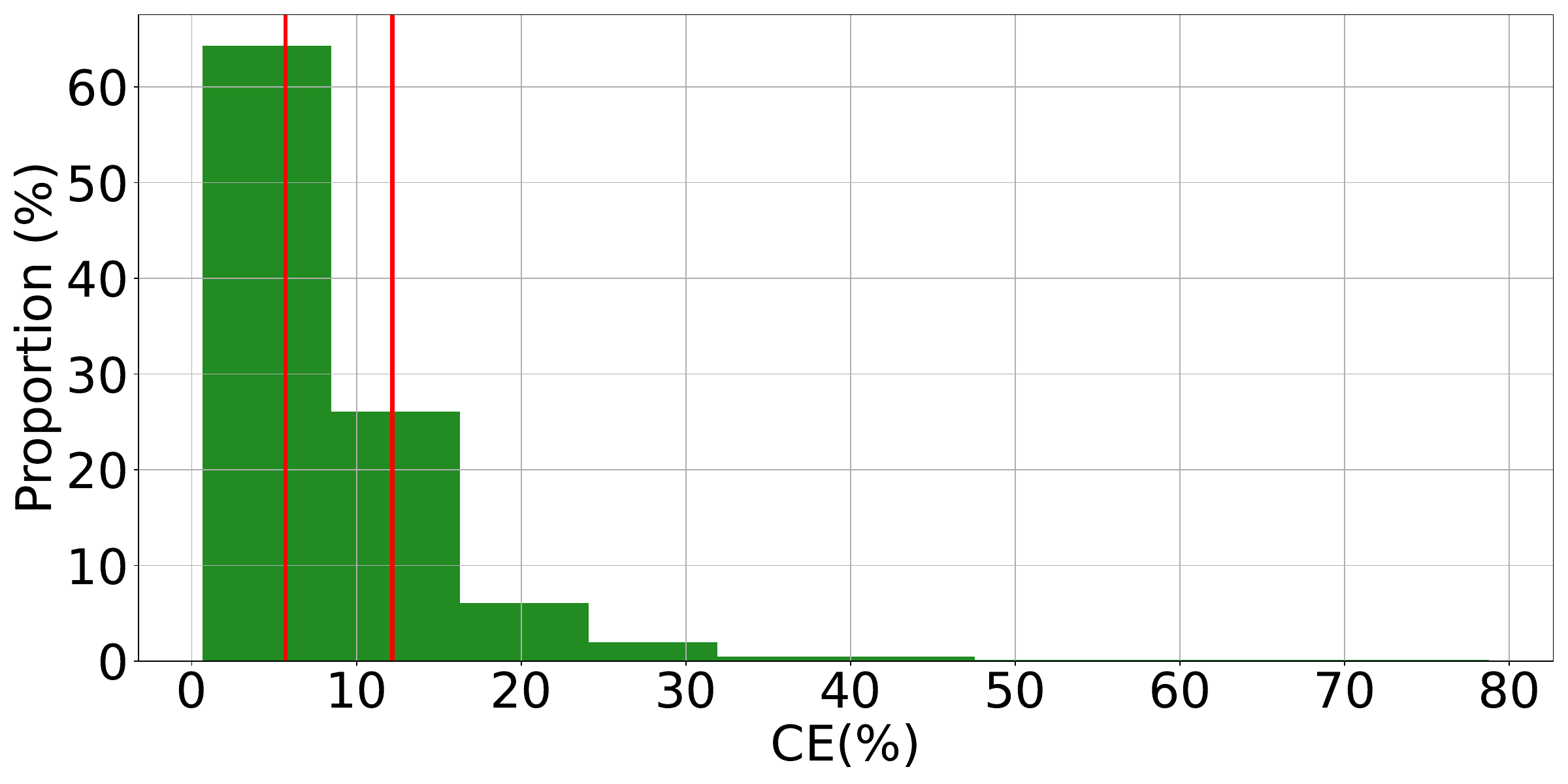}}
	\subcaptionbox{Driver attribute}{\includegraphics[width=0.4\linewidth]{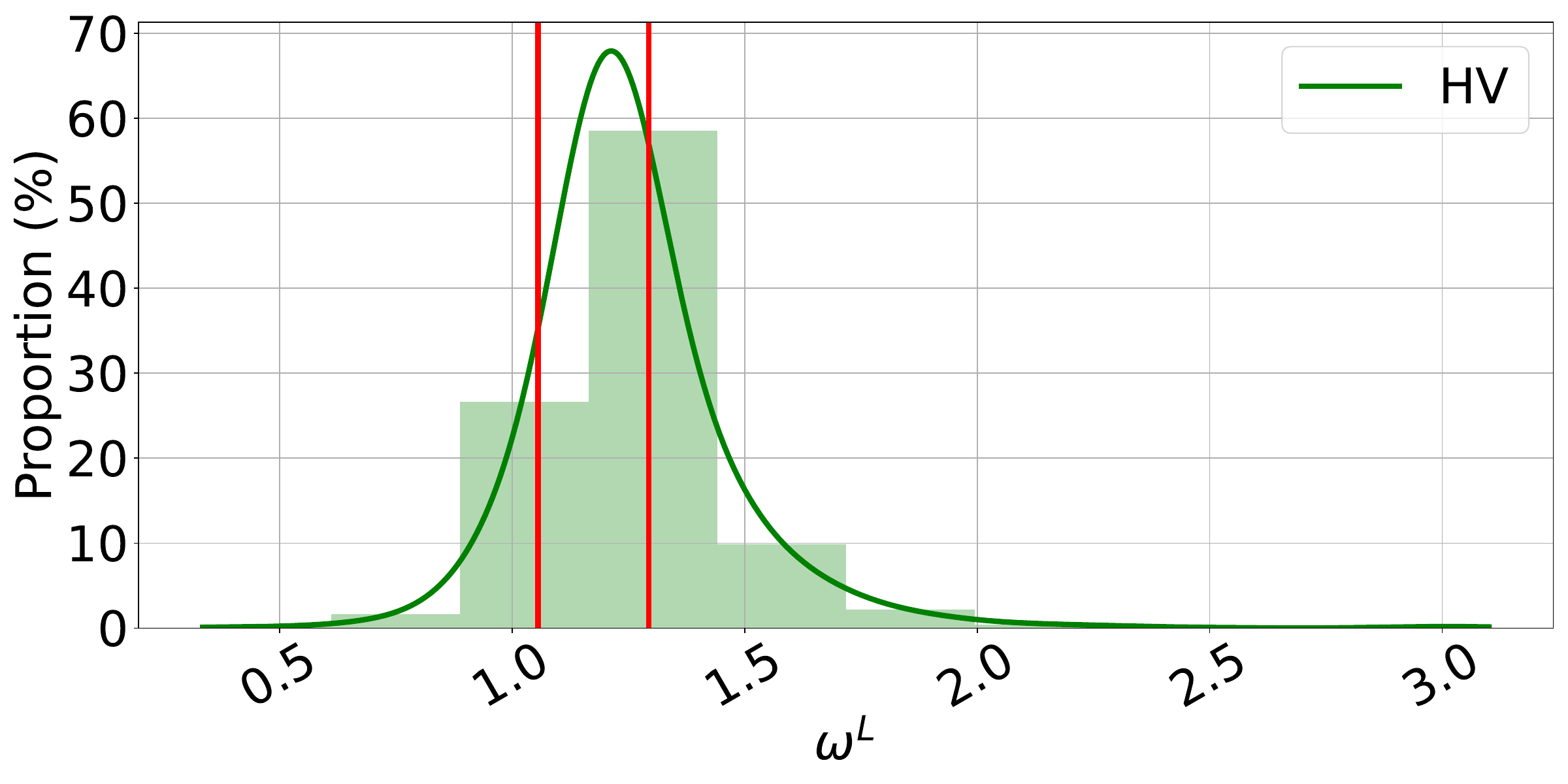}} \\
	I-90 I-94 Movings \\
	\subcaptionbox{Constancy error}{\includegraphics[width=0.4\linewidth]{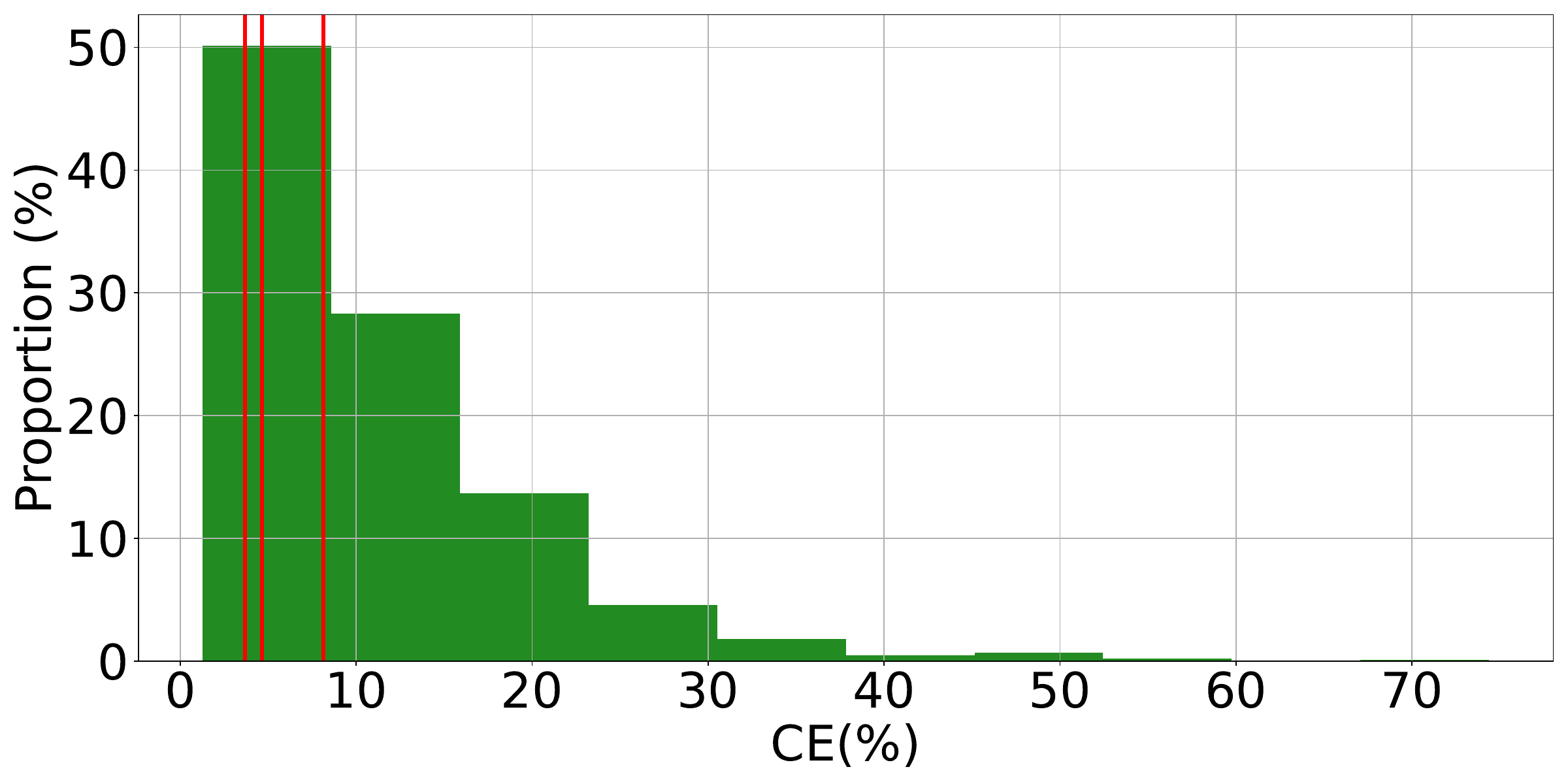}}
	\subcaptionbox{Driver attribute}{\includegraphics[width=0.4\linewidth]{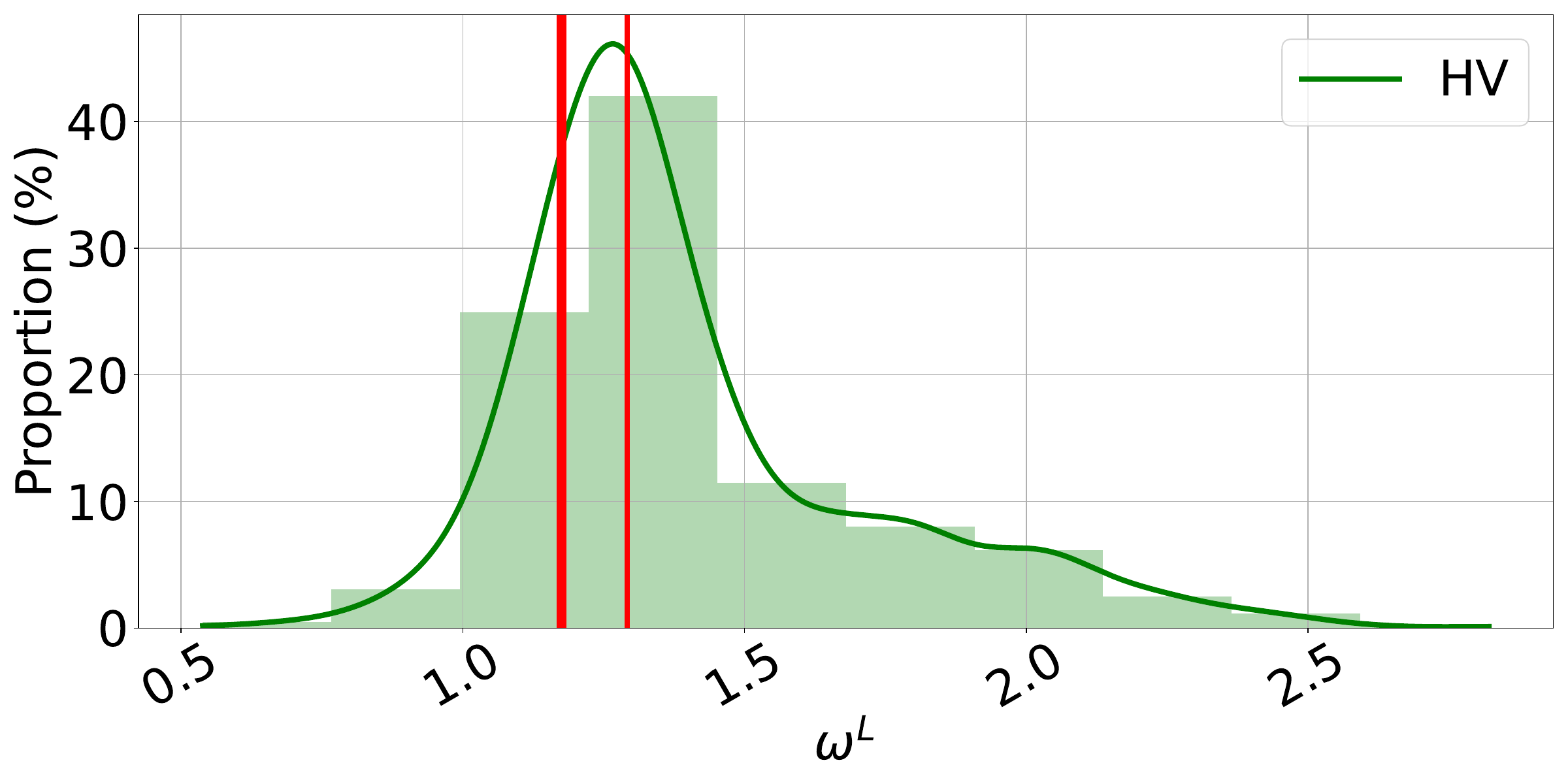}} \\
	I294 L1 \\
	\subcaptionbox{Constancy error}{\includegraphics[width=0.4\linewidth]{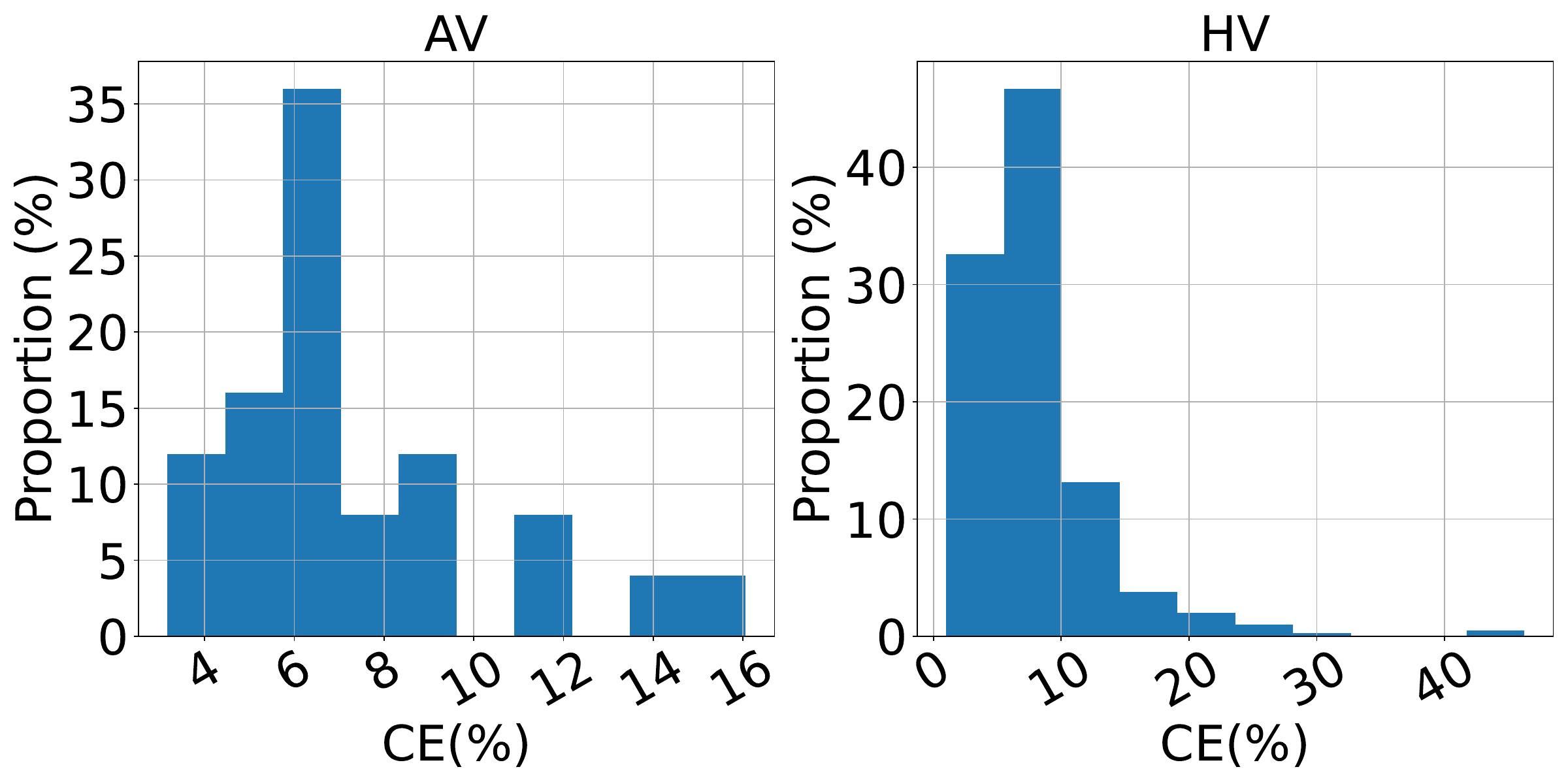}}
	\subcaptionbox{Driver attribute}{\includegraphics[width=0.4\linewidth]{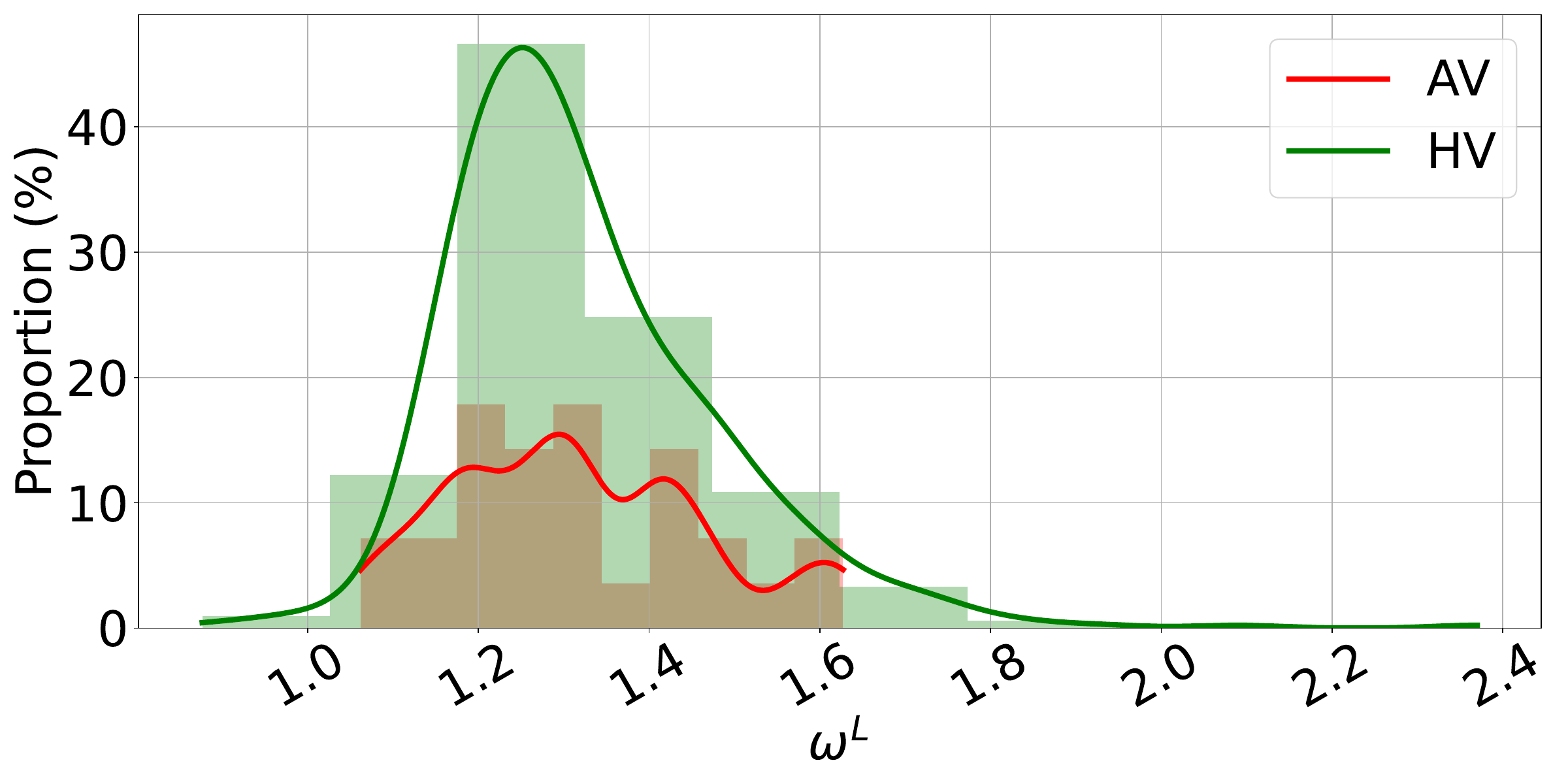}}
	\caption{The vehicle attribute for three sub-datasets in the TGSIM dataset.}\label{fig:TGSIM:jerk}
\end{figure}

In the above analysis for longitudinal traffic, we have selected two values, average of absolute jerk $C^{\mathrm{absave}}$, and standard deviation of jerk $C^{\mathrm{std}}$, to calculate a driver attribute value. We show here that these two values still work if we consider lateral dynamics. We select one example AV and one example HV from the TGSIM dataset, and give in~\Cref{fig:TGSIM:jerk example}  the scatter plot of $C^{\mathrm{absave}}$ and $C^{\mathrm{std}}$. As the scatter plot shows, there is still a strong linear relationship between these two values, for both AV and HV. The regression correlation coefficient for the example AV and HV is 0.99 and 0.98 respectively, which indicates a high linear correlation between $C^{\mathrm{absave}}$ and $C^{\mathrm{std}}$. We use the same formulation as for the longitudinal traffic~\Cref{eq:omega L jerk} to get the  $\omega^{\mathrm{L}}$ value. \Cref{fig:TGSIM:jerk example}(b) and~\Cref{fig:TGSIM:jerk example}(d) give the $\omega^{\mathrm{L}}$ during the driving process for the example AV and HV respectively. As the figures show, even with lateral motions, the proposed $\omega^{\mathrm{L}}$ still remains approximately the same during the driving process.   In \Cref{fig:TGSIM:jerk}, we further give the constancy error (CE) of all vehicles in the dataset. For the I290-I294 Stationary dataset in \Cref{fig:TGSIM:jerk}(a) and I290-I294 Movings dataset in  \Cref{fig:TGSIM:jerk}(c), the histogram gives the distribution of CE for all HVs, and the three vertical red line shows the CE for the three AVs. We see that for the three AVs, their CE are lower than 10~\%. And for HVs, although some HVs present a larger CE around 60~\%, over 90~\% of drivers still have the CE lower than 30~\%. For the I294-L1 dataset in~\Cref{fig:TGSIM:jerk}(e), we see that the CE for all HVs are lower than 50~\%, and the CE for all AVs are lower than 16~\%.  Therefore, the proposed $\omega^{\mathrm{L}}$  is a driver-specific attribute that remains approximately constant during the driving process.

We  compare the distribution of the driver-level stochasticity-related attribute $\omega^{\mathrm{L}}$ for all HVs and AVs, as shown in~\Cref{fig:TGSIM:jerk}(b)(d)(f). Similar to the findings for longitudinal dynamics, it is evident that both inter-class (between AVs and HVs) and intra-class (within AVs and within HVs) heterogeneity exist.
For the I290-I294 Stationary dataset in \Cref{fig:TGSIM:jerk}(b) and I290-I294 Movings dataset in  \Cref{fig:TGSIM:jerk}(d), if we reduce the mixed traffic to a two-class AV–HV model, the average $\omega^{\mathrm{L}}$ values for AVs and HVs are approximately equal, both around 1.2.
In the I-294 L1 dataset shown in~\Cref{fig:TGSIM:jerk}(f), substantial overlap is observed between AVs and HVs, with most $\omega^{\mathrm{L}}$ values within the range of 1.0 to 1.6 for both classes. 
The intra-class heterogeneity further validates that, in mixed traffic, the major source of heterogeneity arises not between AVs and HVs, but among individual vehicles. As a result, a continuous traffic attribute variable should be used instead of a discrete vehicle-type classification.

\subsection{Reconstruction of traffic heterogeneity from data}

\begin{figure}[!t]
	\centering
	I90 I94 Stationary \\
	\includegraphics[width=0.4\linewidth]{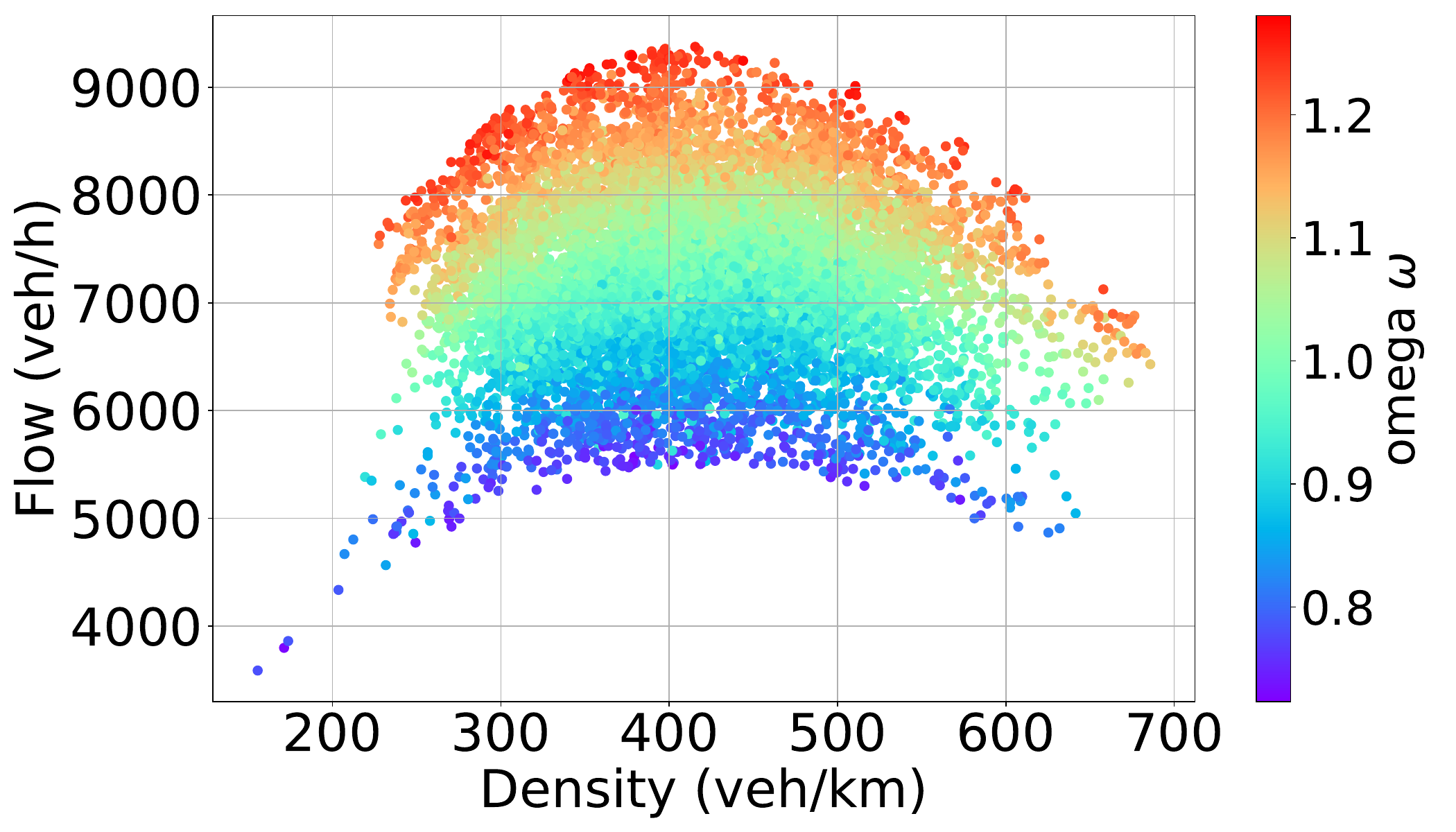}
	\includegraphics[width=0.4\linewidth]{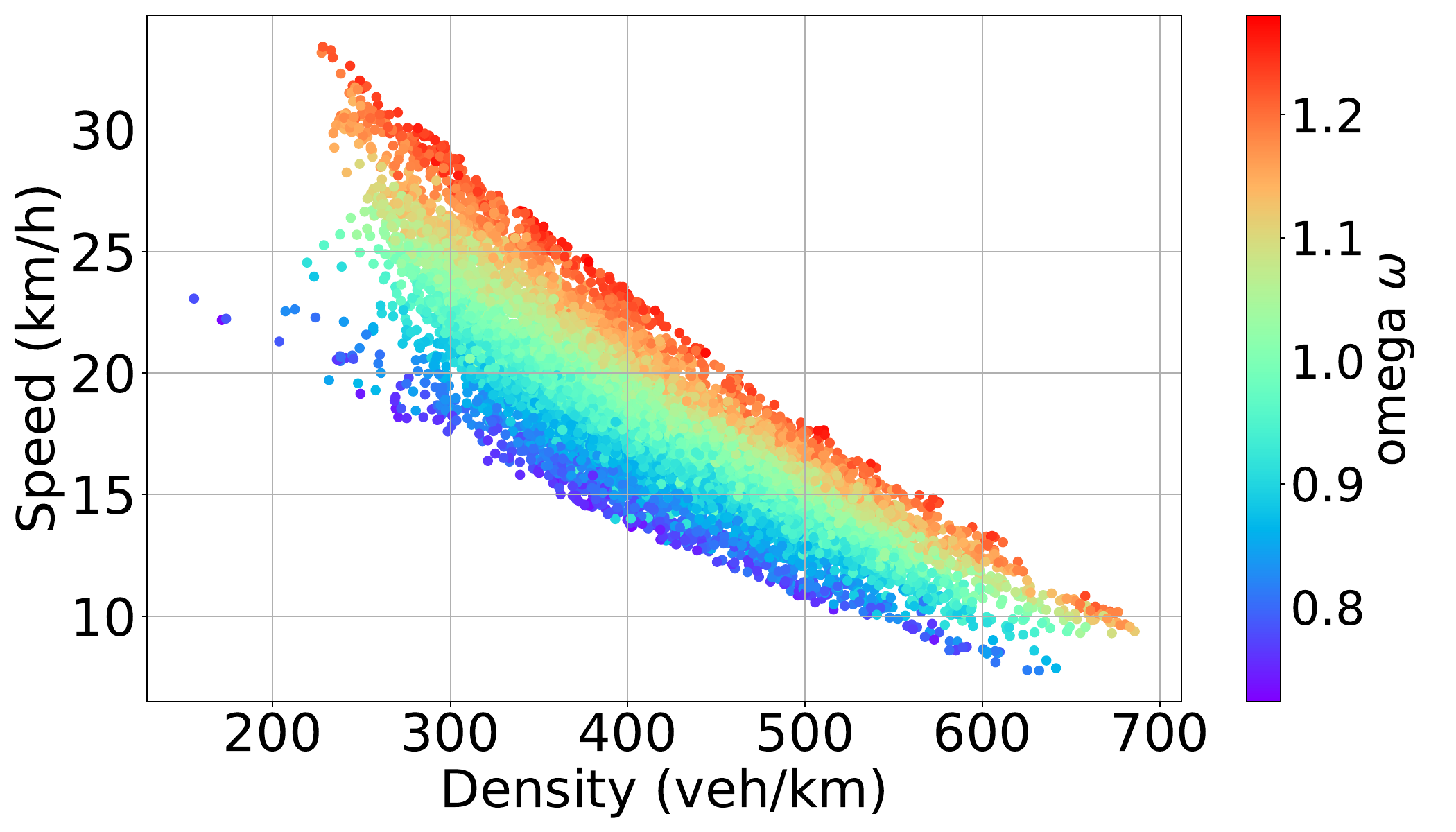} \\
	I90 I94 Movings \\
	\includegraphics[width=0.4\linewidth]{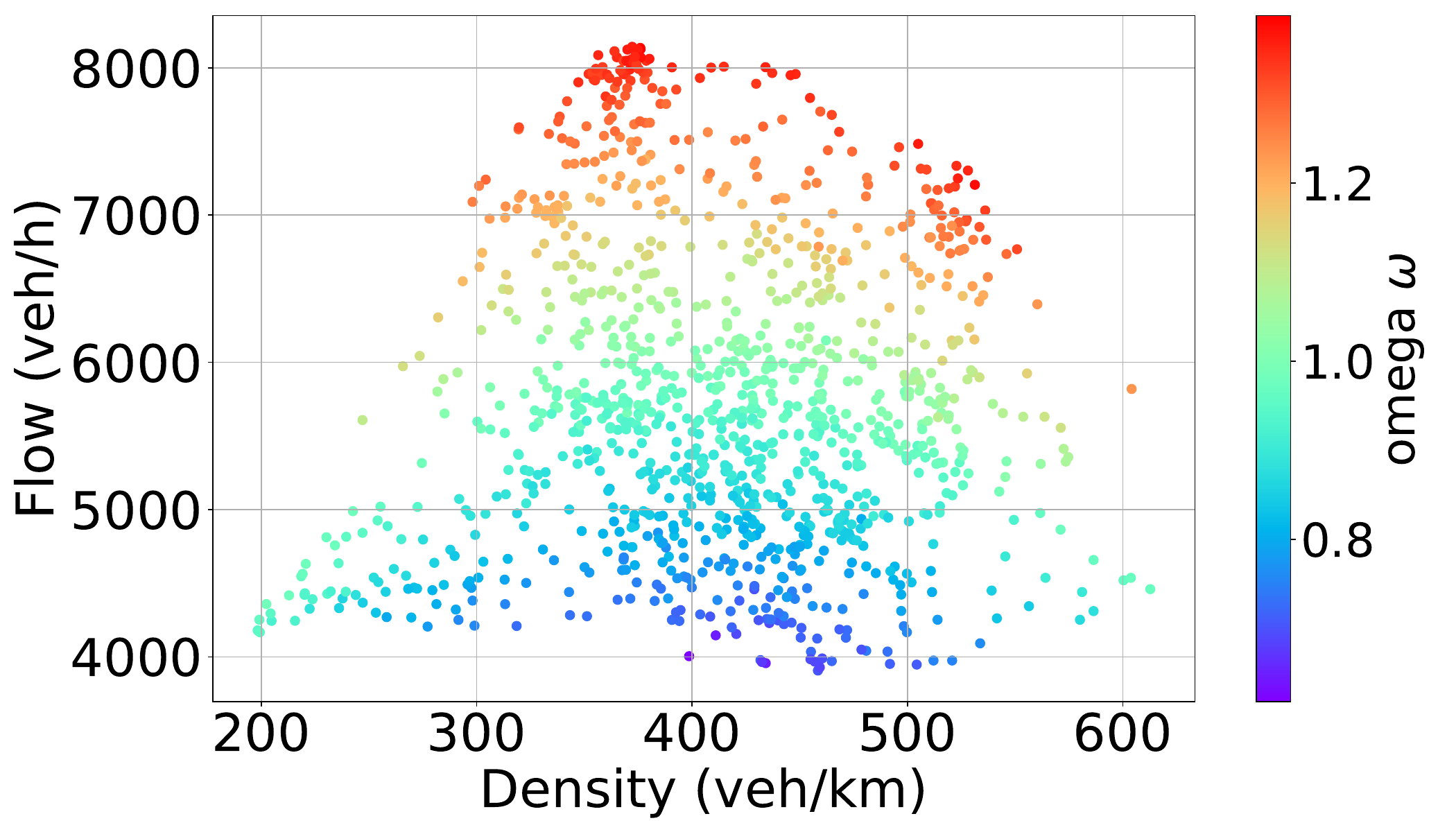}
	\includegraphics[width=0.4\linewidth]{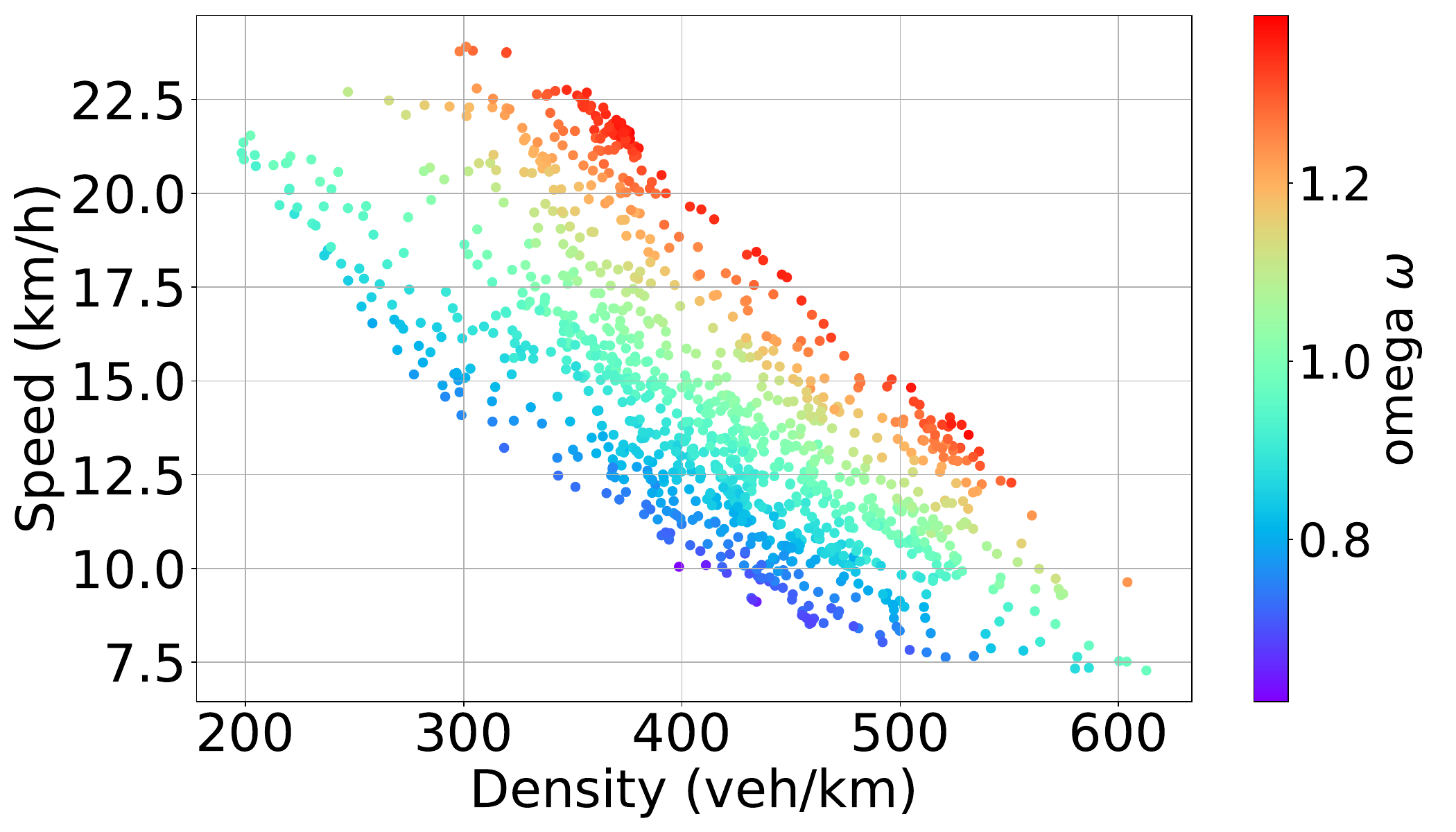} \\
	I 294 L1\\
	\includegraphics[width=0.4\linewidth]{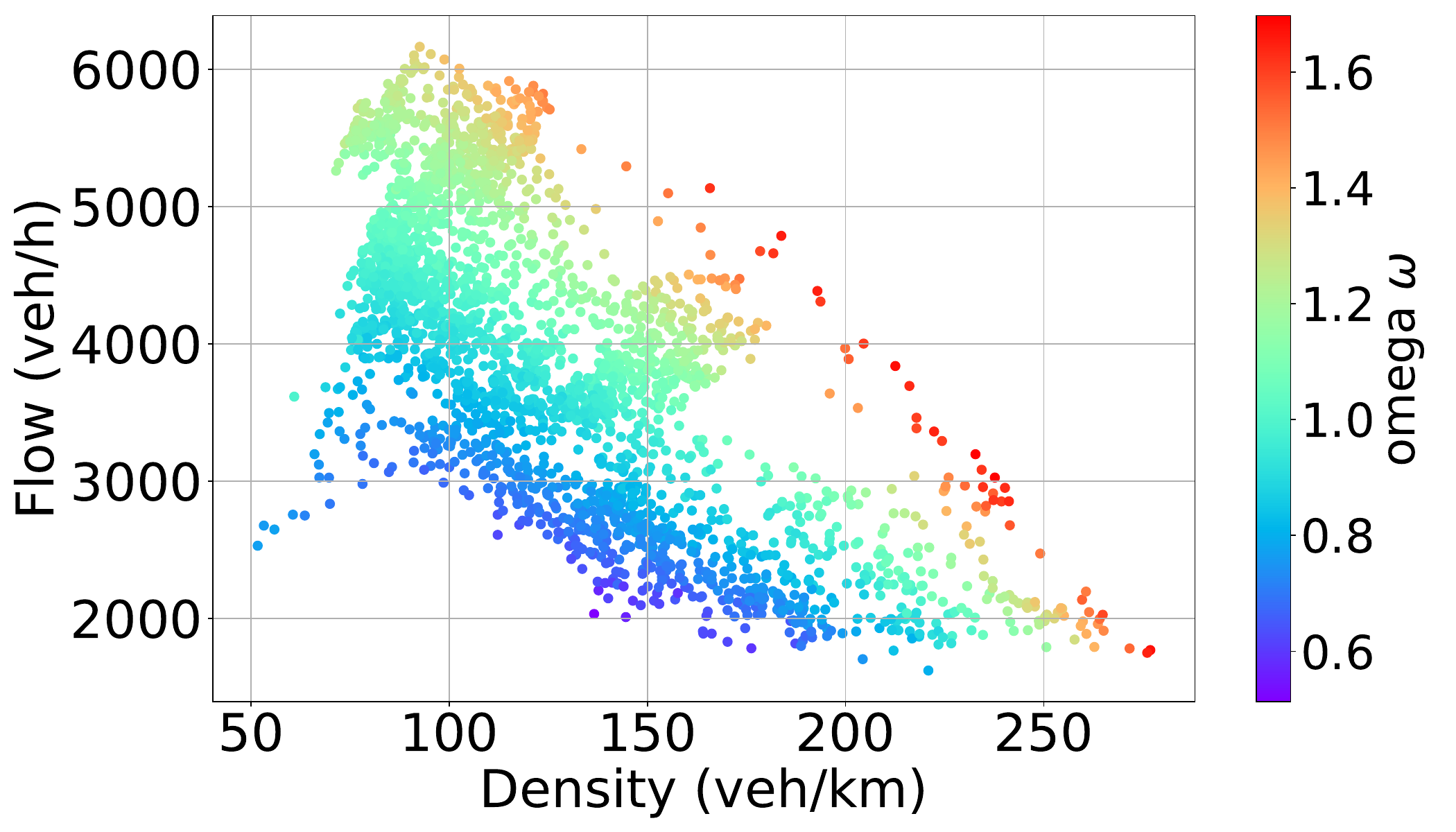}
	\includegraphics[width=0.4\linewidth]{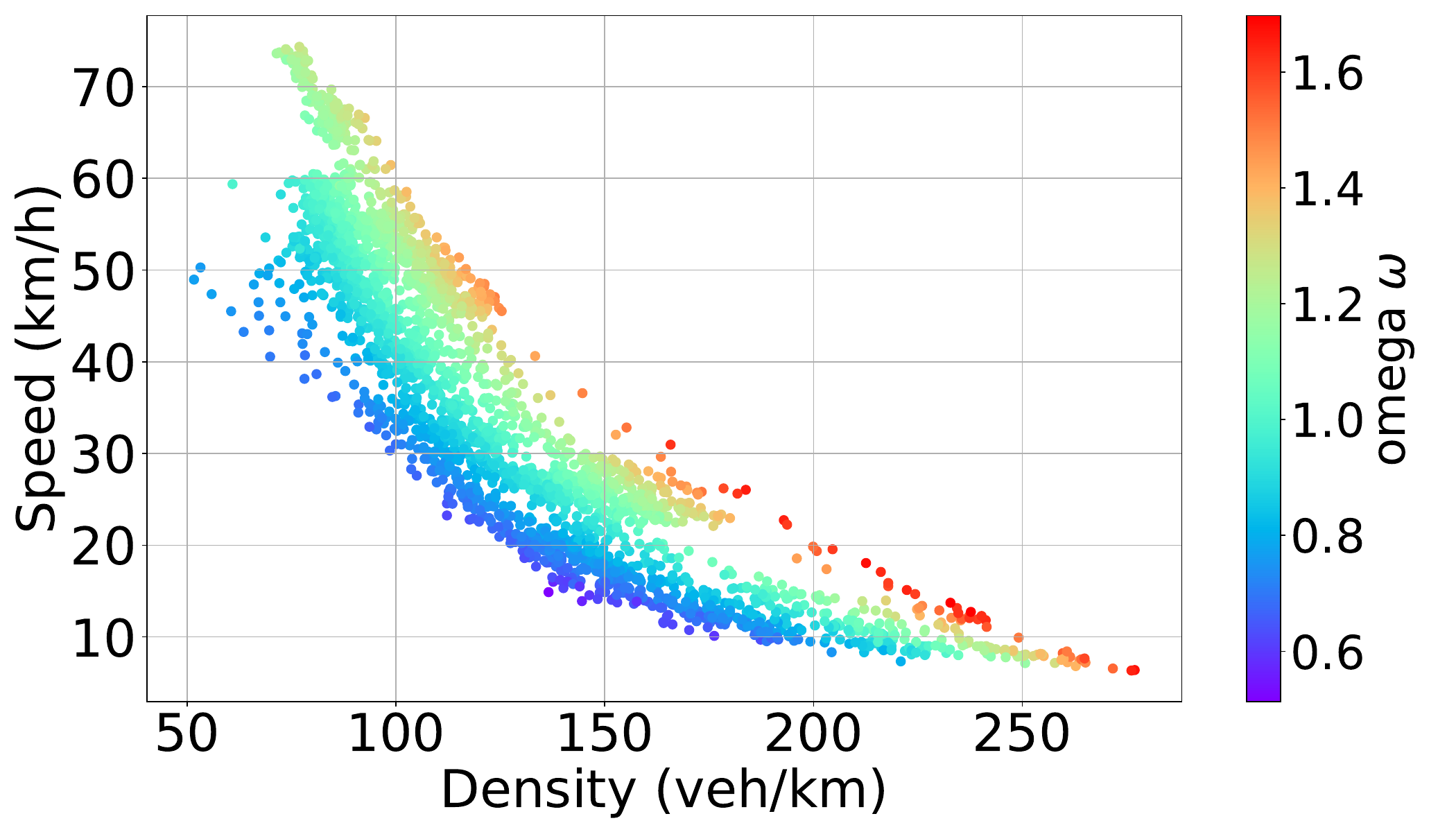}
	\caption{The fundamental diagram separated by the learned traffic attribute for the TGSIM dataset.}
	\label{fig:TGSIM:FD}
\end{figure}

\begin{figure}[!t]
	\centering
	04:00 - 04:15 \\
	\includegraphics[width=0.4\linewidth]{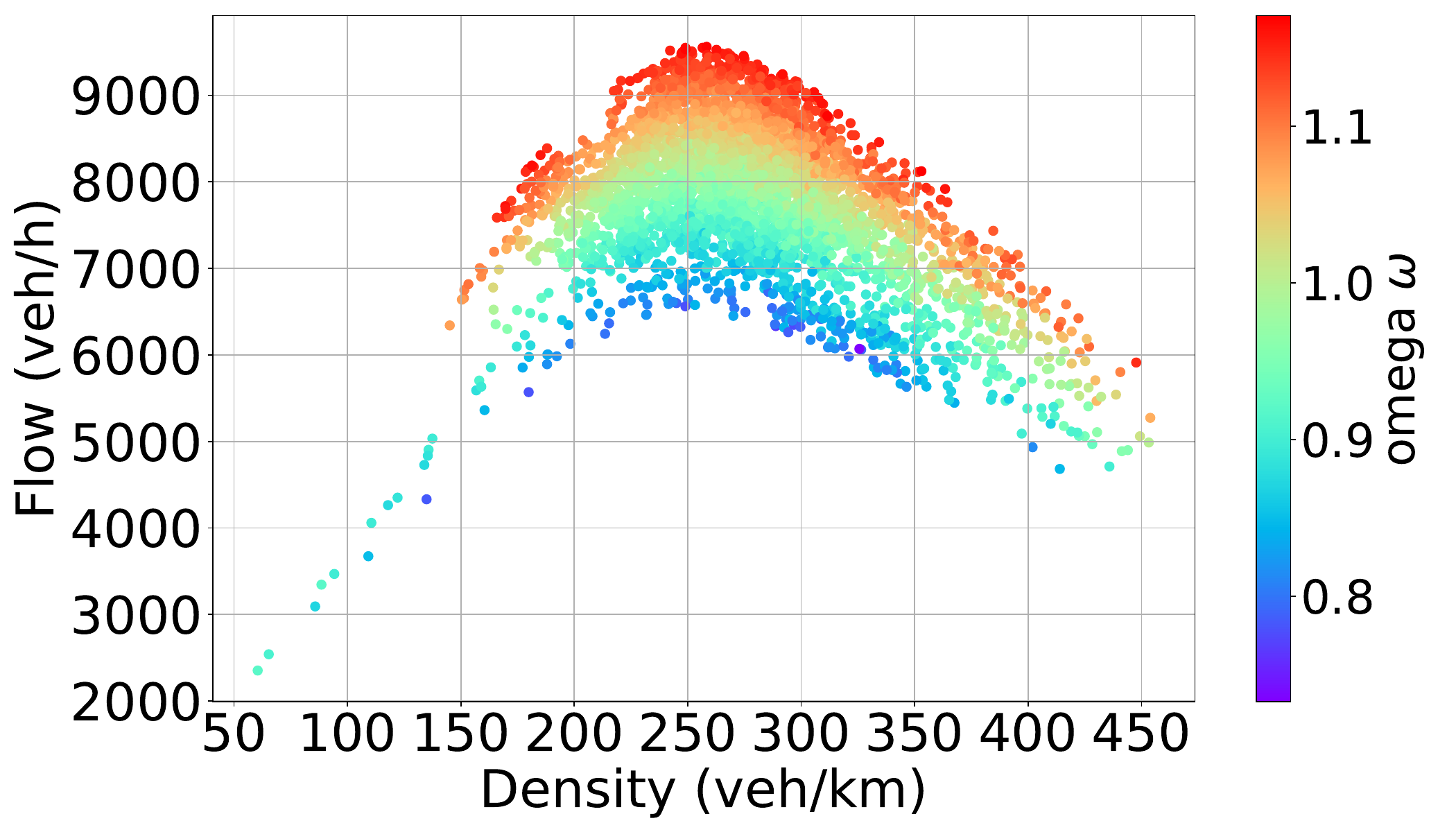}
	\includegraphics[width=0.4\linewidth]{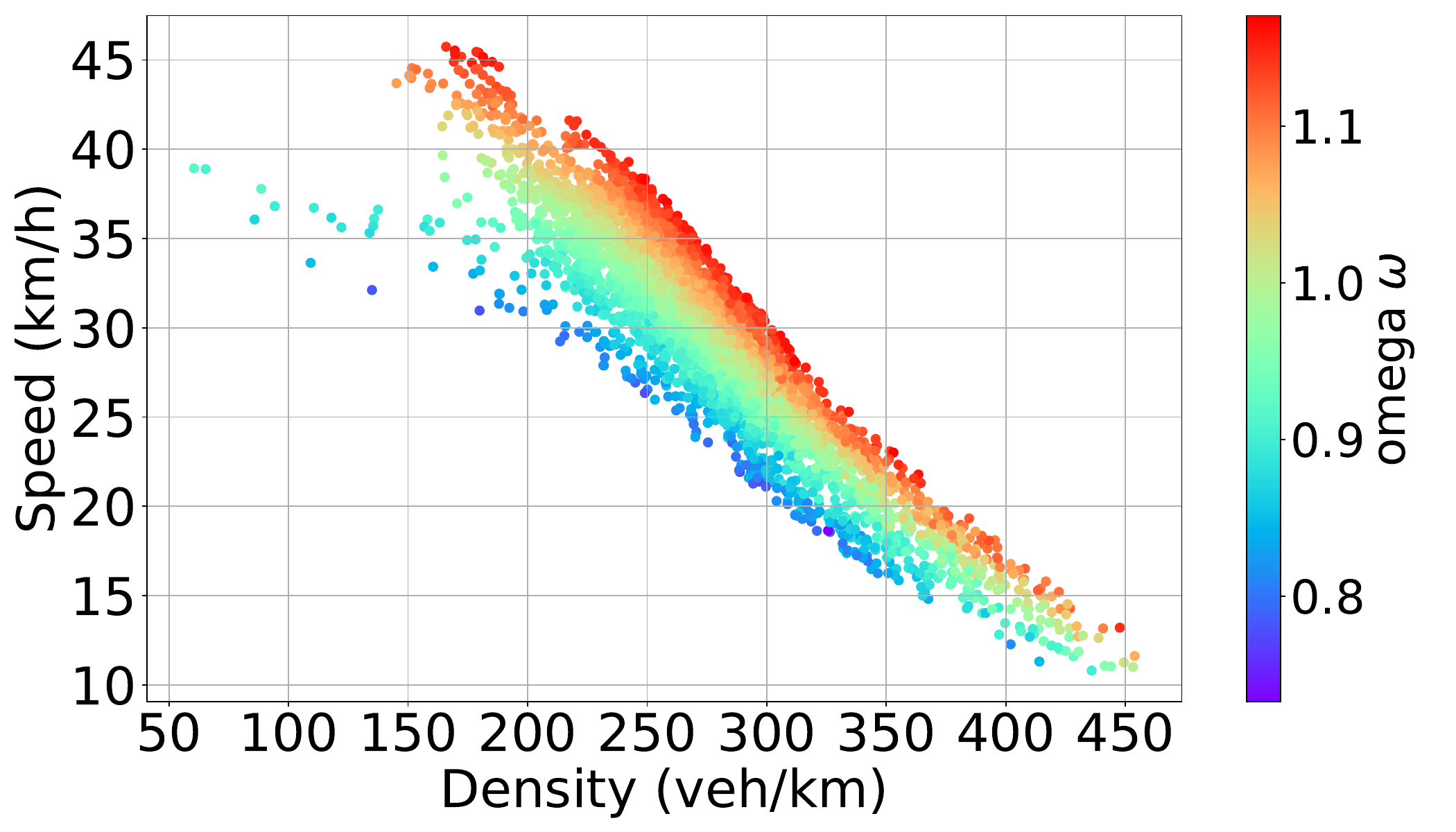} \\
	05:00 - 05:15 \\
	\includegraphics[width=0.4\linewidth]{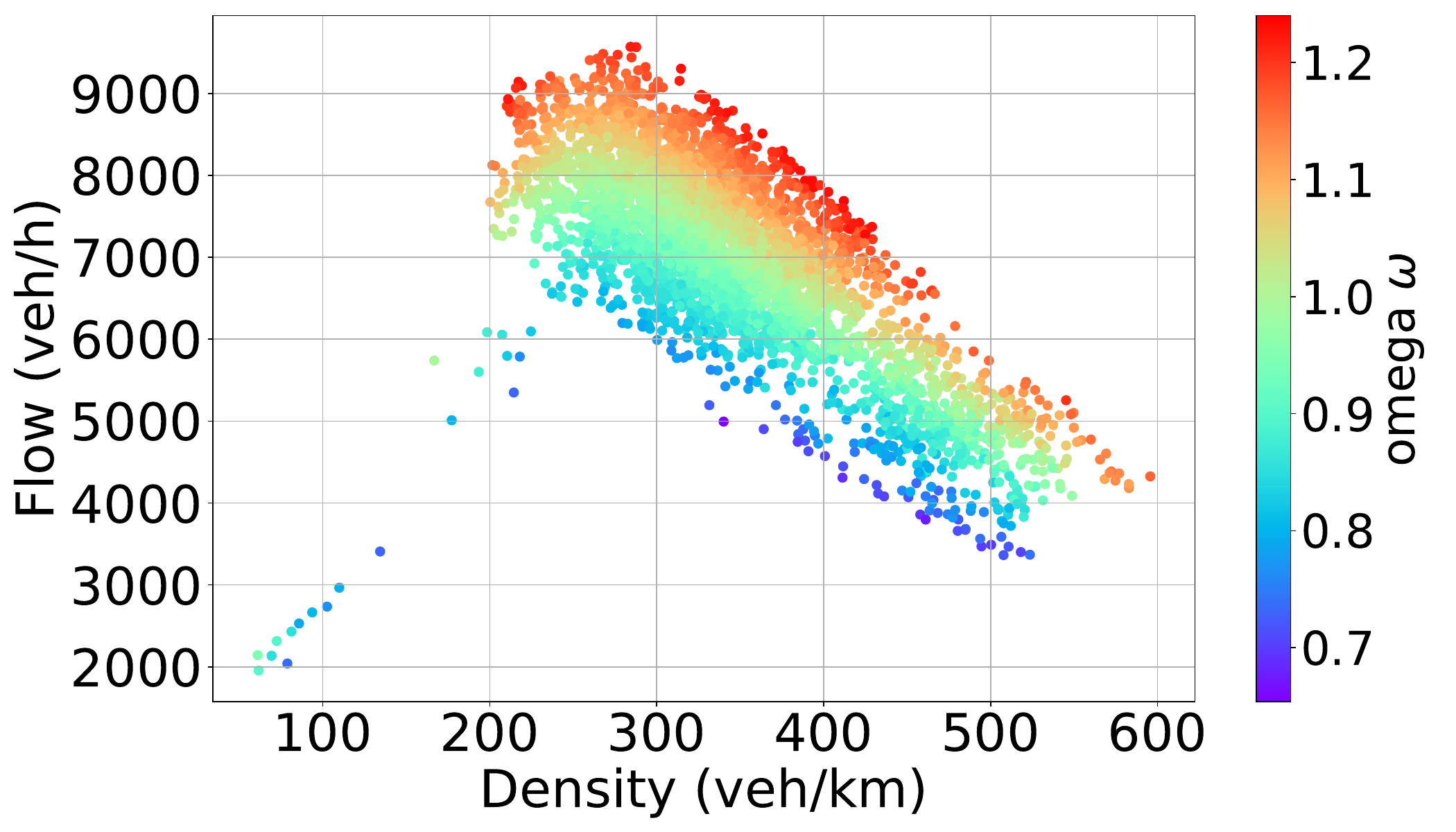}
	\includegraphics[width=0.4\linewidth]{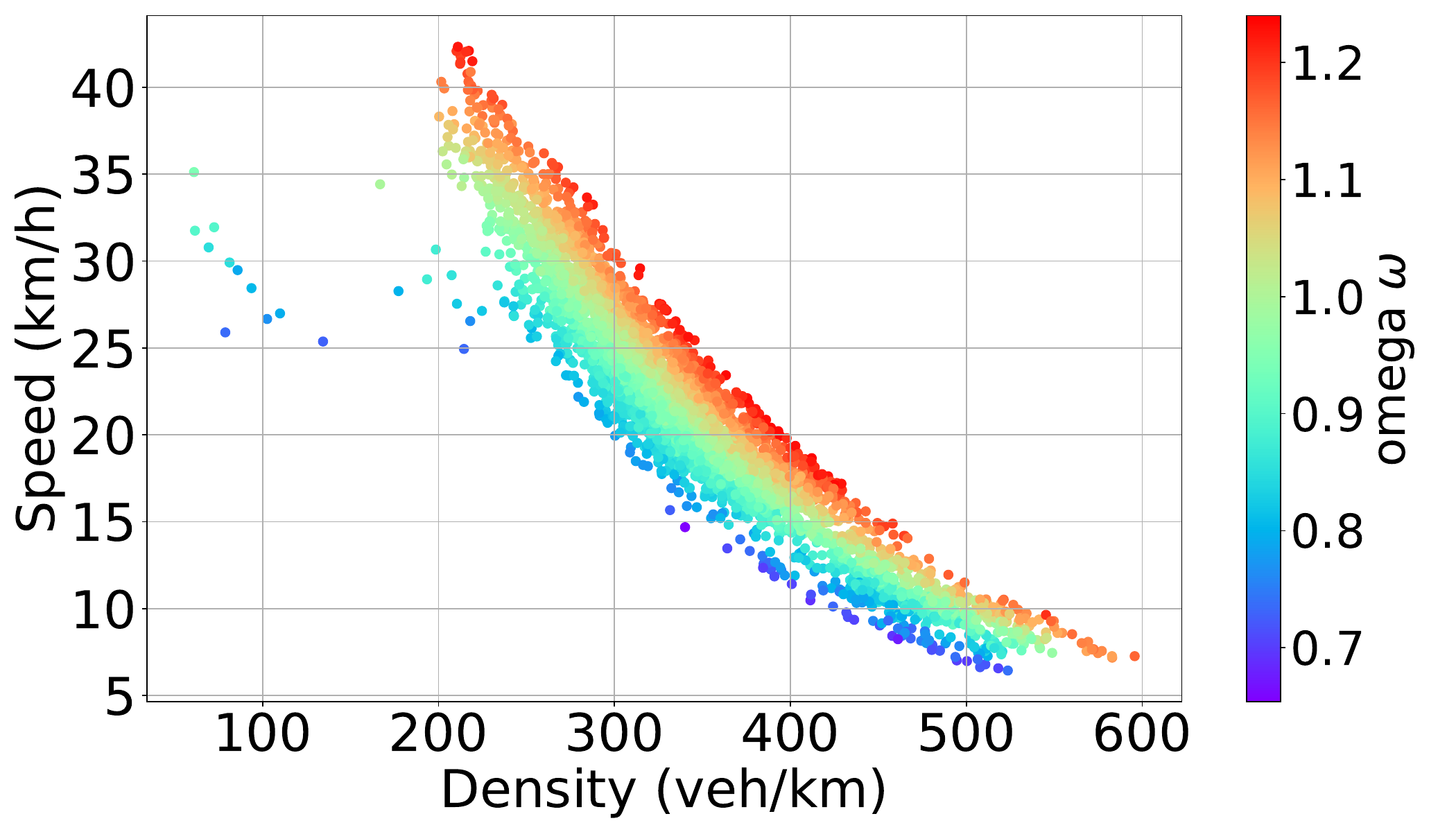} \\
	05:15 - 05:30\\
	\includegraphics[width=0.4\linewidth]{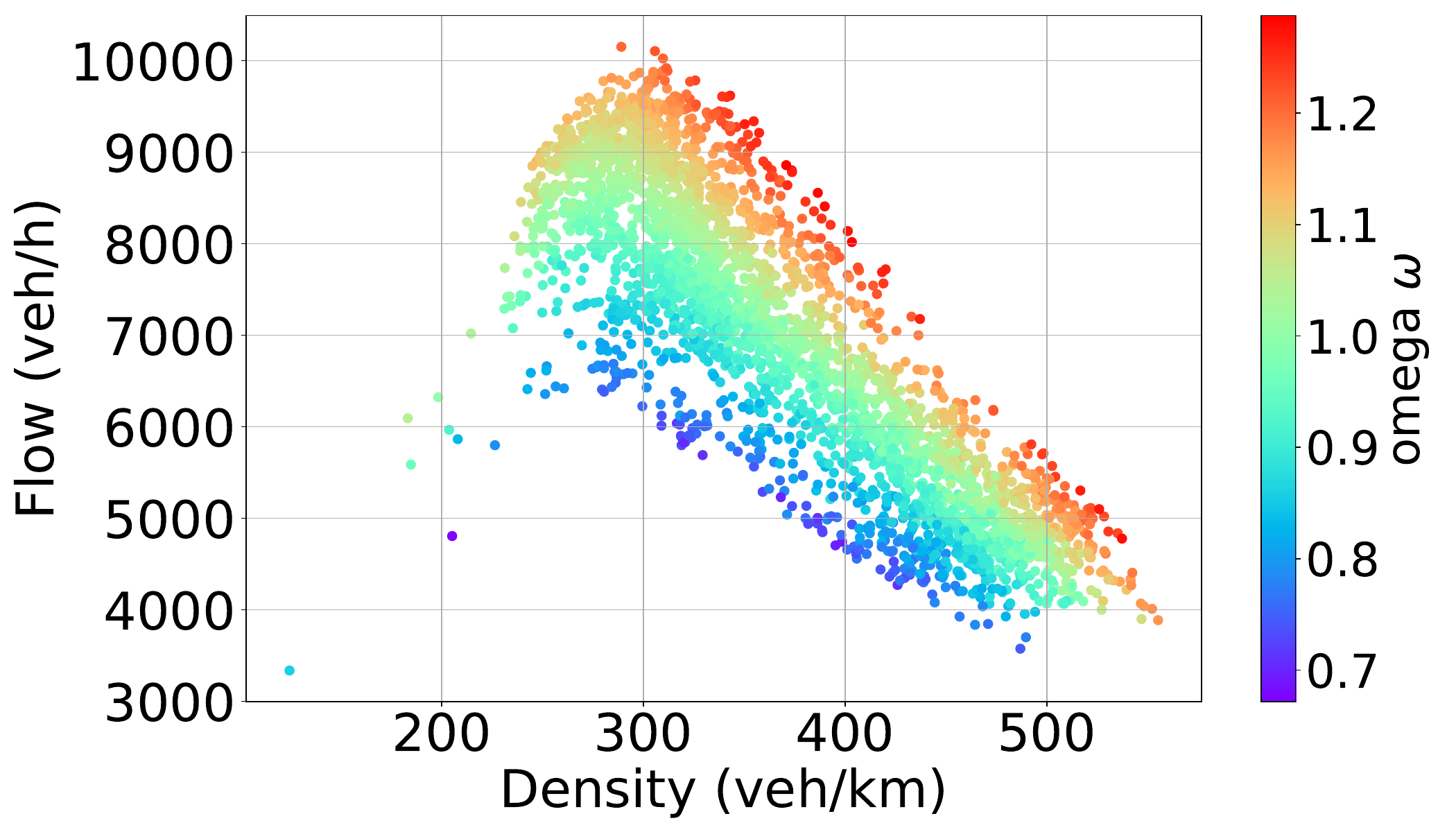}
	\includegraphics[width=0.4\linewidth]{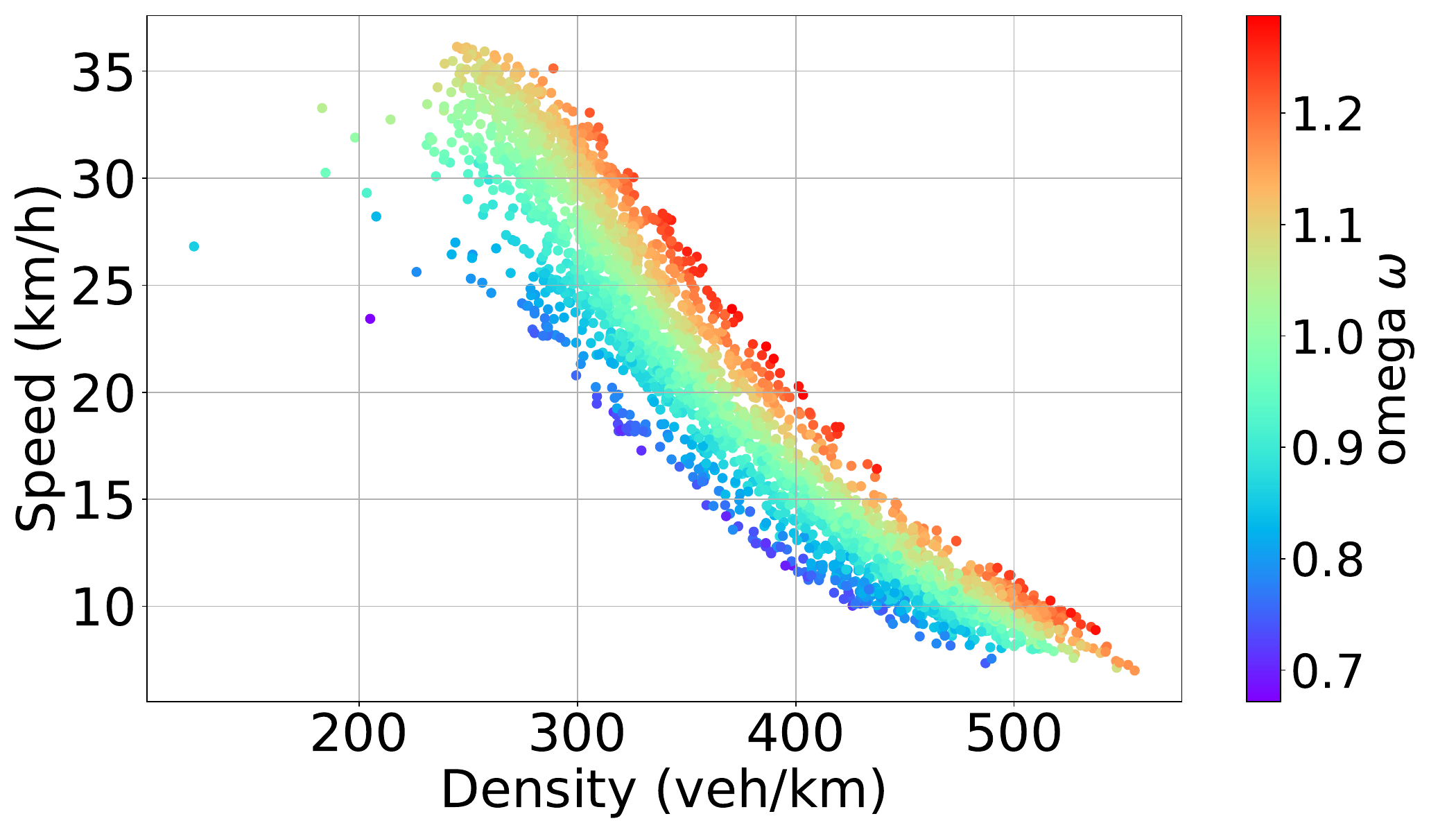}
	\caption{The fundamental diagram separated by the learned traffic attribute for the NGSIM dataset.}
	\label{fig:NGSIM:FD}
\end{figure}

\begin{table}[!t]
    \centering
    \caption{The calibration error (\%) for one-variable fundamental diagram and two-variable fundamental diagram}
    \label{tab:error 1FD 2FD}
    \begin{tabular}{c|c|c|c||c|c|c}
    \multirow{2}{*}{} & \multicolumn{3}{c||}{NGSIM}  & \multicolumn{3}{c}{TGSIM} \\ \cline{2-7} 
         &  04:00-04:15 & 05:00-05:15 & 05:15-05:30 & I90-I94 Stationary & I90-I94 Movings & I-294\\ \hline
     $V(\rho)$ & 8.50 & 11.16 & 12.19 & 11.10 & 18.21 & 21.74 \\ \hline
     $\mathcal{V}(\rho,\omega)$ & 1.46 & 1.65 & 1.86 & 2.45 & 2.11 & 2.02 \\
    \end{tabular}
    
\end{table}

In this part, we design a neural network to learn a mapping from vehicle's trajectory to macroscopic traffic attribute. We get such a neural network via the following two-steps. First, we get $\omega$ value from collected $\rho$-$v$ data. Second, we build and learn a neural network that maps vehicle's behavior to traffic's attributes. In the training process, the method follows a macro-then-micro procedure, i.e., we first analyze macroscopic traffic and then learn how microscopic driving behavior maps to macroscopic traffic attributes. But after we have finished the training process, the  neural network works in a micro-then-macro step, i.e., we collect vehicles' driving trajectory and then get a traffic attribute value via forward propagation of the trained neural network.  The detailed procedures of the two steps are introduced as follows.

In the first step,  we assume that the $\omega$ value implies the free-flow speed, i.e., there exists a one-dimensional $V(\rho)$ such that
\begin{align} \label{eq:FD V=w*v}
	\mathcal{V}(\rho,\omega) = \omega V(\rho).
\end{align}
We solve the ~\Cref{eq:FD optimization} and get  the one-dimensional fundamental diagram $V(\rho)$.  
From the calibrated fundamental diagram $V(\rho)$, we get the traffic attribute value $\omega$ as:
\begin{align}
	\omega_{i,j} = v_{i,j}/V(\rho_{i,j}).
\end{align}
The $\omega$ value is used as the ground-truth value to train a neural network that maps from microscopic vehicle-related values to macroscopic traffic-related values.

In the second step, we build a neural network with trainable parameter $\theta_{\omega}$. The NN takes trajectory-related values as input and outputs the reconstructed traffic attribute $\omega$  value. For each cell $i,j$, we denote $\mathcal{N}_{i,j}$ as the set of vehicles that travel within the cell. For each vehicle $k \in \mathcal{N}_{i,j}$, we extract a trajectory attribute  vector $f_{i,j,k}$ from the collected trajectory. For each vehicle, the NN outputs a value $ \hat{\omega}(f_{i,j,k};\theta_{\omega})$ that reflects how these vehicle attributes affect the traffic flow dynamics. The overall traffic attribute value in a cell $i,j$ is the sum of  attributes of  all vehicles within the cell, i.e., the NN estimated traffic attribute is
 $\hat{\omega}_{i,j} (\theta_{\omega}) = \sum_{k\in\mathcal{N}_{i,j}} \hat{\omega} (f_{i,j,k};\theta_{\omega}) $.
The NN is trained with the loss function:
\begin{align}
	L(\theta_{\omega}) = \sum_{(i,j)} \left( \hat{\omega}_{i,j} (\theta_{\omega}) - \omega_{i,j} \right)^2.
\end{align}

To get the mapping from vehicle trajectory to traffic attribute,  we set the vehicle feature $f_{i,j,k}$ as a 8-dimensional vector: cell density,  the average value of its speed profile in the cell, the standard deviation of its speed profile in the cell, the average value of its acceleration profile in the cell, the standard deviation of its acceleration profile in the cell, the average of the absolute value of is jerk profile in the profile, the standard deviation of its jerk profile in the cell. For the first  feature, the cell density is calculated via the Edie's formula, and it reflects the overall traffic condition within the cell. For the last seven features, they are calculated directly from the vehicle's trajectory within the cell, without any  information about the vehicle's historical motion. We note that the last two features are the two values $C^{\mathrm{absave}}$ and $C^{\mathrm{std}}$ that we have used to get driver's attribute $\omega^{\mathrm{L}}$. We directly use $C^{\mathrm{absave}}$ and $C^{\mathrm{std}}$ instead of 
$\omega^{\mathrm{L}}$. This is because that for a new-arriving vehicle, its $\omega^{\mathrm{L}}$ value is unavailable, but the $C^{\mathrm{absave}}$ and $C^{\mathrm{std}}$ values are available directly from its trajectory. To summarize, all elements in the vehicle feature vector $f_{i,j,k}$ are obtained only using vehicle trajectories within the current cell $(i,j)$.  We use a feed-forward fully connected neural network with 7 hidden layers and 50 neurons in each hidden layer. The output of the NN is the one-dimensional vehicle-related attribute.

In \Cref{fig:TGSIM:FD}, we give the  scatter plot of $\rho$-$v$ and $\rho$-$q$ separated by the learned $\hat \omega$ value on the TGSIM dataset.  We see that the proposed method presents a clear separation for the fundamental diagram. We also test the proposed method using the NGSIM dataset, and we plot the result in~\Cref{fig:NGSIM:FD}. We give the calibration error on these two datasets in~\Cref{tab:error 1FD 2FD}. The results show that the proposed method learns an accurate mapping from microscopic vehicle features to macroscopic traffic attributes.

\begin{remark}[Comparison of the two reconstruction procedures with abundant data or scarce data]
{
Although the modeling procedures differ between~\Cref{sec:reconstruct model} and~\Cref{sec:data driven}, both aim to construct a unified pipeline that obtains the traffic attribute $\omega$ from vehicle trajectories and calibrates the two-dimensional fundamental diagram $\mathcal{V}(\rho,\omega)$  as in~\Cref{fig:intro}.  In~\Cref{sec:reconstruct model}, we directly obtain $\omega$ value from vehicle trajectories and then fit the fundamental diagram $\mathcal{V}(\rho,\omega)$. In contrast, neither $\omega$ nor the functional form of $\mathcal{V}(\rho,\omega)$ is directly observable in this section. Directly learning both simultaneously results in an ill-posed problem with infinite  equivalent solutions. For example, if $\mathcal{V}(\rho,\omega)$ is a valid solution, then any reparameterization $\tilde{\omega} = g(\omega)$ leads to an equivalent solution $\mathcal{V}(\rho,g^{-1}(\tilde{\omega}))$, where $g^{-1}$ is the inverse mapping of $g$. To address this ambiguity, we assign a physical interpretation to $\omega$ based on prior analysis in~\Cref{sec:reconstruct model}, where the heterogeneity is primarily attributed to variations in the maximum speed. Accordingly, we assume a multiplicative form of the fundamental diagram as in~\Cref{eq:FD V=w*v} and obtain  $\omega$ from observed $(\rho,v)$ data.  In both cases, the proposed heterogeneity-dependent fundamental diagram $\mathcal{V}(\rho,\omega)$ consistently captures the macroscopic traffic behavior and ensures coherence across modeling. 
}
\end{remark}

\subsection{Mixed traffic analysis}

\begin{figure}[!t]
    \centering
    I90-I94 Stationary\\
    \includegraphics[width=0.4\linewidth]{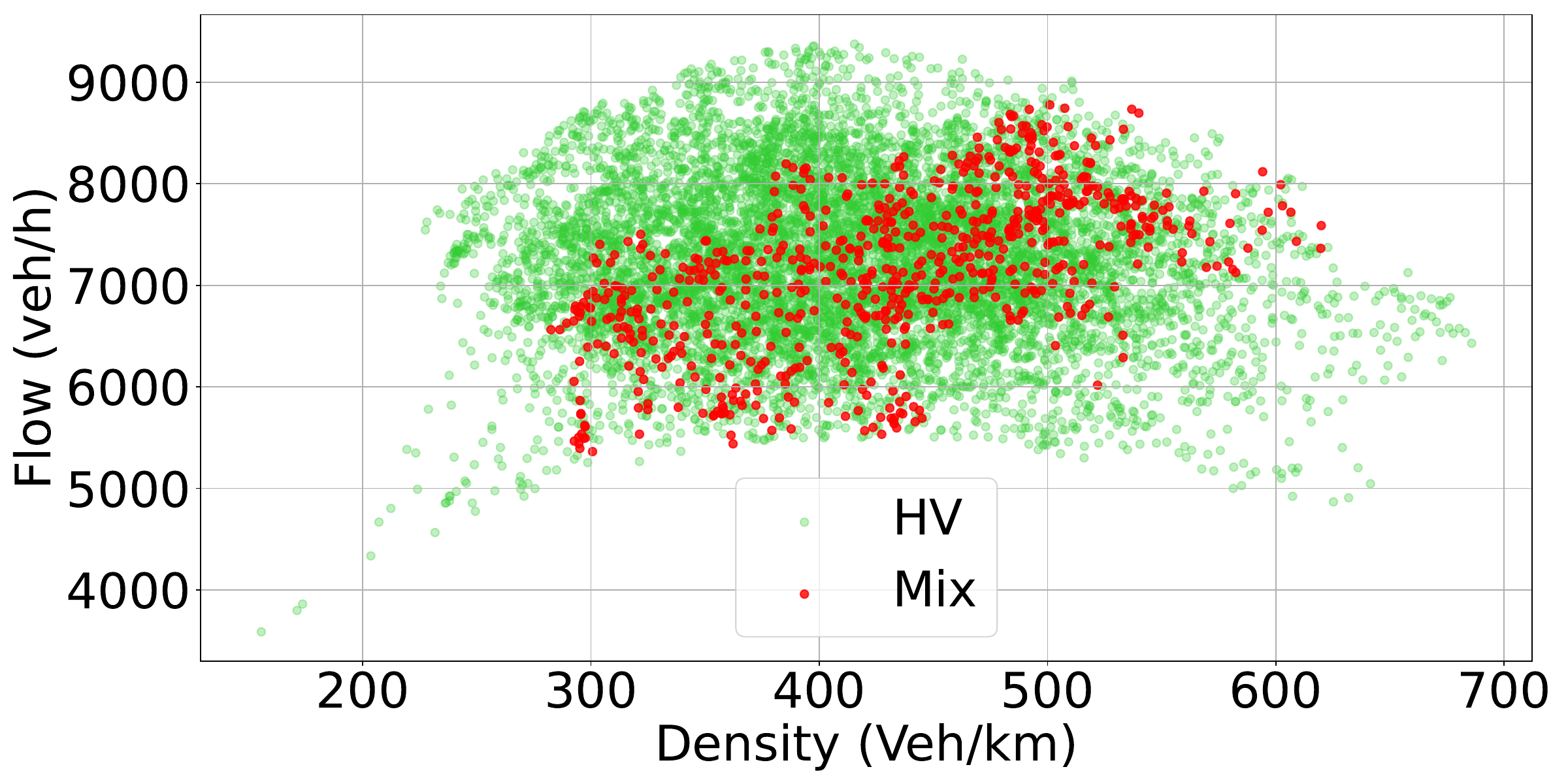}
    \includegraphics[width=0.4\linewidth]{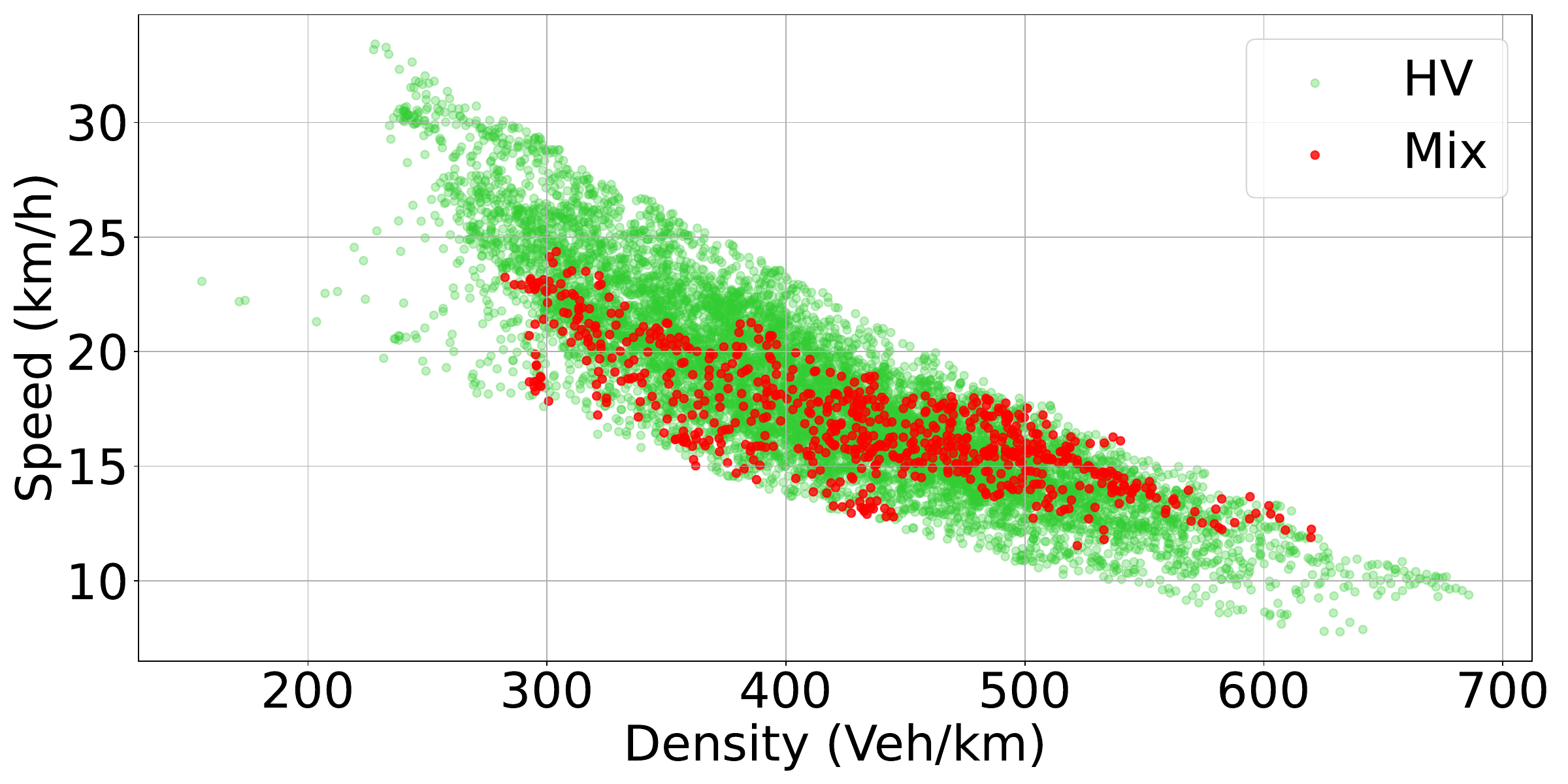} \\
    I-294 L1\\
    \includegraphics[width=0.4\linewidth]{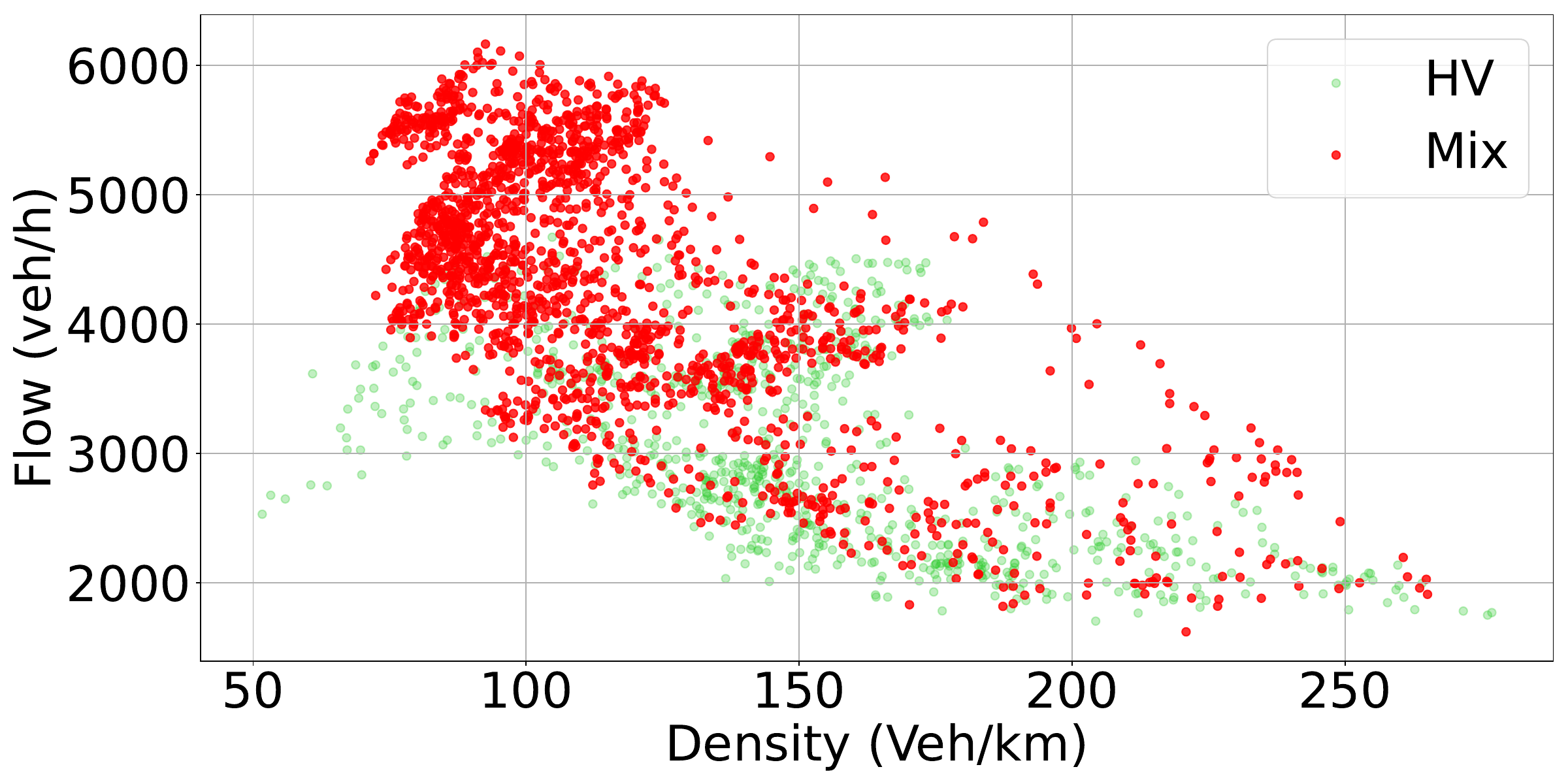}
    \includegraphics[width=0.4\linewidth]{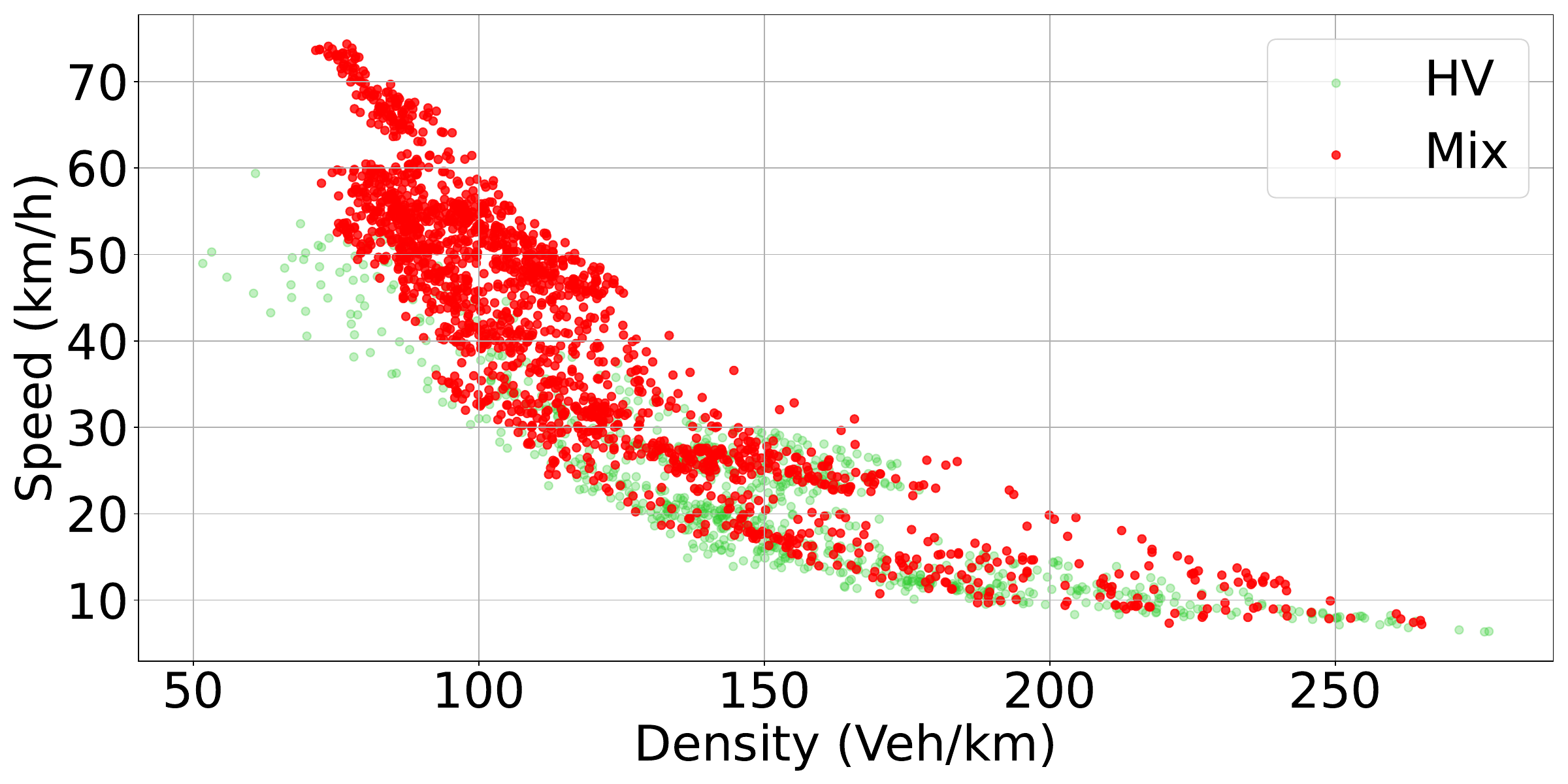}
    \caption{The fundamental diagram of the TGSIM dataset. The green points correspond to the cells that only contain HVs, and the red points are from the traffic states for cells that contain both AVs and HVs.}
    \label{fig:FD:TGSIM:AVHV}
\end{figure}

In the TGSIM dataset, the experiment AVs travel within HVs, which provide mixed traffic data. In this part, we analyze the macroscopic traffic attributes to show how AVs affect traffic flow.

In \Cref{fig:FD:TGSIM:AVHV}, we present the scatter plot of density-speed and density-flow. Green points represent cells containing only HVs, while red points represent mixed cells containing both AVs and HVs.  The first row corresponds to the I90-I94 Stationary dataset. (We neglect the I90-I94 Moving dataset for this part, since it collects traffic data around the probe AVs. Therefore, almost all data corresponds to mixed traffic.) As shown by the microscopic attributes in \Cref{fig:TGSIM:jerk}, AVs exhibit a narrower distribution range. The fundamental diagram in \Cref{fig:FD:TGSIM:AVHV} shows that mixed cells also exhibit a reduced scatter range, suggesting lower macroscopic heterogeneity.  For the I-294 dataset, as shown in \Cref{fig:TGSIM:jerk}, AVs and HVs exhibit approximately the same distribution range of microscopic attributes. Correspondingly, the fundamental diagram in the second row in \Cref{fig:FD:TGSIM:AVHV} indicates that pure HV traffic and mixed traffic exhibit a similar scatter range. These findings suggest that heterogeneity serves as a bridge between microscopic and macroscopic traffic dynamics. Reduced heterogeneity in microscopic vehicle motion leads to reduced heterogeneity in macroscopic traffic flow. Moreover, heterogeneity does not lie only between AVs and HVs, but rather among individual vehicles.

\subsection{Sensitivity analysis}
In this part, we conduct sensitivity analysis to analyze the  performance of  the  designed reconstruction method under varying conditions. 

\subsubsection{The effect of training data}

\begin{figure}[!t]
    \centering
    \includegraphics[width=0.7\linewidth]{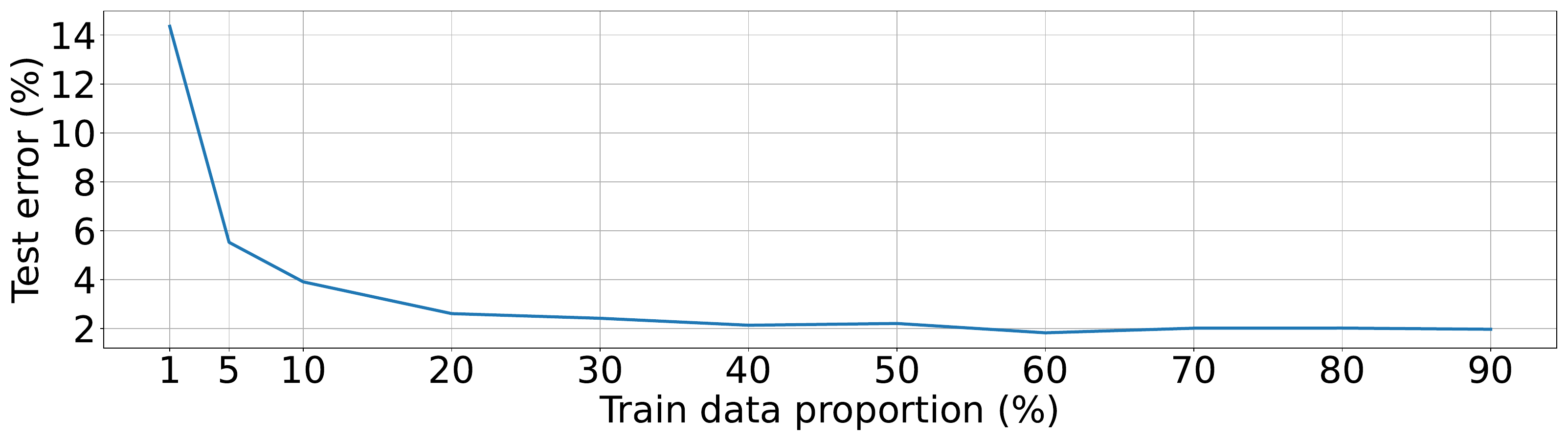}
    \caption{The test error when using different proportion data as the training data.}
    \label{fig:TGSIM:train ratio vs error}
\end{figure}

\begin{figure}[!t]
    \centering
    \subcaptionbox{All data}{\includegraphics[width=0.3\linewidth]{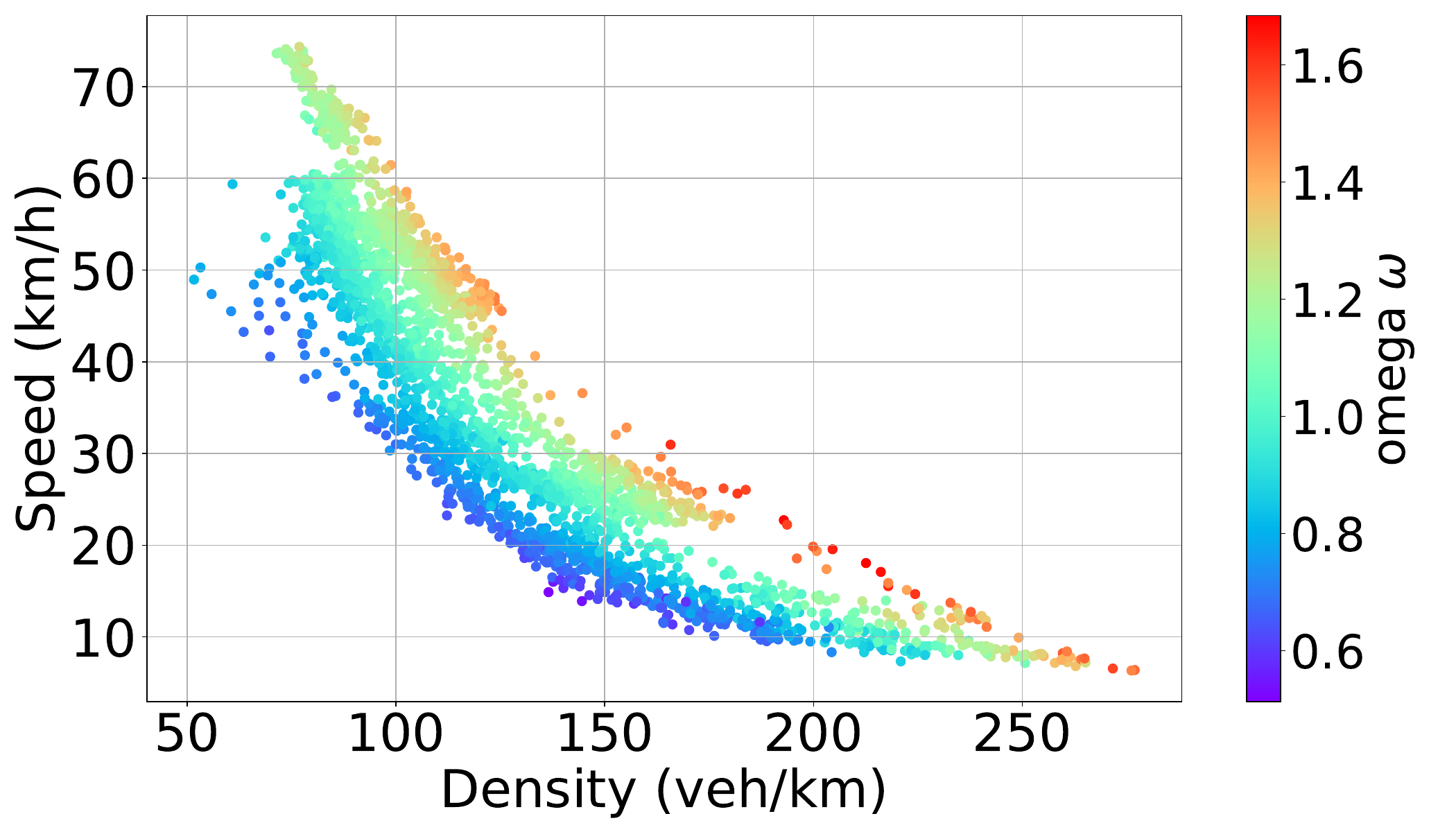}}
    \subcaptionbox{Train data}{\includegraphics[width=0.3\linewidth]{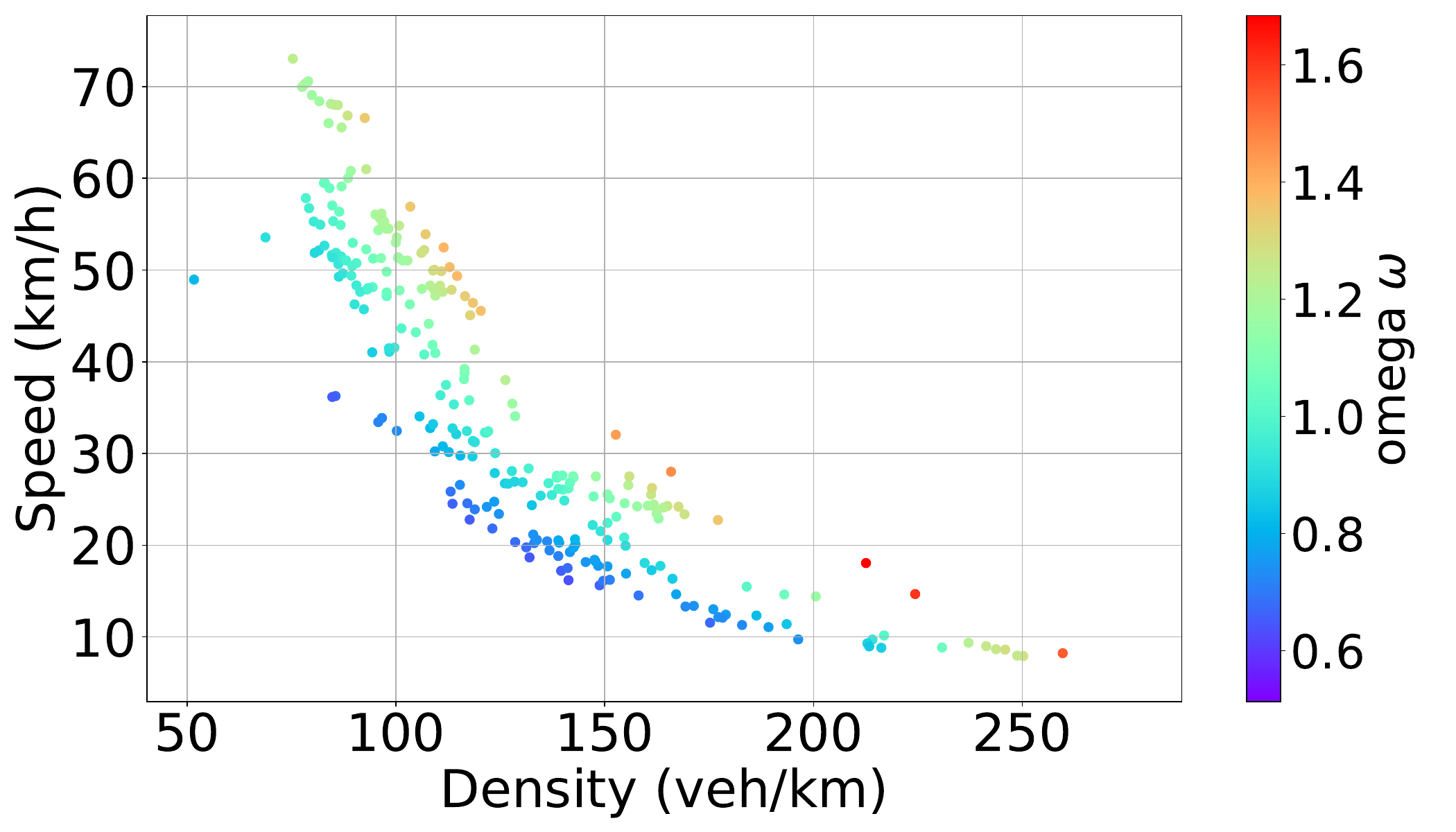}}
    \subcaptionbox{Test data}{\includegraphics[width=0.3\linewidth]{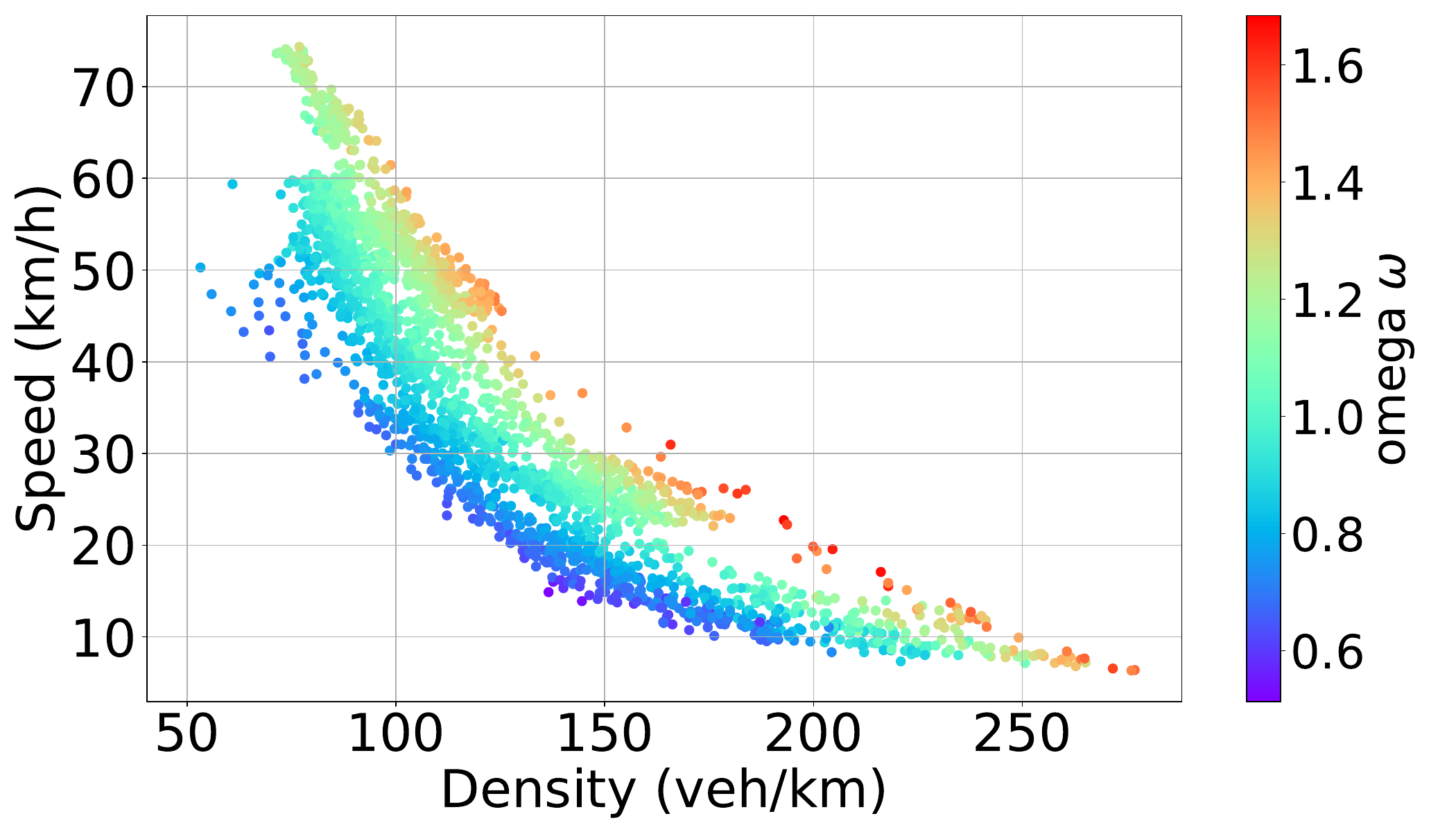}}
    \caption{The density-speed scatter plot when we only select 10 \% as the training data.}
    \label{fig:TGSIM:train ratio FD}
\end{figure}

In the previous simulation, we randomly select 80~\% data from all cells as the training data and other 20~\% data is used as the test data. In~\Cref{fig:TGSIM:train ratio vs error}, we vary the proportion of training data from 1~\% to 90~\% and plot the calibration error evaluated on the test data. The single FD gives a calibration error around 20~\%. We see that with a scarce data, the proposed reconstruction method still present a calibration error lower than 14~\%.  In~\Cref{fig:TGSIM:train ratio FD}, we give the scatter plot when we only use 10 \% data as the training dataset. The low calibration error and clear separation in the fundamental diagram demonstrate that the proposed algorithm can effectively learn a micro-to-macro mapping   even with scarce trajectory data.

\begin{figure}[!t]
    \centering
    \subcaptionbox{All data}{\includegraphics[width=0.3\linewidth]{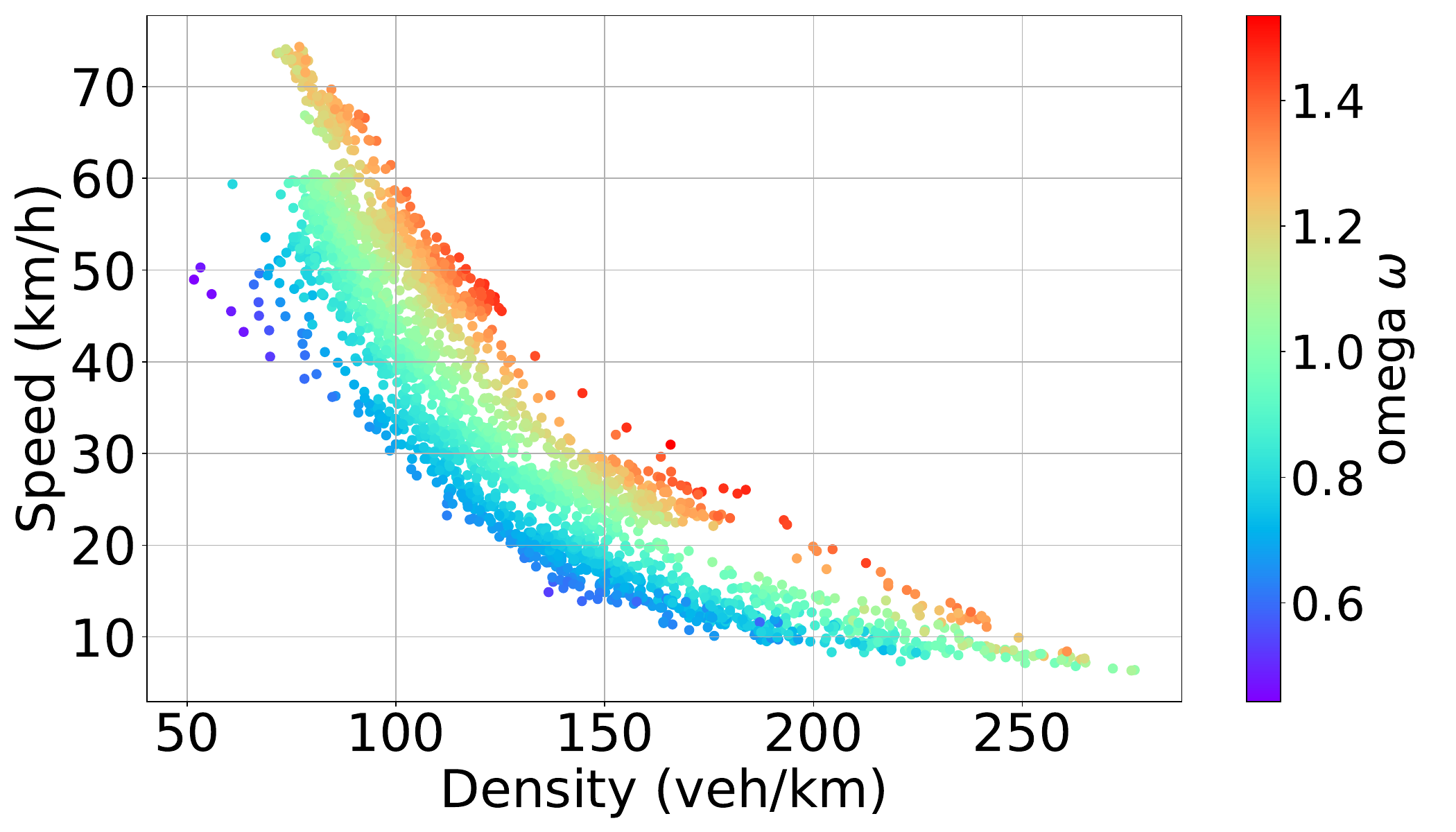}}
    \subcaptionbox{Train data}{\includegraphics[width=0.3\linewidth]{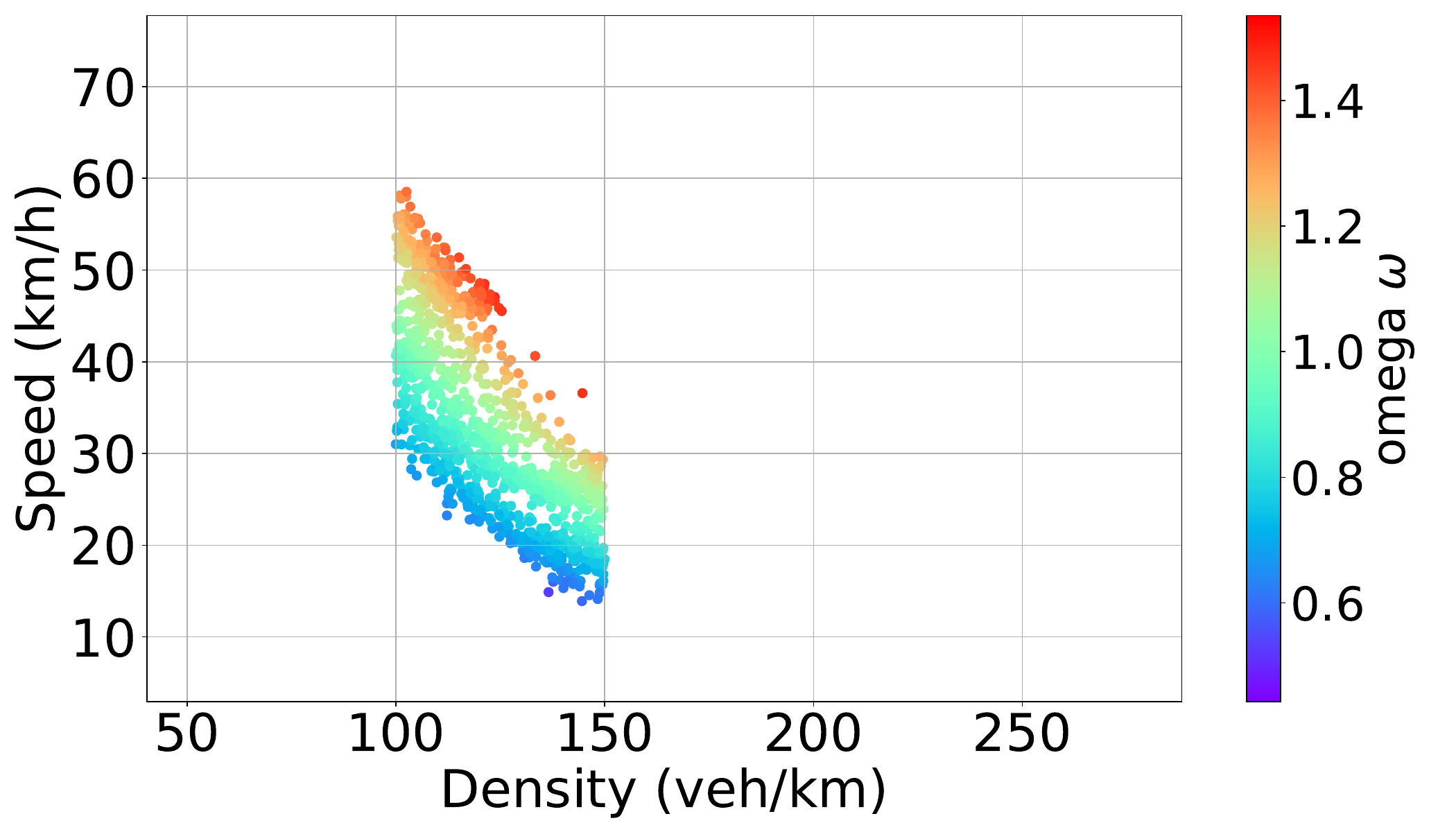}}
    \subcaptionbox{Test data}{\includegraphics[width=0.3\linewidth]{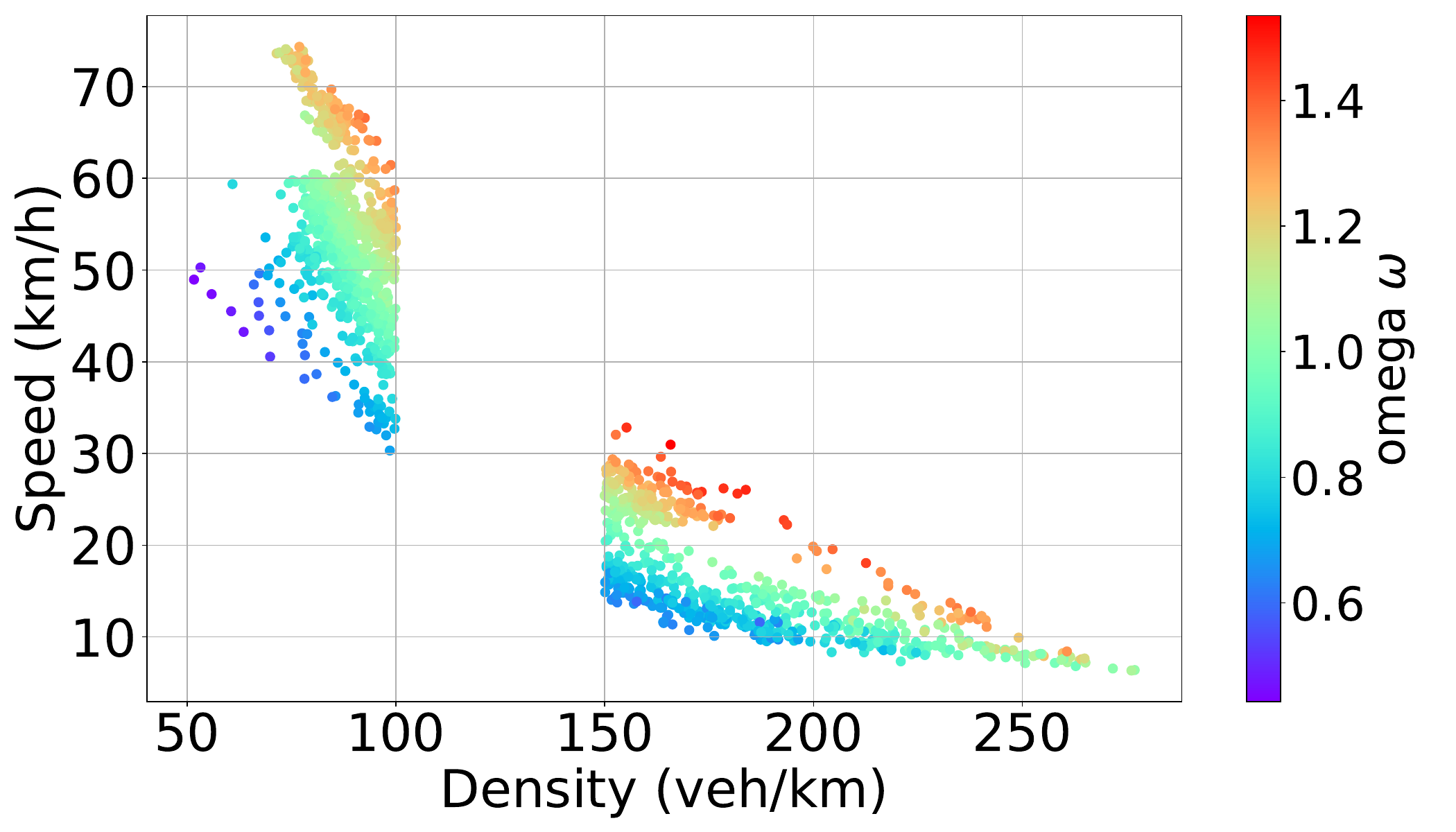}}
    \caption{The density-speed scatter plot when we only select these data within a given range of density as the training data.}
    \label{fig:TGSIM:train density FD}
\end{figure}

\begin{figure}[!t]
    \centering
    \subcaptionbox{Probability density}{\includegraphics[width=0.4\linewidth]{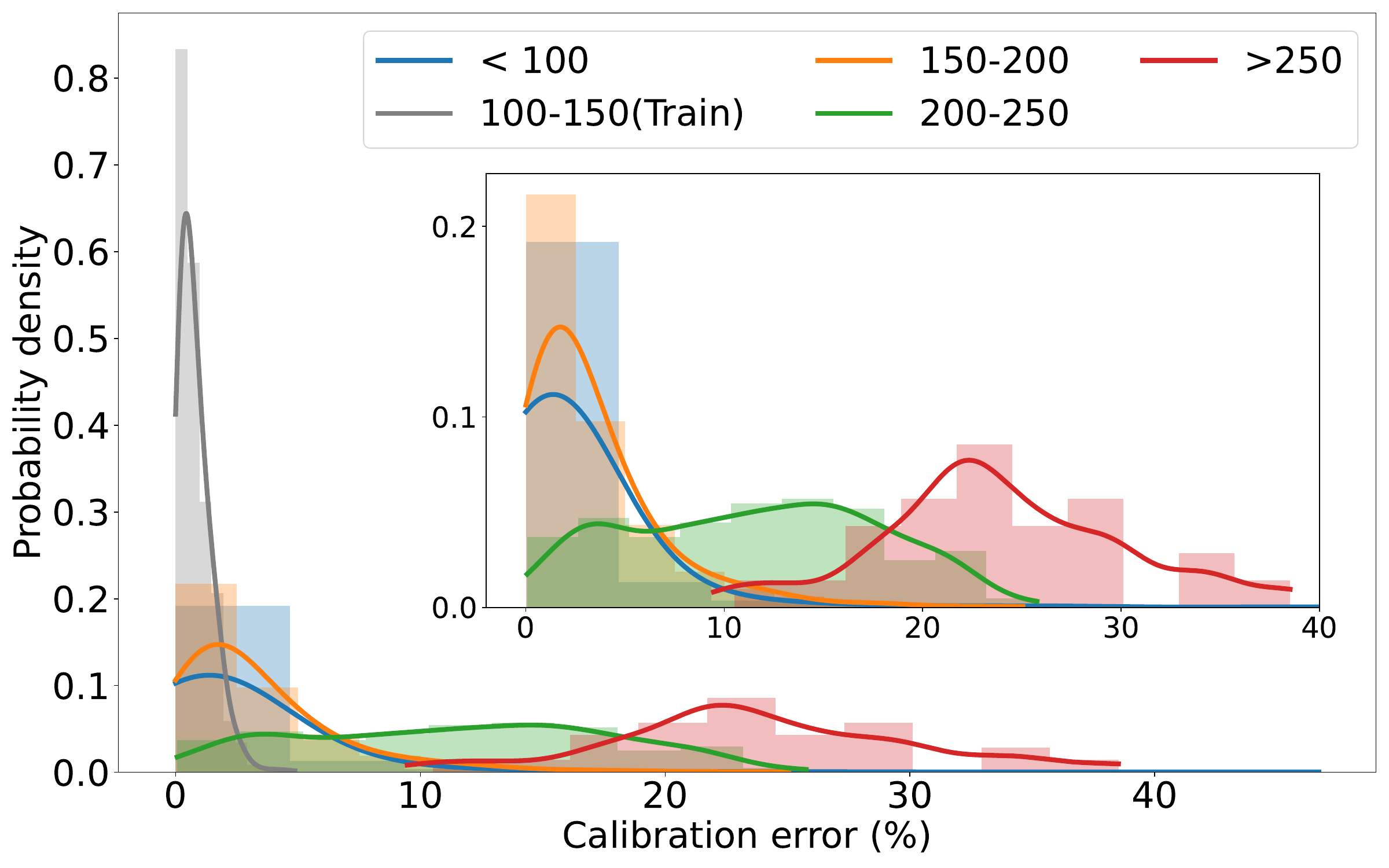}}
    \subcaptionbox{Cumulative distribution}{\includegraphics[width=0.4\linewidth]{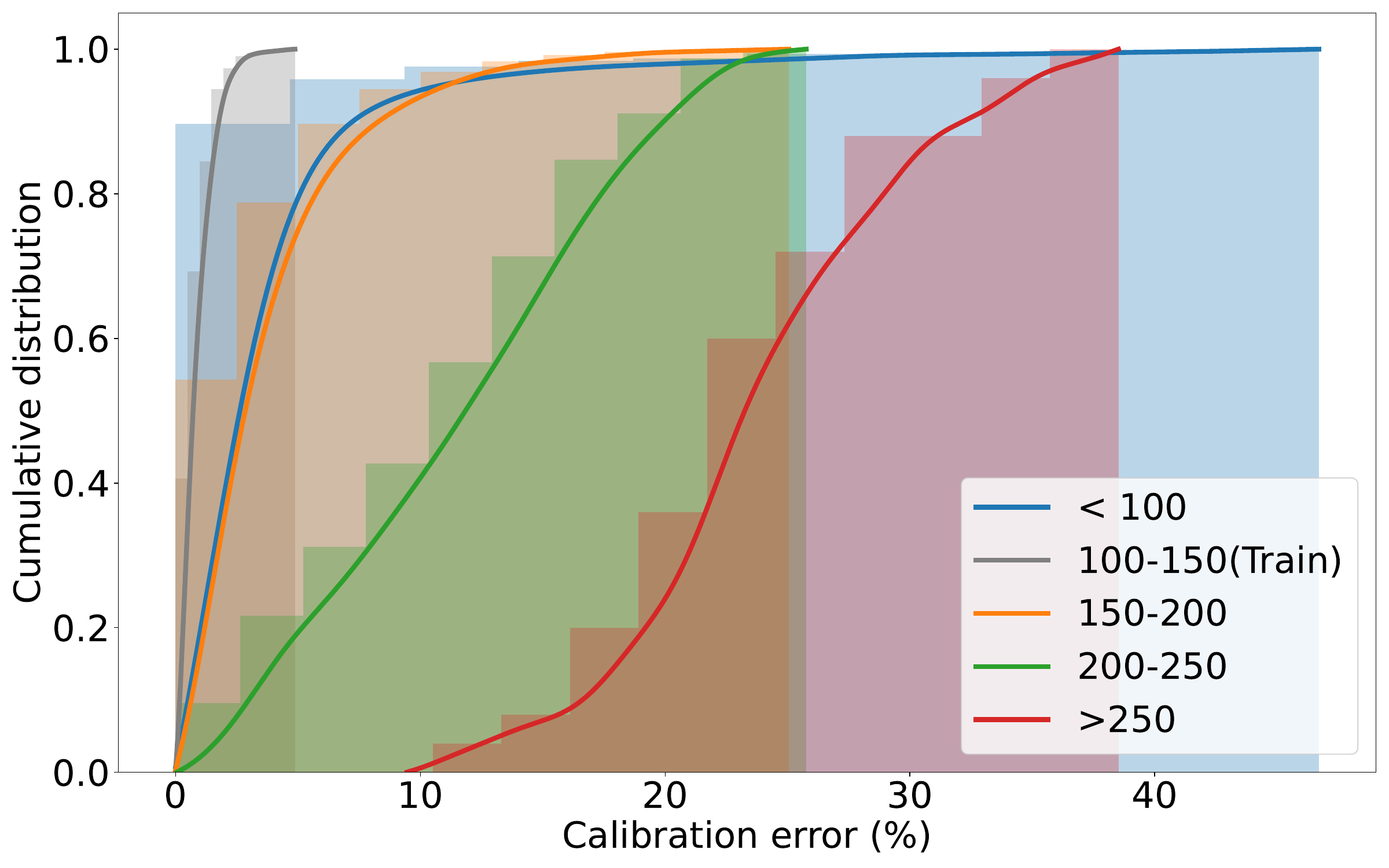}}
    \caption{Distribution of the fundamental diagram calibration error under varying traffic conditions. The legend gives the range of traffic density (in veh/km). We use traffic data whose density is within 100 veh/km to 150 veh/km as the training data. More free traffic (density smaller than 100 veh/km) and more congested traffic (density higher than 150 km/h) are used as the test data.}
    \label{fig:train density error probability}
\end{figure}
\begin{figure}[!t]
    \centering
    \subcaptionbox{All data}{\includegraphics[width=0.3\linewidth]{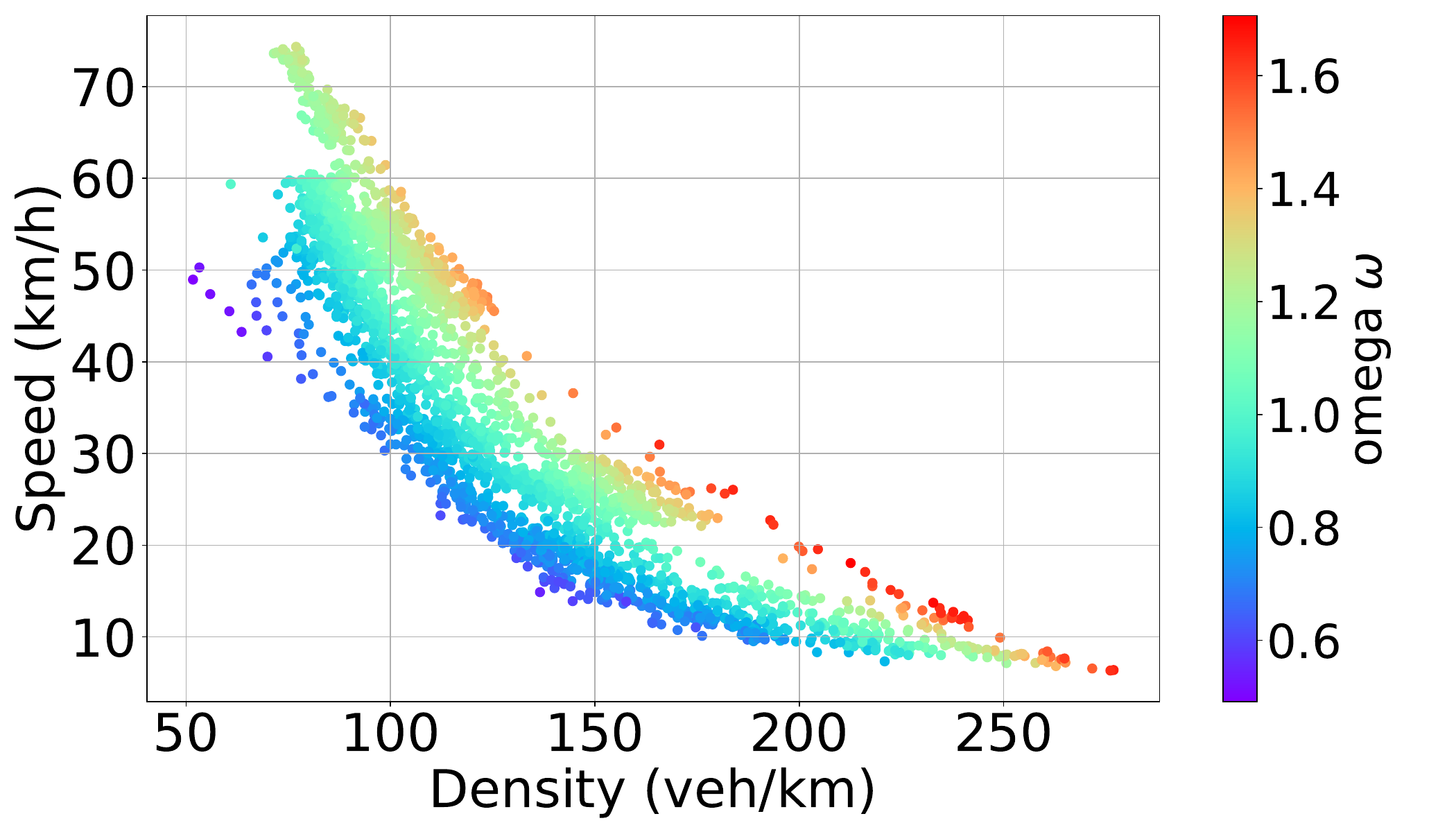}}
    \subcaptionbox{Train data}{\includegraphics[width=0.3\linewidth]{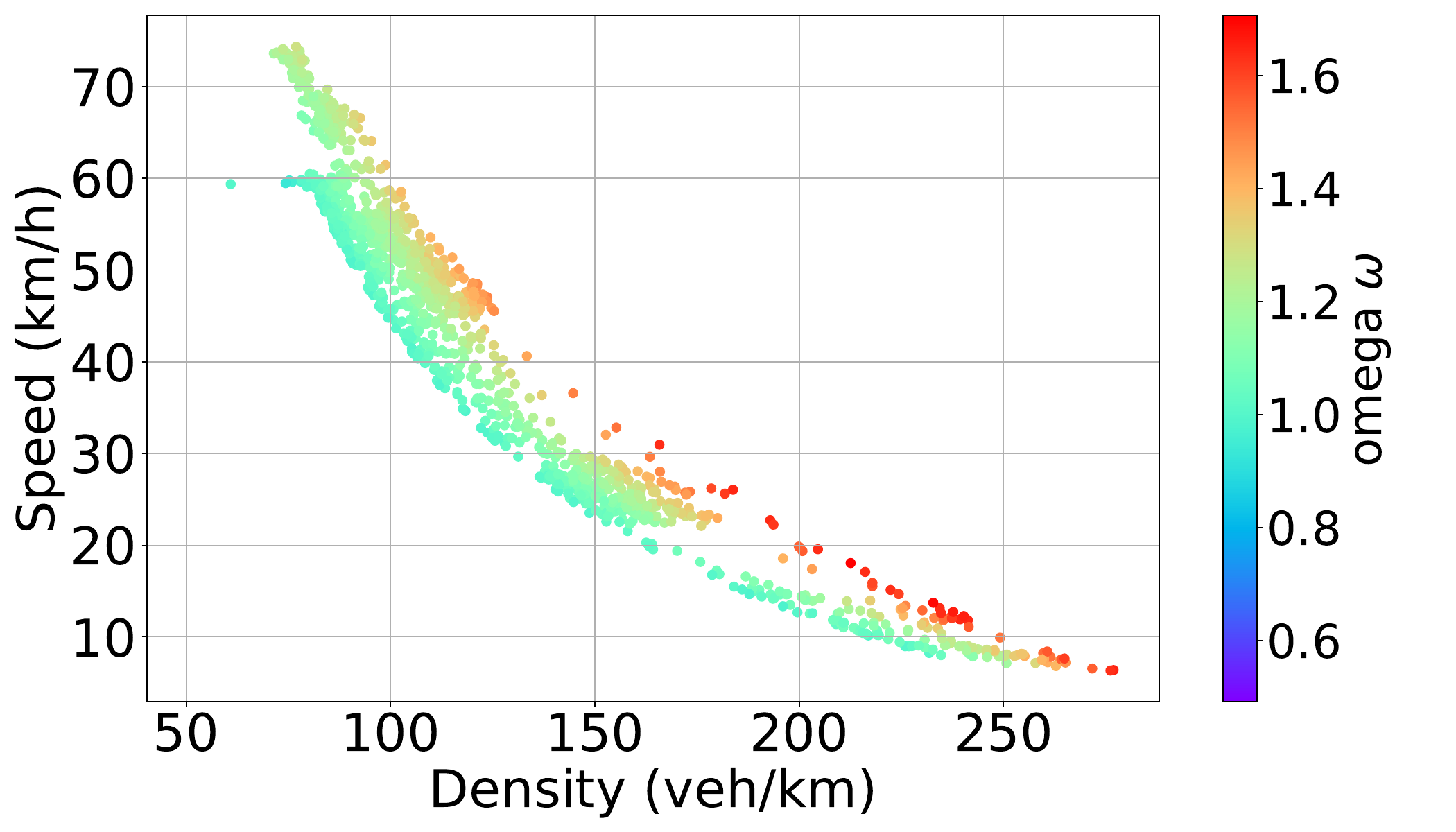}}
    \subcaptionbox{Test data}{\includegraphics[width=0.3\linewidth]{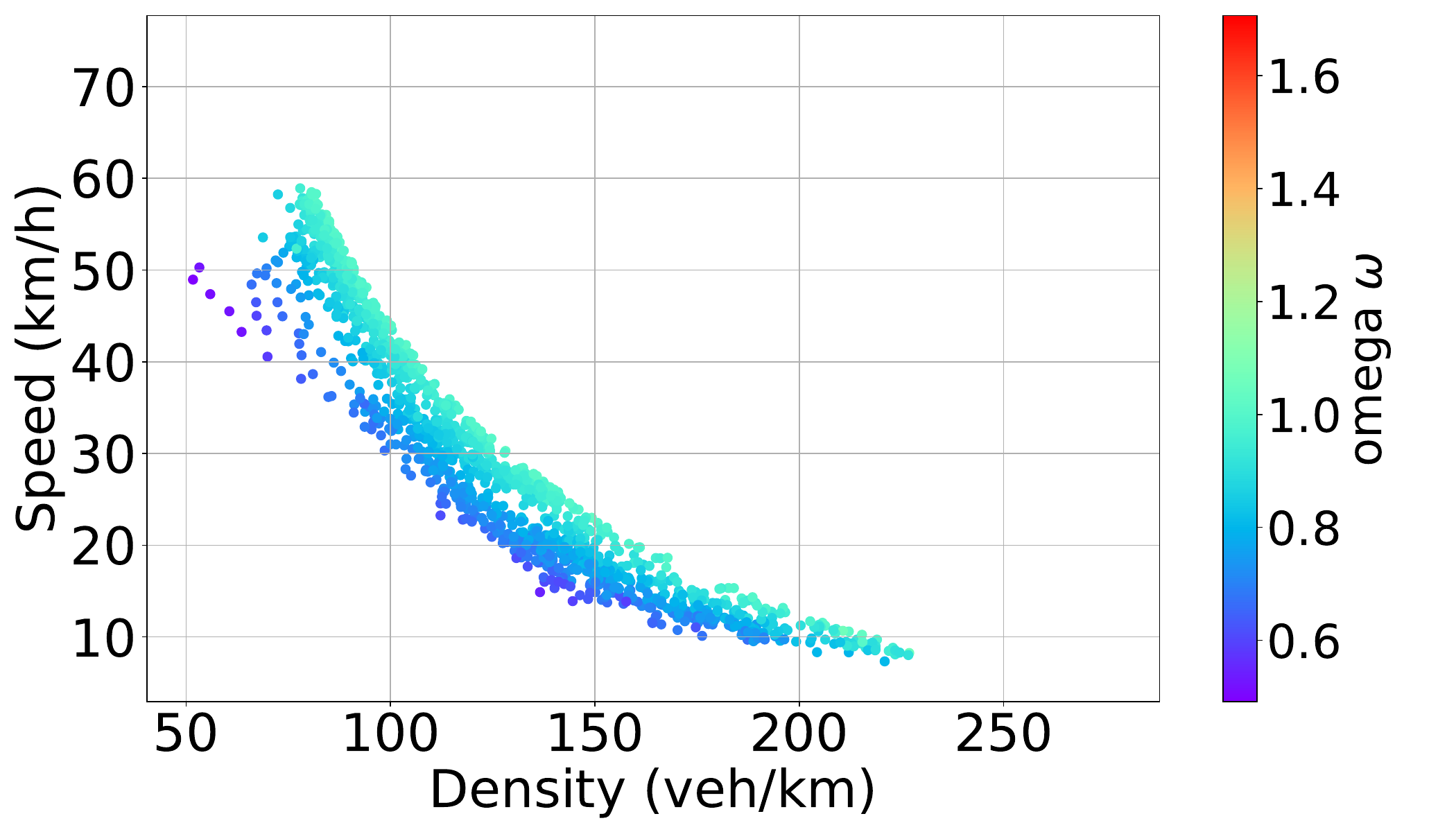}}
    \caption{The density-speed scatter plot when we only select these data with $\omega<1$ as the training data.}
    \label{fig:TGSIM:train omega FD}
\end{figure}

We also test the generalization ability of the proposed method on different traffic conditions. We generate the training data from these cells whose density is within the given range of 100 veh/km to 150 veh/km. And other cells are used as test data. \Cref{fig:TGSIM:train density FD} gives the scatter fundamental diagram on both the training data and test data. The calibration error on the test data is 2.65 \%. We see that the proposed method can generalize to unseen both more free traffic conditions (density is smaller than 100 veh/km) and more congested traffic (density is larger than 150 veh/km). We further shows the probability distribution of the calibration error under varying traffic conditions in~\Cref{fig:train density error probability}. We see that when applying the proposed micro-macro mapping to unseen traffic conditions, a majority of the calibration error is smaller than 20~\%, which is approximately the calibration error when using a single FD. For example, when the free traffic with density smaller than 100 veh/km (blue color in \Cref{fig:train density error probability}), we see that nearly 90~\% of the calibration error is less than 10~\%. For the congested traffic given in orange color corresponding to density between 150 veh/km to 200 veh/km, a majority of the calibration error is also lower than 10~\%. We also note that, as the difference between the training data and test data increases, the test error also increases. For example, when the test density is higher than 200 veh/km, the calibration error has a wider range of distribution, but still with a majority smaller than 20~\%. These results indicate that the proposed method can also build an accurate micro-macro mapping with unseen traffic conditions.

In \Cref{fig:TGSIM:train omega FD},  we generate the training dataset by only keeping these cells whose $\omega$ value is smaller than one. And those cells with $\omega>1$ are taken as the test data.  The calibration error on the test data is 2.84~\%. \Cref{fig:TGSIM:train omega FD} gives the result on the training and test data. We see that there is also a clear separation on the test data. Therefore, when there are more aggressive drivers, the proposed reconstruction method still brings an accurate mapping from microscopic trajectories to macroscopic attributes.

\subsubsection{The effect of vehicle features}

\begin{table}[!t]
    \centering
    \caption{Calibration error on the test data (\%) }
    \label{tab:error features}
    \begin{tabular}{c|c||c|c}
       Name  &  Error  &  Name & Error \\ \hline
         $\backslash$ & $\backslash$ & MACRO & 13.59 \\   \hline
        MICRO-v & 14.52 & MACRO-v & 1.93 \\ \hline
        MICRO-j & 14.99 & MACRO-j & 9.96 \\ \hline
        MICRO-a & 15.09 & MACRO-a & 11.17 \\ \hline
        MICRO & 10.79 & ALL &  2.02
    \end{tabular}
\end{table}
\begin{figure}[!t]
    \centering
    \subcaptionbox{}{\includegraphics[width=0.4\linewidth]{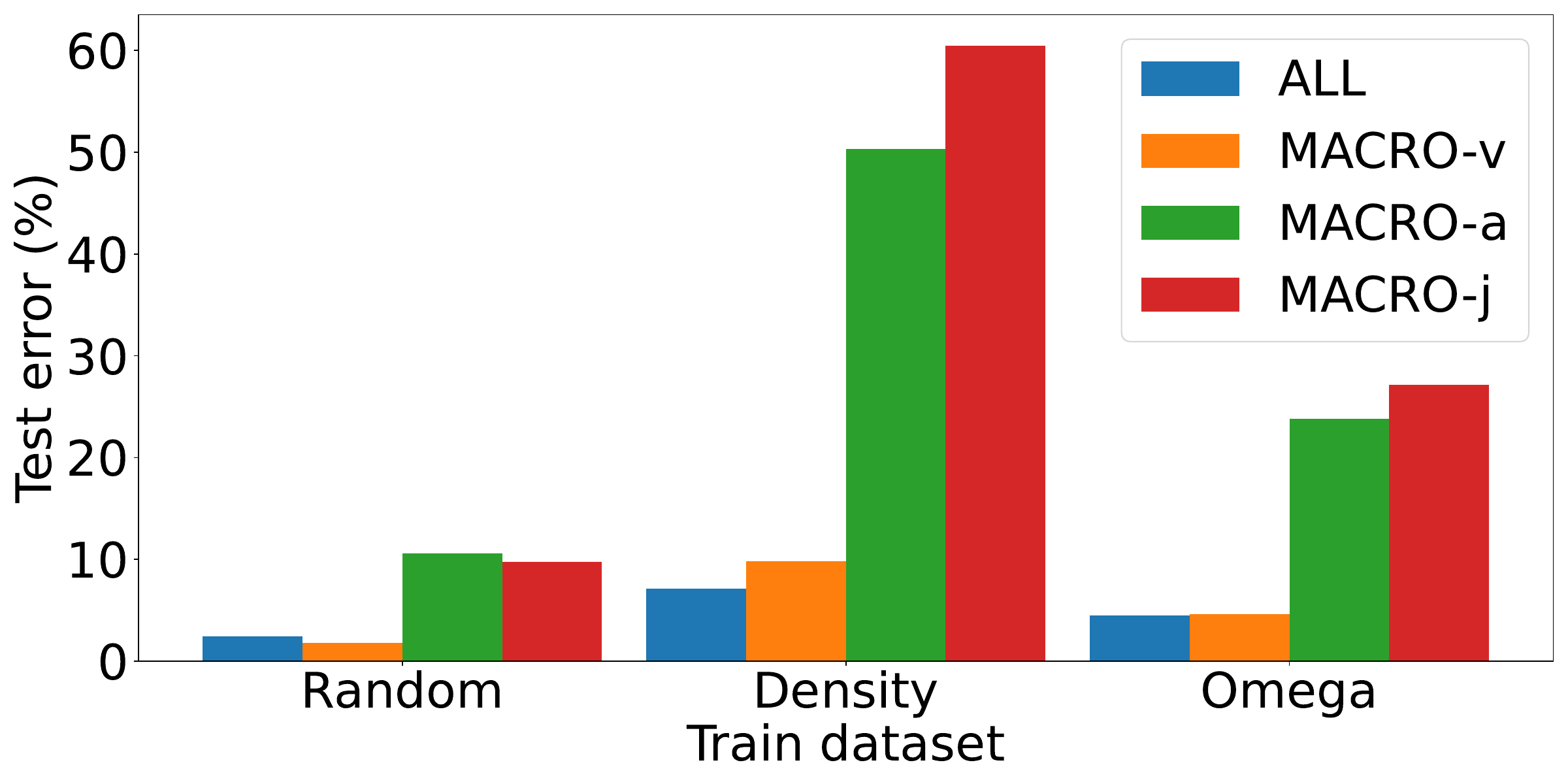}}
    \subcaptionbox{}{\includegraphics[width=0.4\linewidth]{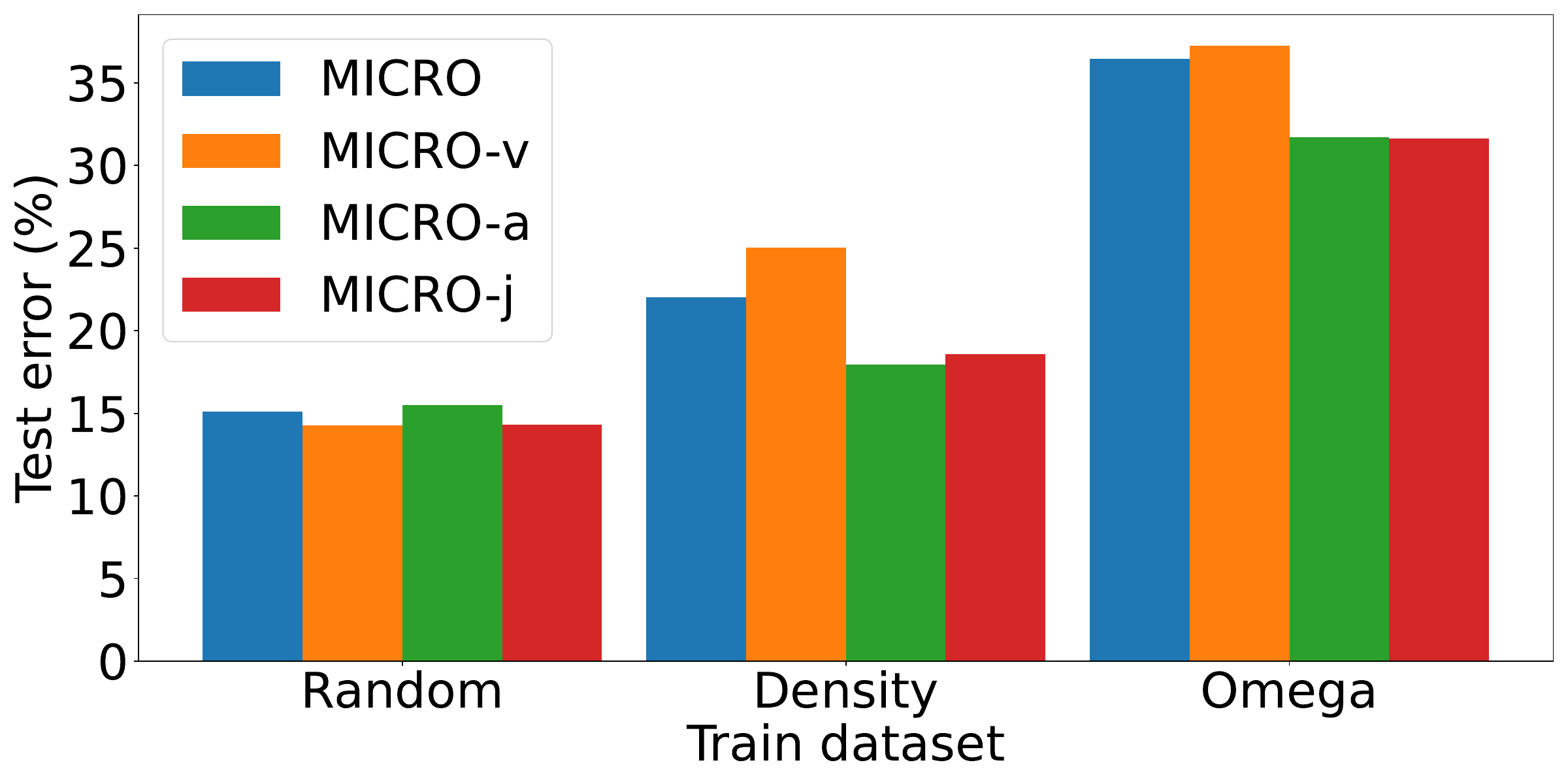}}
    \caption{The fundamental diagram calibration error under different training datasets and different choices of features that used to map to the macroscopic traffic attribute.}
    \label{fig:TGSIM:feature train error}
\end{figure}

In previous simulation, the vehicle feature include mainly four parts:  traffic density, vehicle speed, vehicle acceleration, and vehicle jerk. In this part, we analyze how the choice of vehicle features affect the calibration error. We consider three types of variants:
\begin{itemize}
    \item MACRO: we only use traffic density as the feature, and the feature is a thus a scalar value.
    \item We only use those features from the single vehicle trajectory.
    \begin{itemize}
        \item MICRO-v: we only use the average and standard deviation of the vehicle's speed profile.
        \item MICRO-a: we only use the average and standard deviation of the vehicle's acceleration profile.
        \item MICRO-j: we only use the average, average of absolute, and standard deviation of the vehicle's speed profile.
    \end{itemize}
    \item Mixed with both macro density and micro values:
    \begin{itemize}
        \item MACRO-v: we use the cell density, average of the vehicle's speed, and the  standard deviation of its speed.
        \item MACRO-a: we use the cell density, average of the vehicle's acceleration, and the  standard deviation of its acceleration.
        \item MACRO-j: we use the cell density, average of the vehicle's jerk, average of the absolute values of its jerk, and the  standard deviation of its jerk.
        \item ALL: we use all features, i.e., the feature is  eight dimension.
    \end{itemize}
\end{itemize}

We randomly select 80~\% data as the training data and report in~\Cref{tab:error features} the calibration error on the test data for different variants of the proposed method. In each row, we use the same trajectory feature and compare the calibration error for micro-only vs micro+macro. As the result show, incorporating surrounding traffic information (i.e., cell density) greatly reduces the test error. For example, in the second row, when we only use speed average and speed standard deviation as vehicle features, i.e., MICRO-v, the calibration error is around 15~\%. But when we also incorporate traffic density into the vehicle feature, the calibration error reduces to around 2~\%. Comparing the errors for the first column, we see that the MICRO, which incorporates features from speed, acceleration, and jerk, gives the lowest calibration error. Comparing results in the third column, we see the MACRO-v, which use three features of density, speed average, speed standard deviation, and the ALL, which uses density and all features from speed, acceleration, and jerk,  give the lowest calibration error. The comparison of calibration error implies that the speed information is the main features when we map from microscopic trajectories to macroscopic traffic attributes.

In~\Cref{fig:TGSIM:feature train error}, we further compare the test error  by different variants on different training datasets. We consider three training datasets: `Random' where 30~\% of data is randomly selected as the training data, `Density' where cells whose density is within 100 veh/km to 150 veh/km are the training data, and `Omega' where cells with $\omega>1$ are selected as training data. We see that in the Random case, the MICRO-v has a slight improvement over ALL. But for the Density and Omega case, where the test data contains unseen traffic conditions, the ALL version gives a lower calibration error than MICRO-v. Therefore, ALL that contains traffic density information has a better generalization ability.

\subsubsection{The effect of the NN size}
\begin{figure}
    \centering
    \subcaptionbox{}{\includegraphics[width=0.4\linewidth]{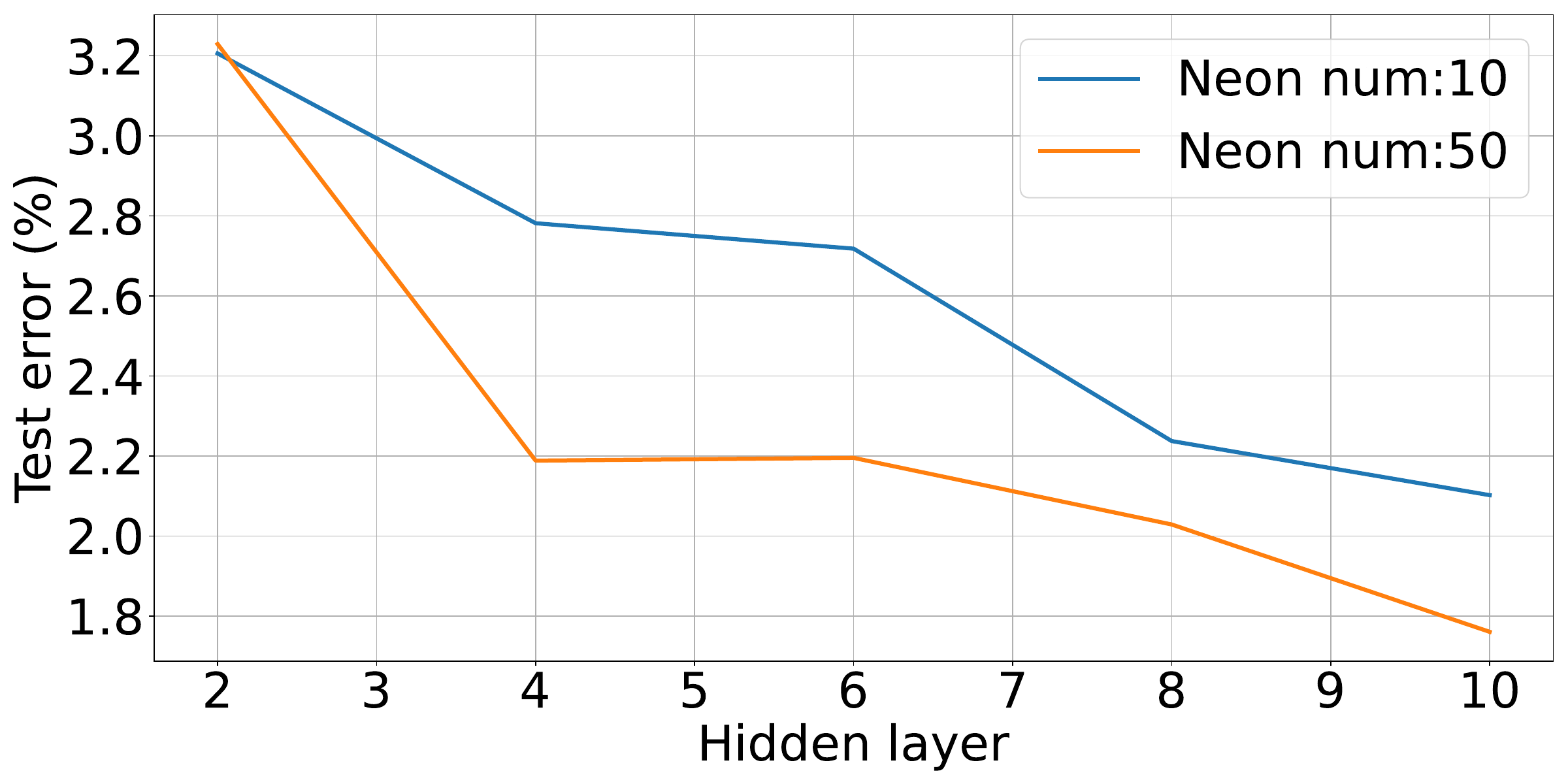}}
    \subcaptionbox{}{\includegraphics[width=0.4\linewidth]{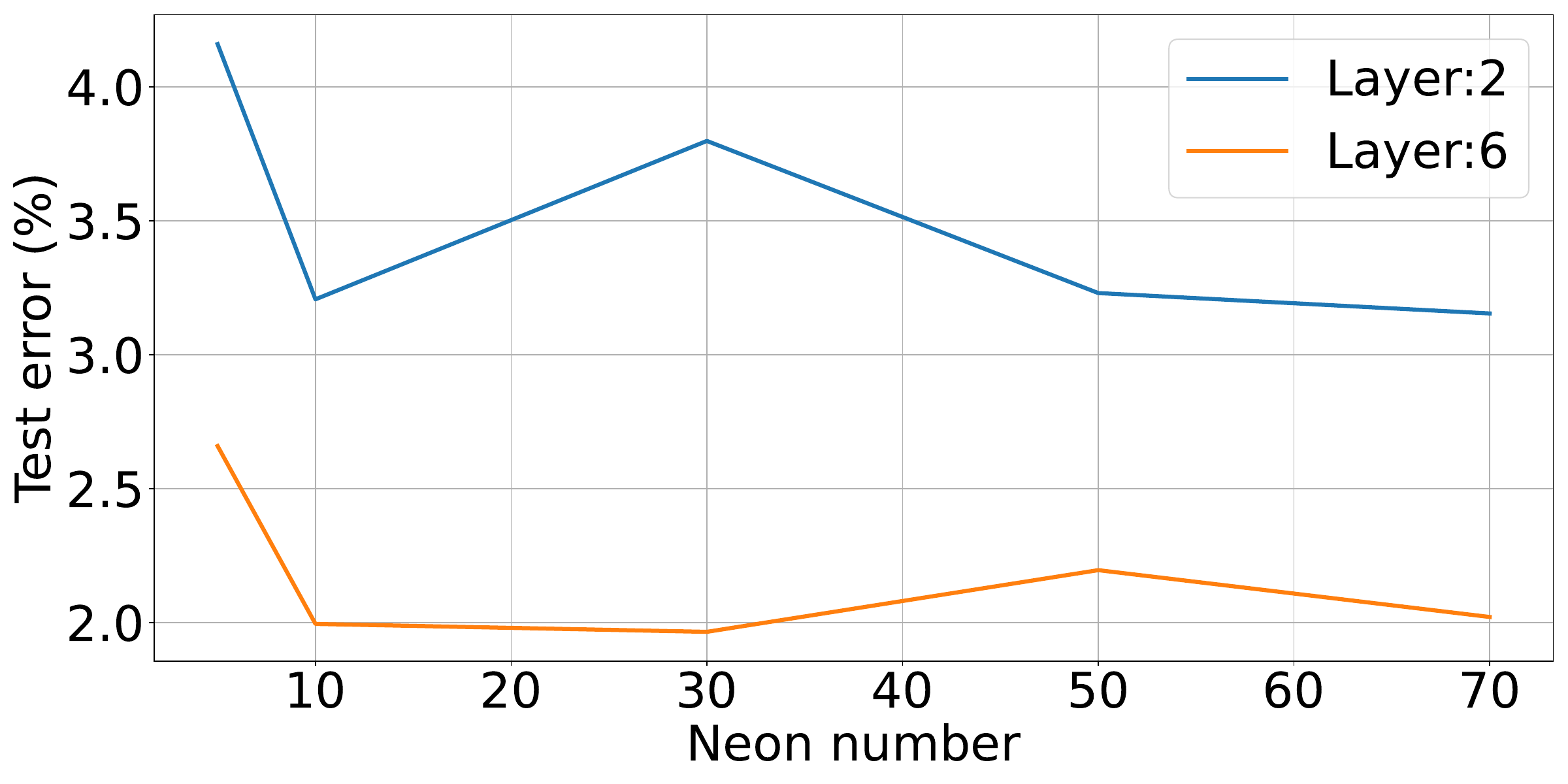}}
    \caption{The fundamental diagram calibration error with different NN size. We see that the proposed method has a low calibration error even with a simple neural network structure.}
    \label{fig:TGSIM:NN error}
\end{figure}

We run simulations with varying numbers of hidden layers and neurons to analyze how the proposed micro-macro mapping method works under different NN sizes. We give the test error in~\Cref{fig:TGSIM:NN error}. In~\Cref{fig:TGSIM:NN error}(a), we fix the neuron numbers in each hidden layers and run simulations when the number of hidden layers changes. In~\Cref{fig:TGSIM:NN error}(b), we fix the number of hidden layers and  change the neuron numbers in each hidden layer.   Increasing the hidden layers or the number of neurons brings a lower error, which agrees with naive intuition.  We also note that even using a very small network  with only 2 hidden layers and 5 neurons in each hidden layer, we still have an accurate calibration with the calibration error lower around 4~\%. With such a straightforward and easy-to-implement neural network structure, the training computation cost is very low, and the proposed method can be easily applied to practical traffic systems.

\section{Conclusion}

In this paper, we utilize heterogeneity in mixed-autonomy traffic and bridge the gap between Lagrangian trajectory data and Euclidean traffic flow models. We demonstrate that heterogeneity exists not only between AVs and HVs but also within each class individually, which motivates us to introduce a traffic attribute variable to reflect  vehicles' heterogeneity.  Based on real trajectory data, we design a micro-to-macro mapping that incorporates drivers' attributes and local behavioral features.  The mapping provides insight into how AV controllers influence mixed traffic flow and offers a macroscopic formulation that can integrate AV-collected data.  We also extend our approach to address the more challenging scenario of scarce data availability by designing a data-driven reconstruction mapping.  Analysis using real trajectory data validates the effectiveness of the proposed mapping in capturing microscopic-macroscopic relations. We also show that the model exhibits strong generalization capability when applied to unseen traffic conditions.

This work provides several promising directions for future research. First, while we adopt a fully connected neural network in this study, exploring more advanced architectures, such as graph-based or attention-based networks, may improve the interpretability of the mapping. Second, alternative methods for extracting vehicle attributes from trajectory data can be investigated. In this work, features are computed using statistical measures (e.g., average and standard deviation) over the full trajectory, which requires complete driving records. Incorporating other features could enhance robustness against data noise or temporary data unavailability caused by communication failure.  Third, while wave propagation prediction is an important application of macroscopic traffic flow models, we only validate the flow pattern prediction in~\Cref{sec:reconstruct model} on heterogeneous HV dataset. The proposed reconstruction methods for mixed traffic dynamics have not been covered in this paper, since the mixed-autonomy TGSIM dataset provides trajectory data only for localized roadway segments, lacking the full spatiotemporal coverage required to analyze traffic-wave propagation. Consequently, we validate our model through the fundamental diagram in~\Cref{sec:data driven}. Using the proposed attribute-based model to capture wave propagation of mixed traffic will be of future interest when large-scale, continuous spatiotemporal datasets become available.


\bibliographystyle{apalike} 
\bibliography{reference}

\appendix
\section*{Appendix}
\subsection*{A. Data collection and features of the datasets}

The Ring dataset was collected in Beijing, China~\citep{zheng2021experimental}.
The authors conducted an experiment on an 800-meter ring road and collected the trajectories of 40 participating human drivers over a period of approximately 20 minutes. 
The measurement errors for position and speed were 1 m and 1 km/h, respectively. The sampling rate was 10 Hz.

he CATS dataset was collected by~\citep{shi2021empirical} using a Lincoln MKZ autonomous vehicle. The AV traveled on SR-56 in Florida, USA. The data was collected at a sampling rate of 1 Hz, with a position accuracy of 0.26 m and a speed accuracy of 0.089 m/s. The OpenACC dataset~\citep{makridis2021openacc} was collected by the European Commission’s Joint Research Center across Hungary, Italy, and Sweden. Various commercial ACC vehicles were used to collect the data, including the Tesla Model 3, BMW X5, Toyota RAV4, and Mercedes A-Class.
The sampling rate ranged from 3 Hz to 100 Hz.
The position accuracy ranged from approximately 0.02 m to 0.5 m, and the speed accuracy was around 1 km/h.
The CentralOhio dataset was collected in Ohio, USA~\citep{xia2023automated,seitz2024advanced}.
The measurement errors for position and speed were approximately 1 m and 0.1 m/s, respectively. 

\begin{figure}[!t]
    \centering
    Speed \\
    \subcaptionbox{Ring}{\includegraphics[width=0.24\linewidth]{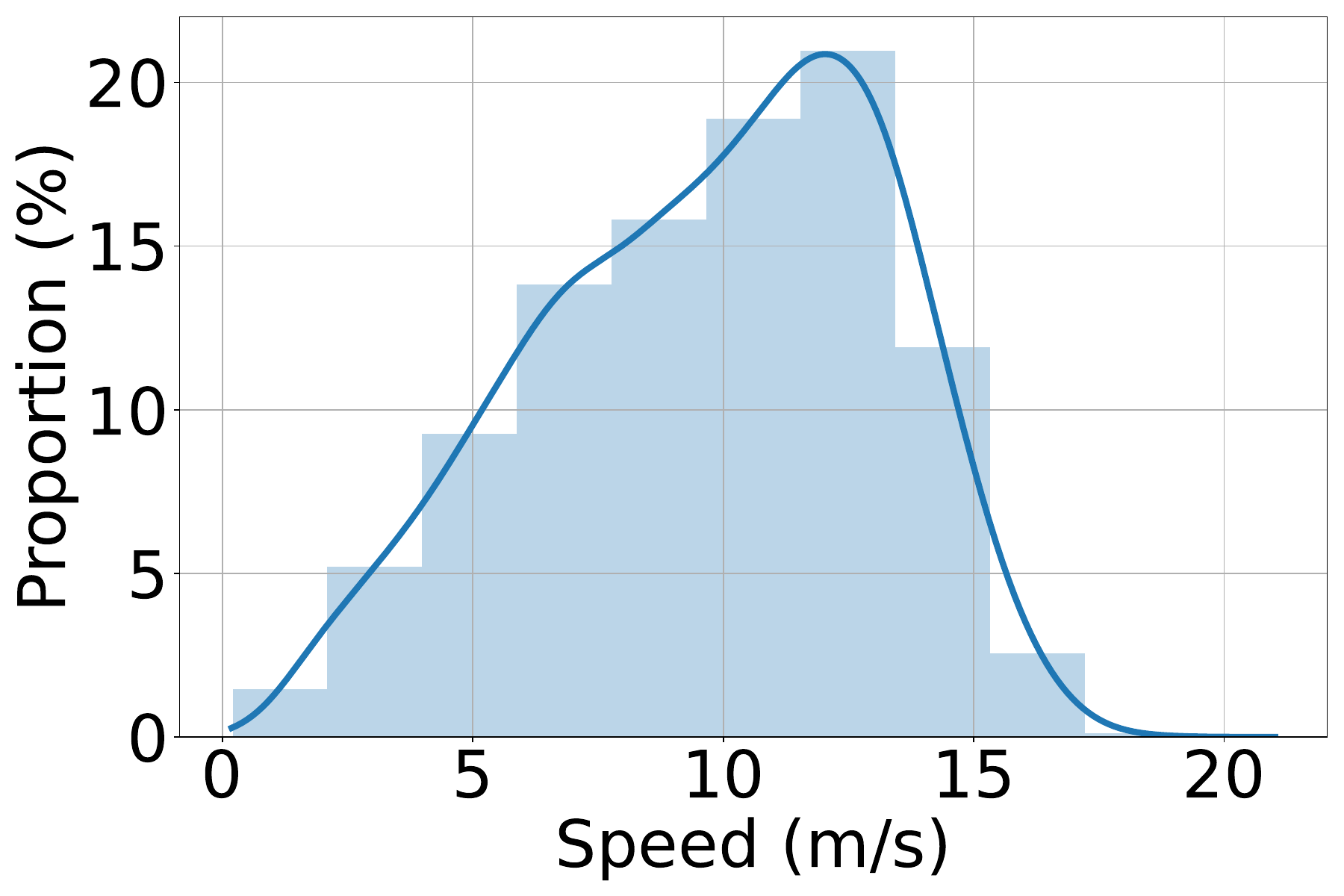}}
    \subcaptionbox{OpenACC}{\includegraphics[width=0.24\linewidth]{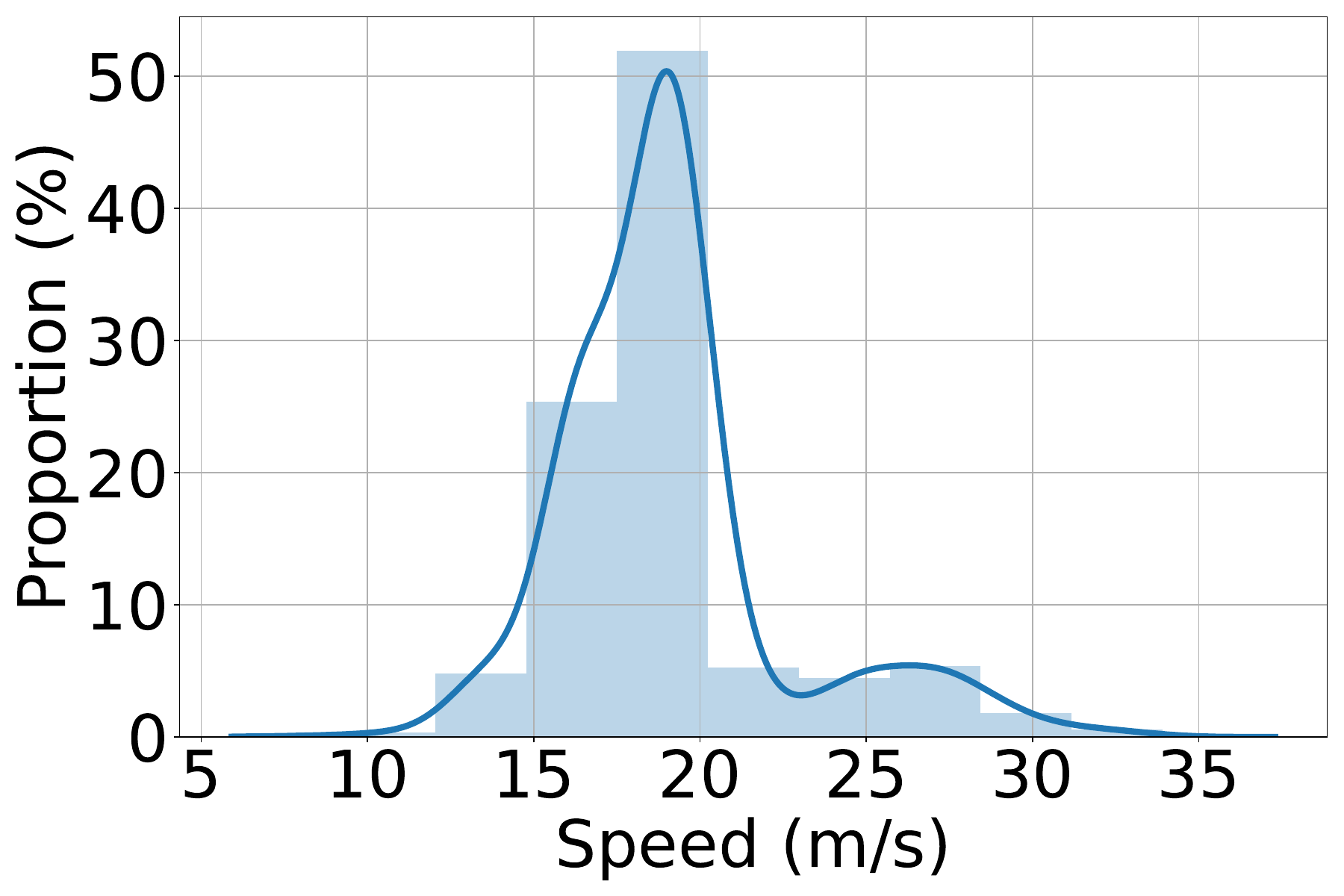}}
    \subcaptionbox{CATS}{\includegraphics[width=0.24\linewidth]{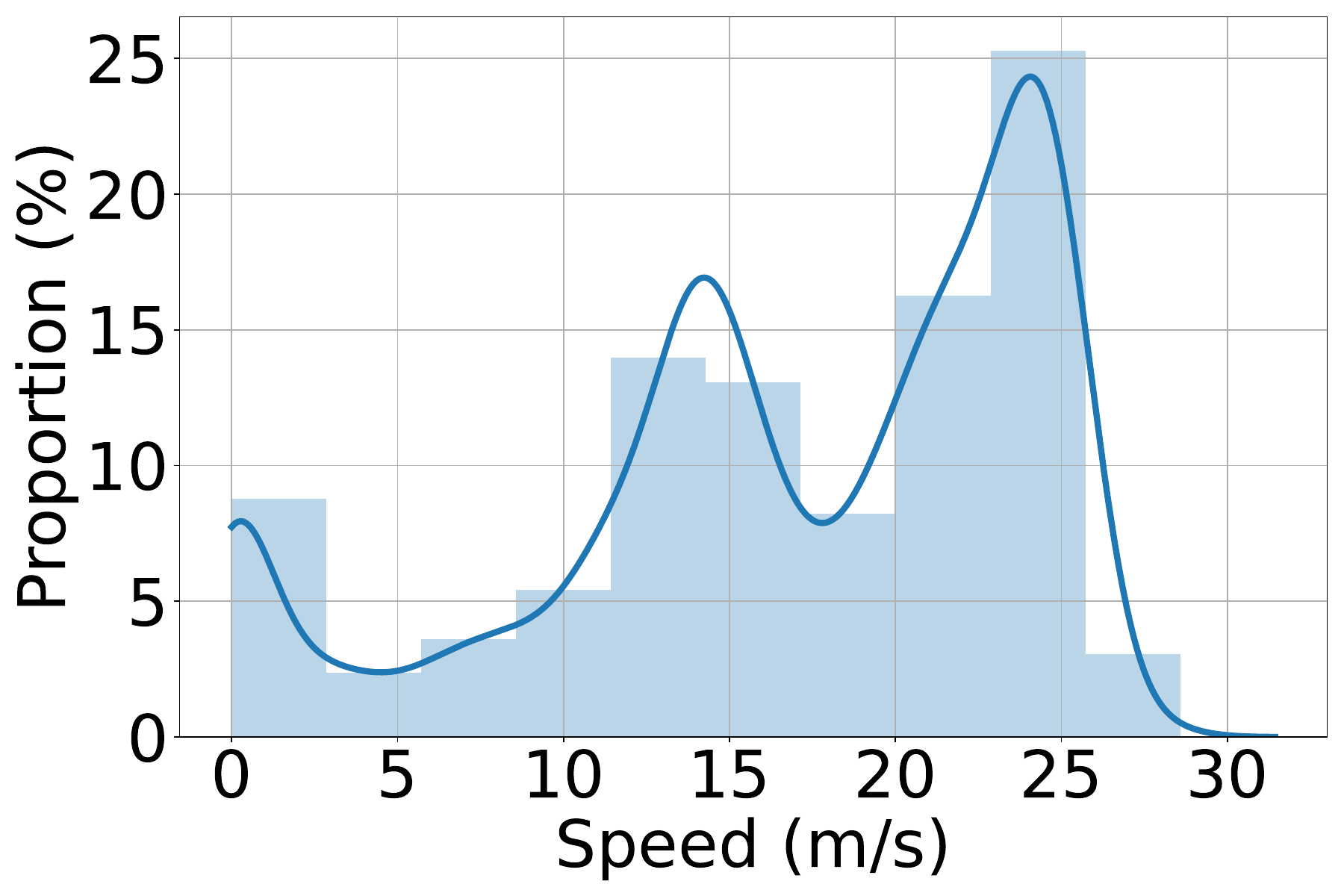}}
    \subcaptionbox{CentralOhio}{\includegraphics[width=0.24\linewidth]{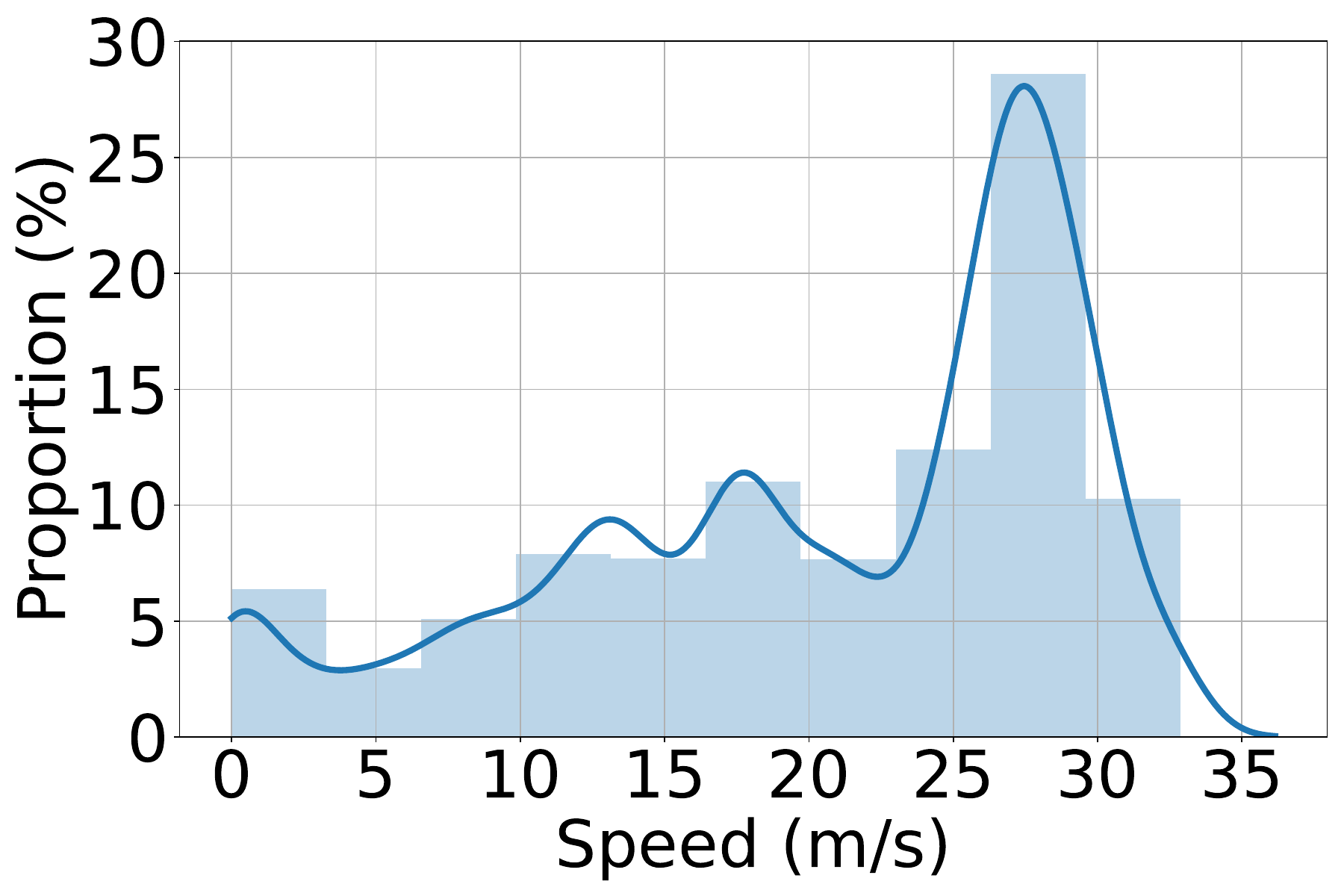}}\\
    Gap \\
    \subcaptionbox{Ring}{\includegraphics[width=0.24\linewidth]{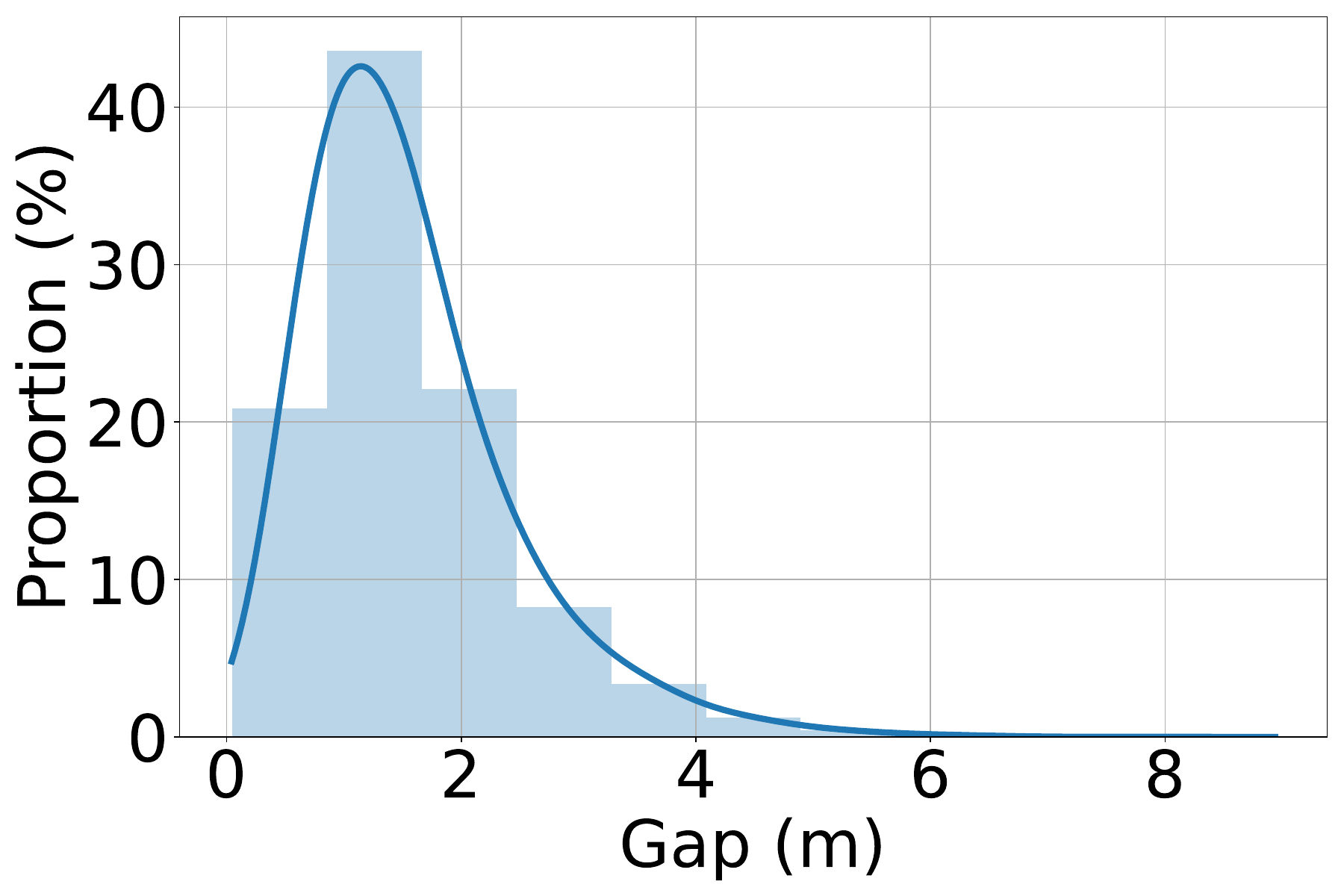}}
    \subcaptionbox{OpenACC}{\includegraphics[width=0.24\linewidth]{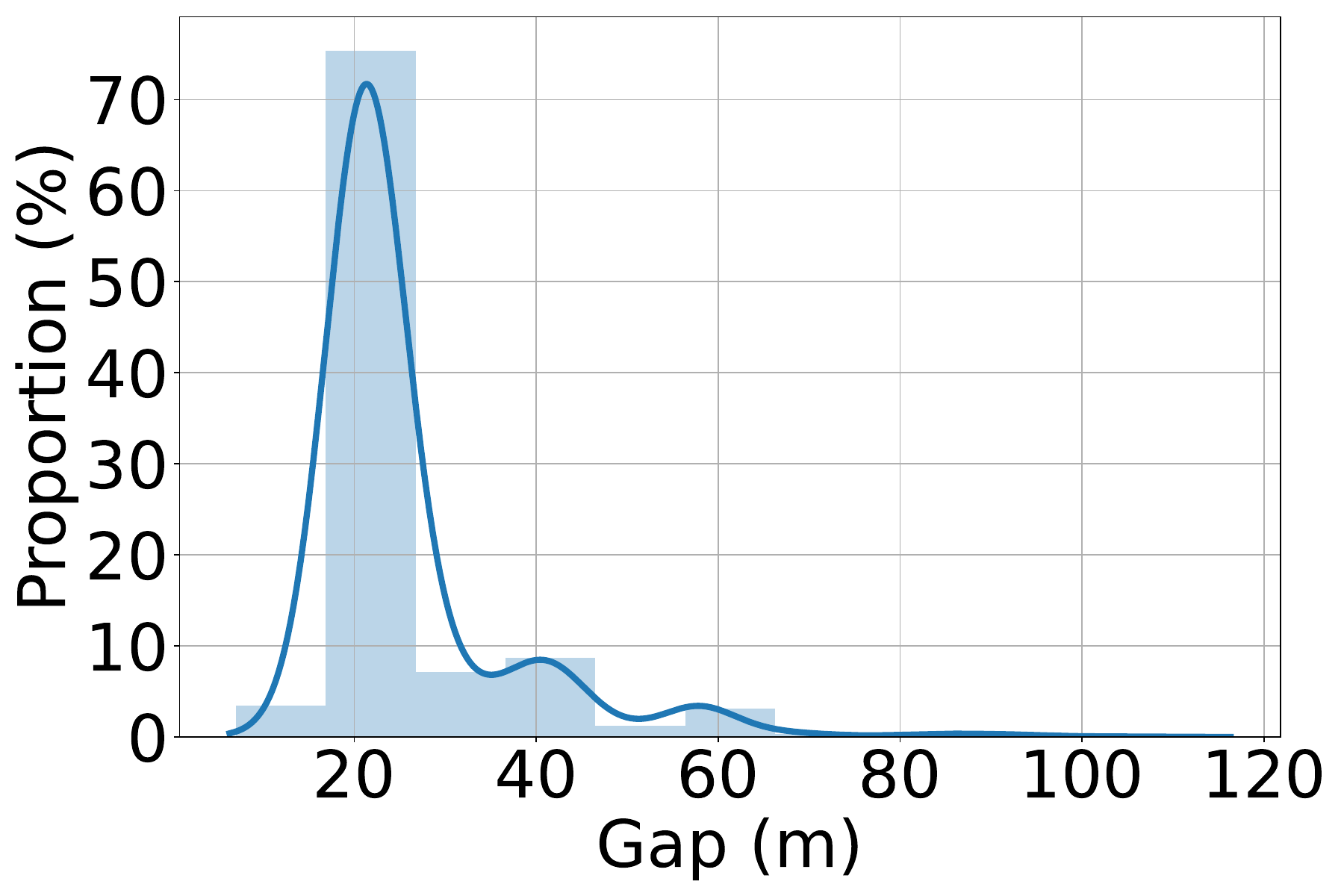}}
    \subcaptionbox{CATS}{\includegraphics[width=0.24\linewidth]{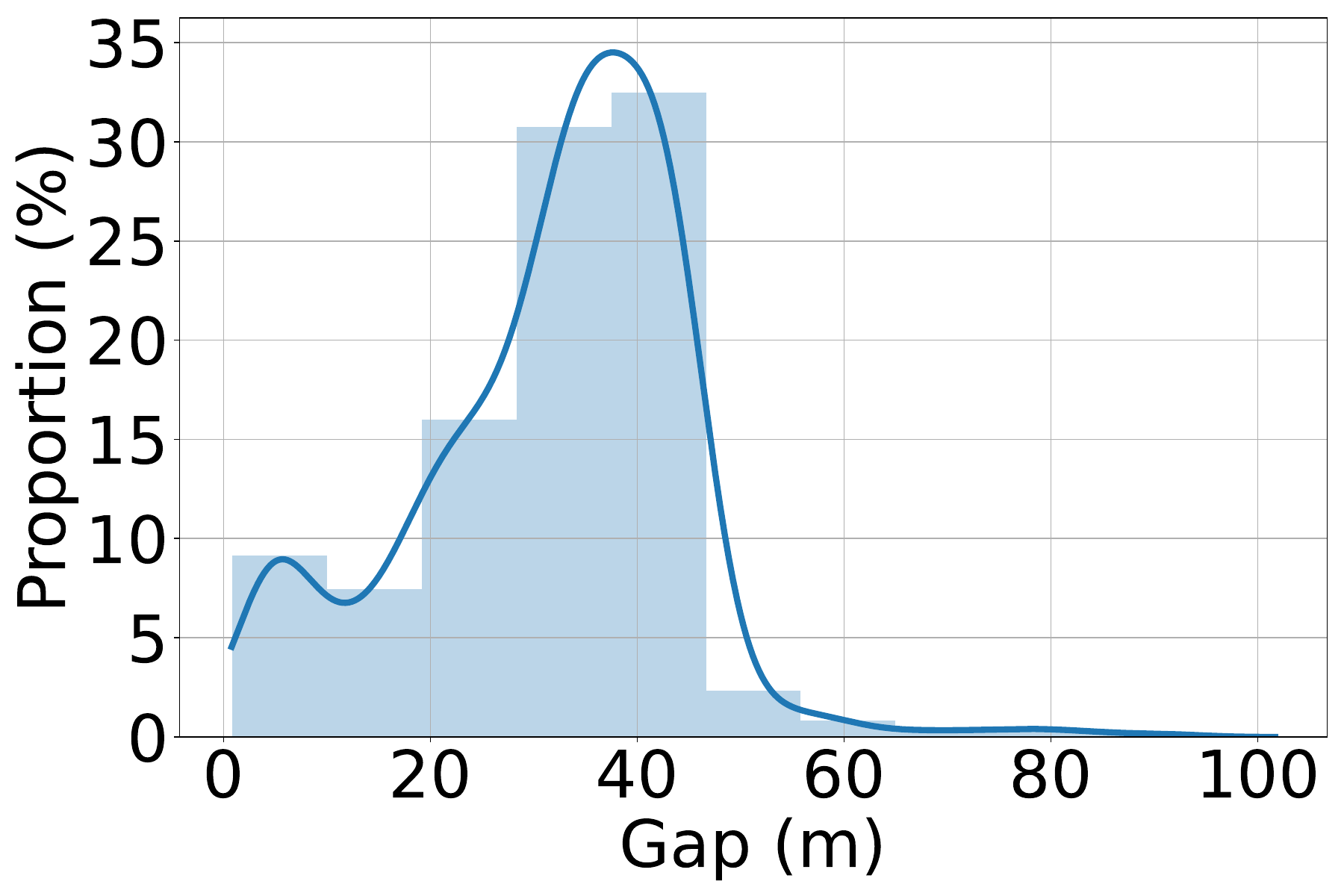}}
    \subcaptionbox{CentralOhio}{\includegraphics[width=0.24\linewidth]{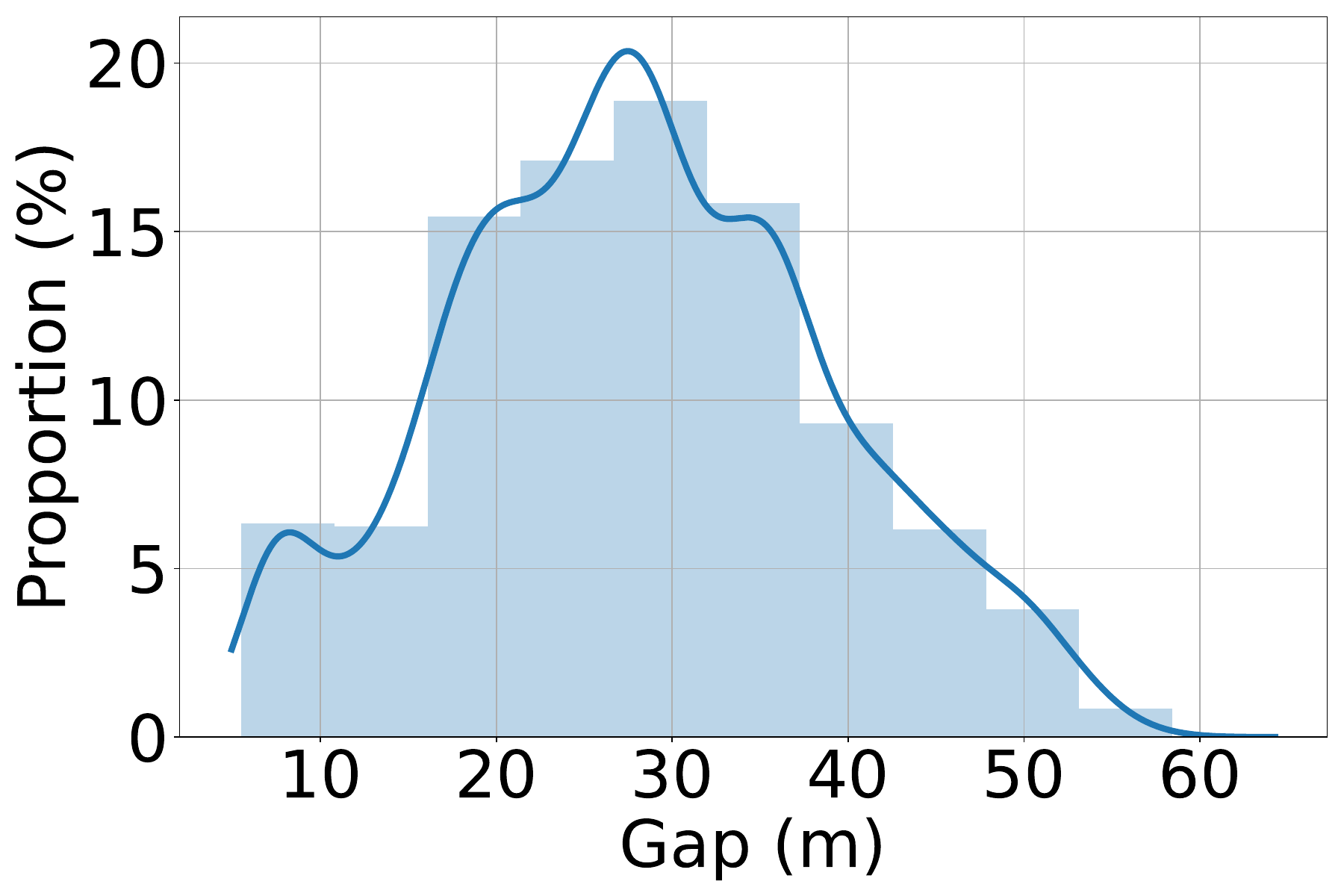}}
    \caption{Speed and gap distribution in datasets.}
    \label{fig:dataset feature distribution}
\end{figure}

The distributions of vehicle speed and gap in the datasets are shown in~\Cref{fig:dataset feature distribution}. The results indicate that the collected data span a wide range of traffic scenarios.  The speed profiles include complete stops (i.e., zero speed), as seen in CATS~\Cref{fig:dataset feature distribution}(c) and CentralOhio~\Cref{fig:dataset feature distribution}(d). They also capture free-flow conditions with high speeds around 30 m/s, as in OpenACC~\Cref{fig:dataset feature distribution}(b). The wide range  in the gap distributions further demonstrates that the datasets cover diverse traffic conditions.

\subsection*{B. OVM calibration}

\begin{figure}[!t]
    \centering
    Speed \\
    \subcaptionbox{Ring}{\includegraphics[width=0.24\linewidth]{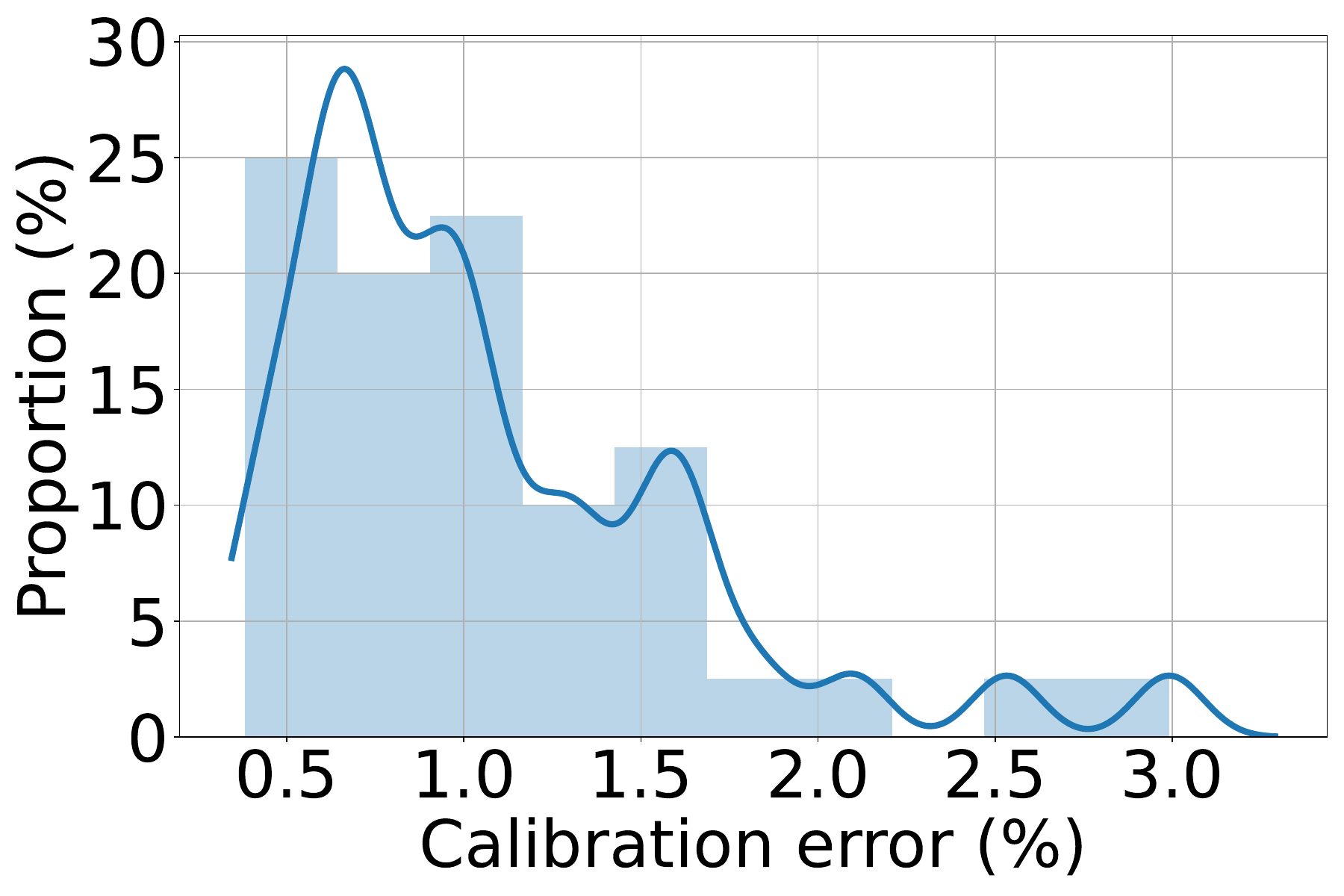}}
    \subcaptionbox{OpenACC}{\includegraphics[width=0.24\linewidth]{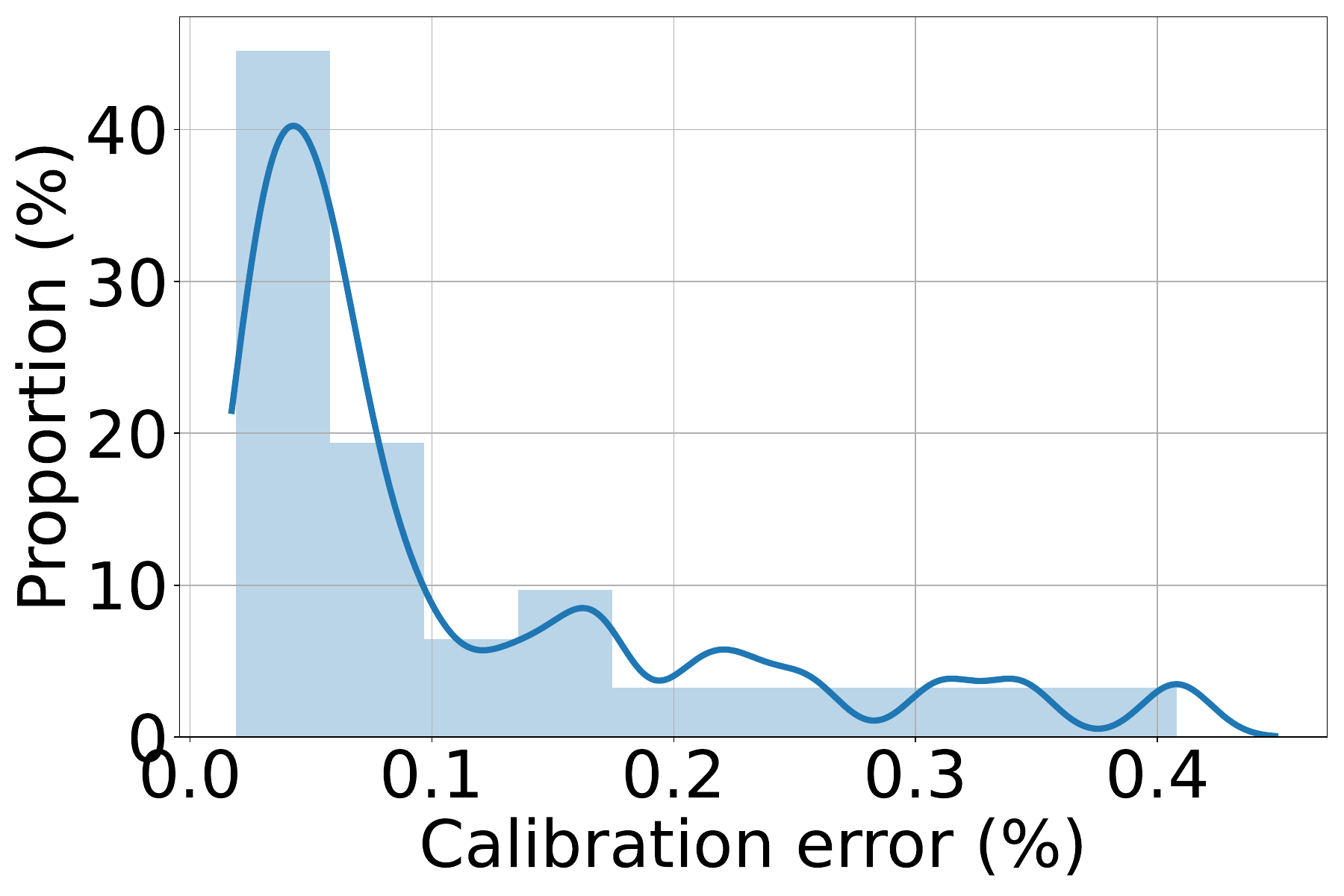}}
    \subcaptionbox{CATS}{\includegraphics[width=0.24\linewidth]{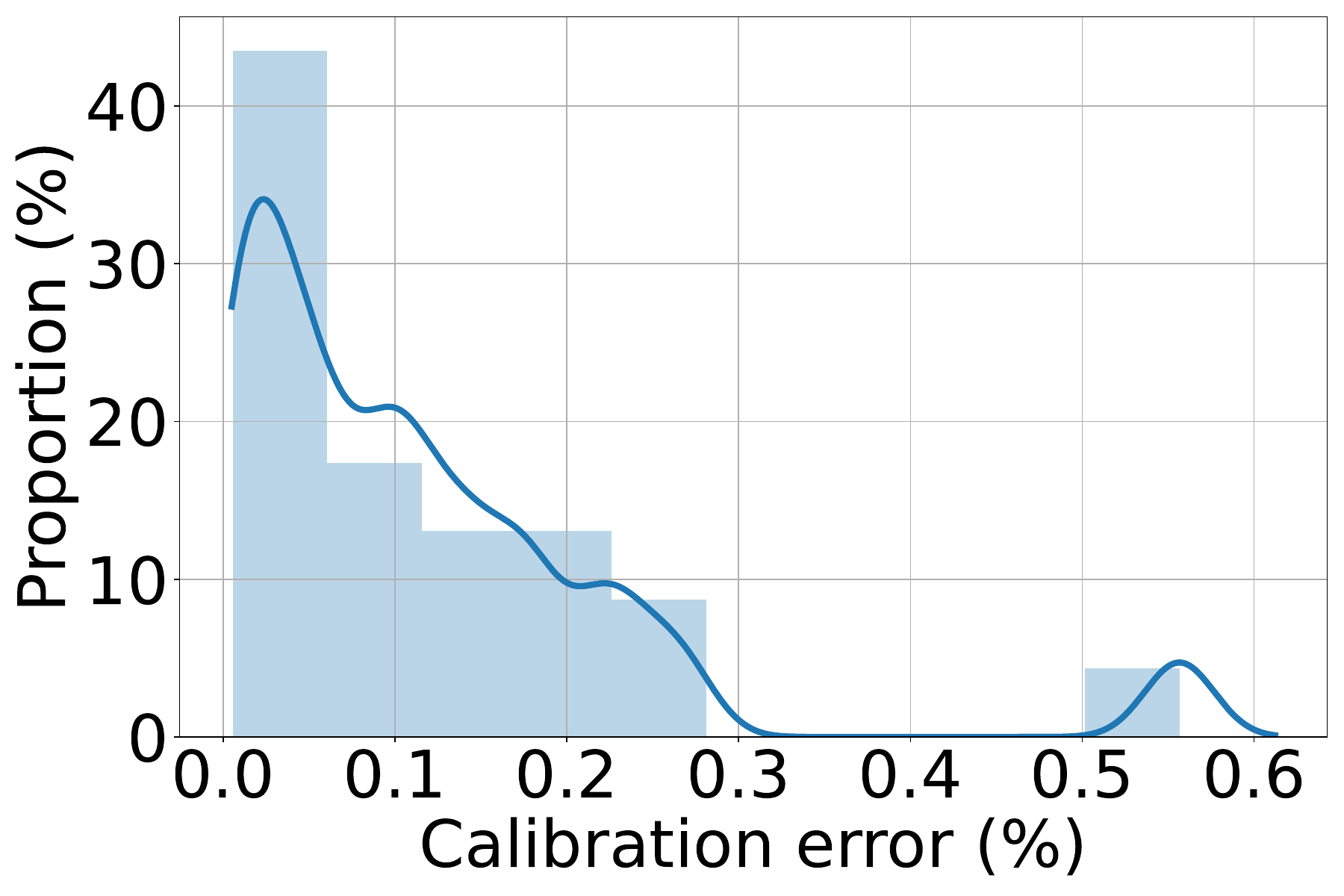}}
    \subcaptionbox{CentralOhio}{\includegraphics[width=0.24\linewidth]{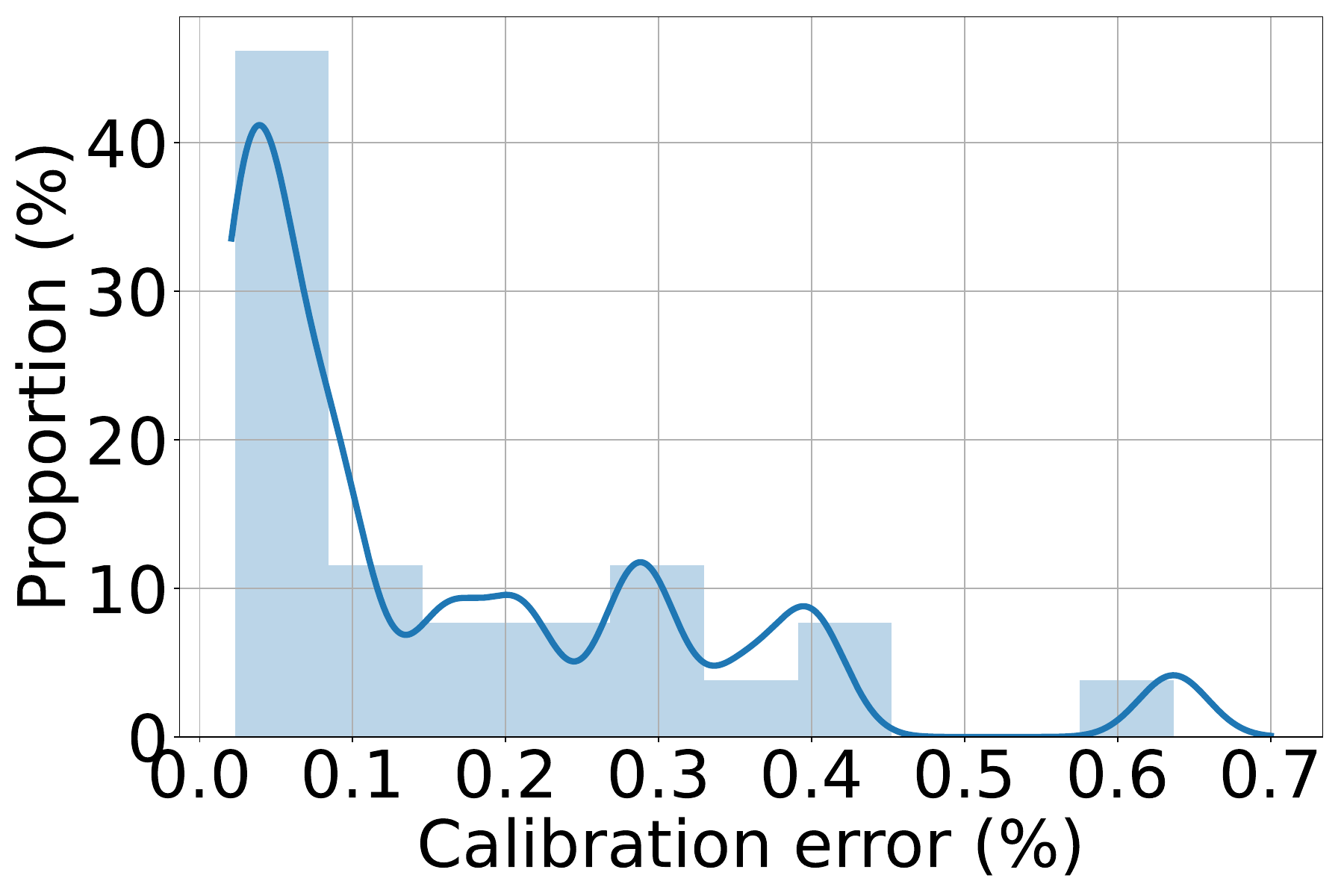}}\\
    Gap \\
    \subcaptionbox{Ring}{\includegraphics[width=0.24\linewidth]{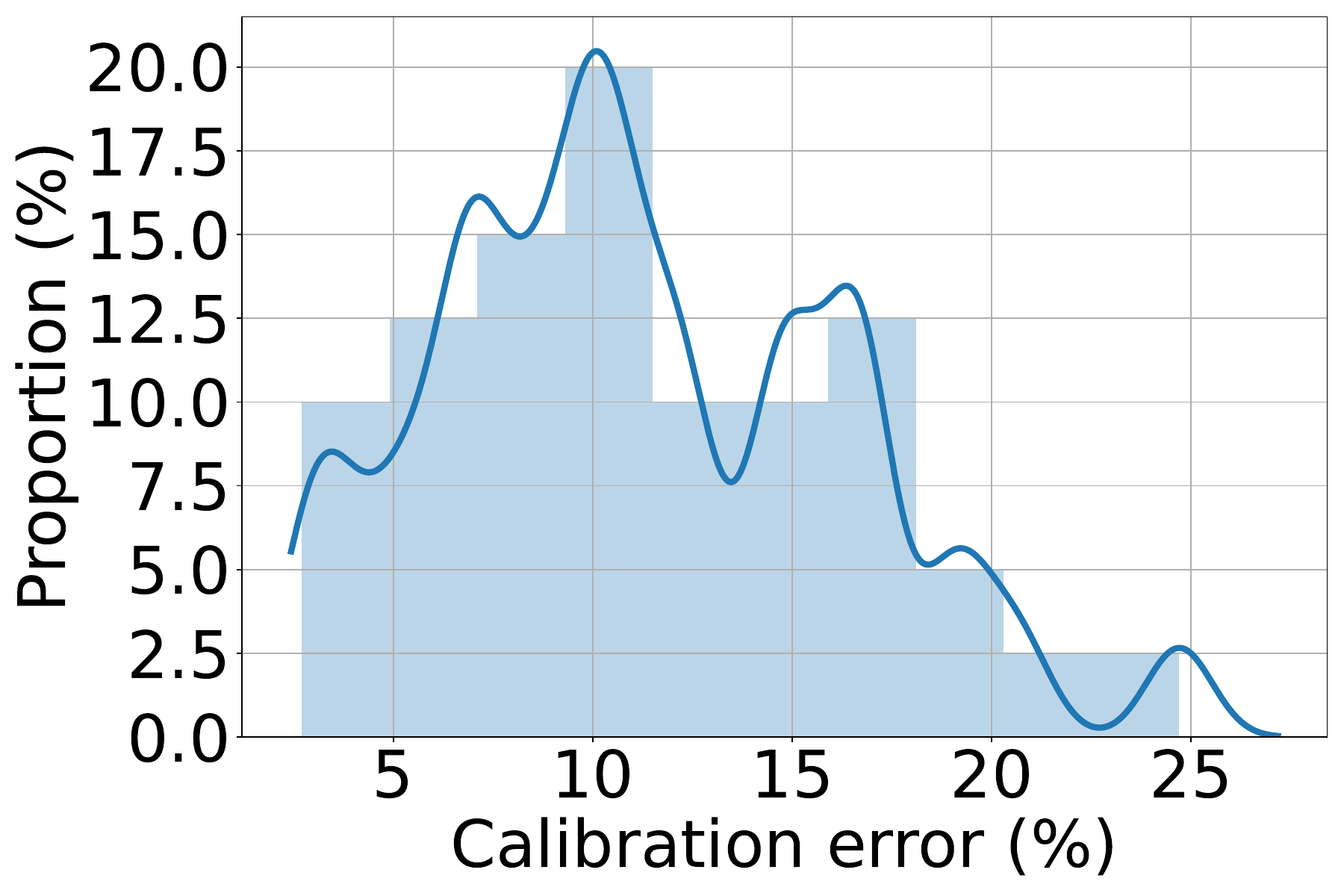}}
    \subcaptionbox{OpenACC}{\includegraphics[width=0.24\linewidth]{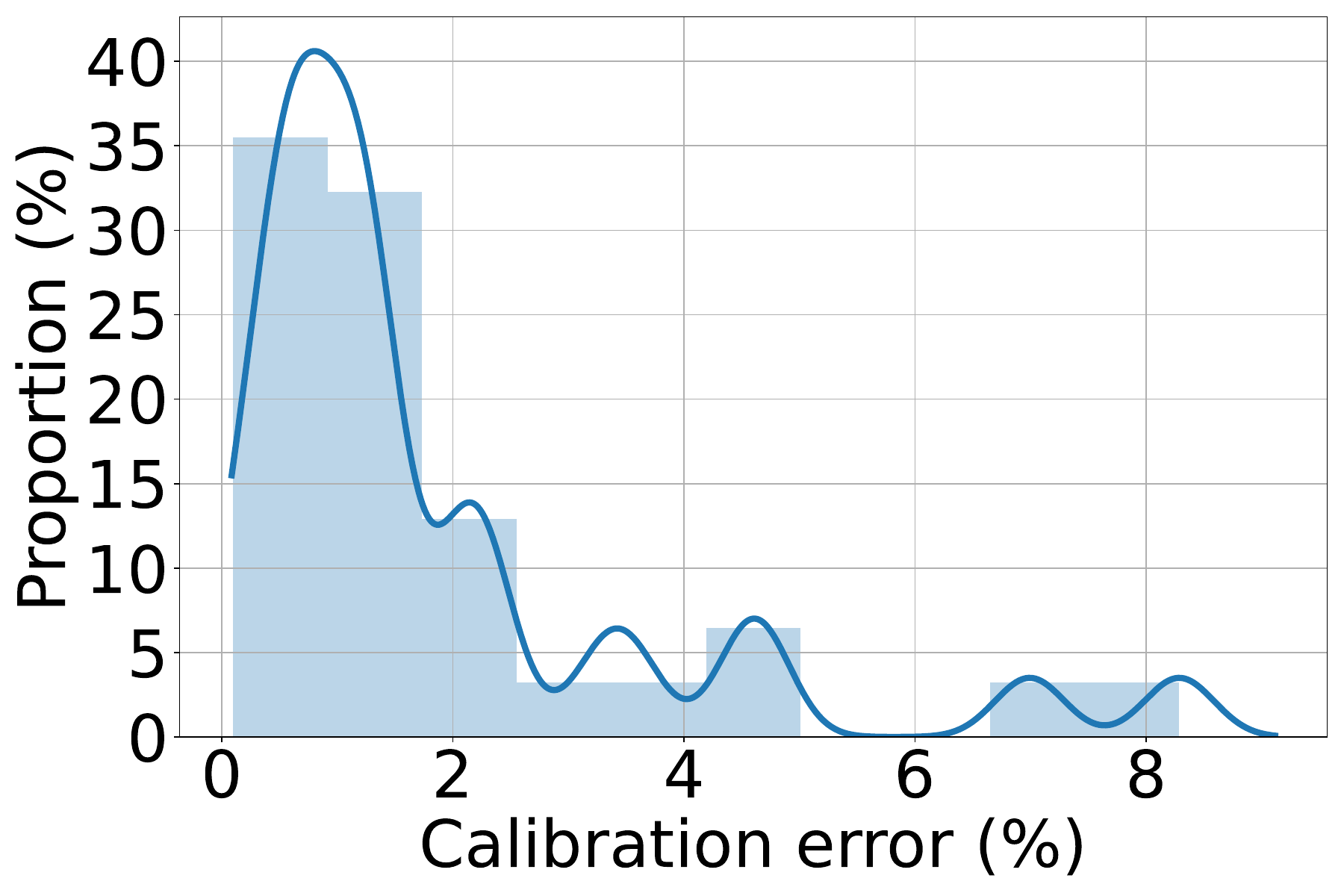}}
    \subcaptionbox{CATS}{\includegraphics[width=0.24\linewidth]{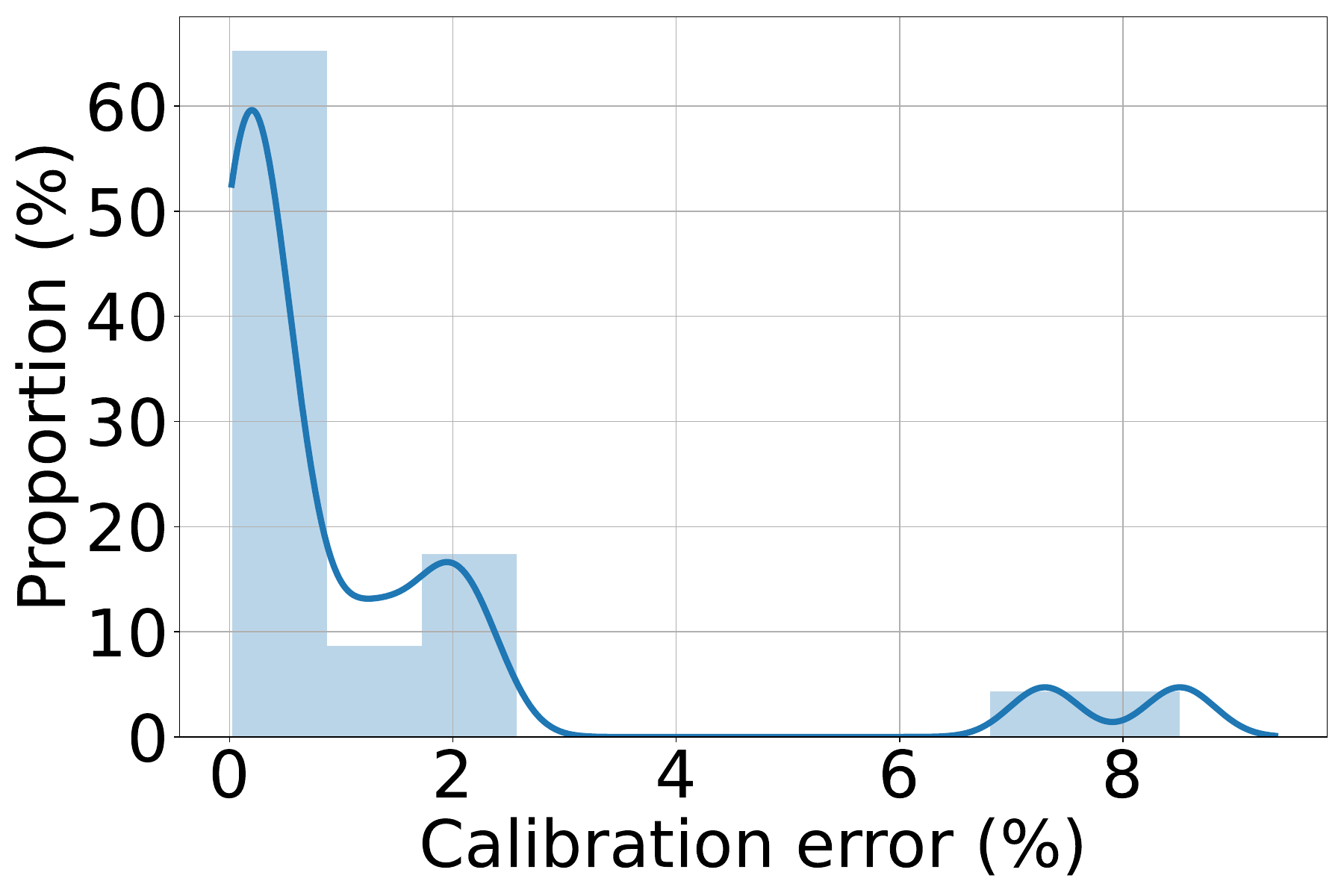}}
    \subcaptionbox{CentralOhio}{\includegraphics[width=0.24\linewidth]{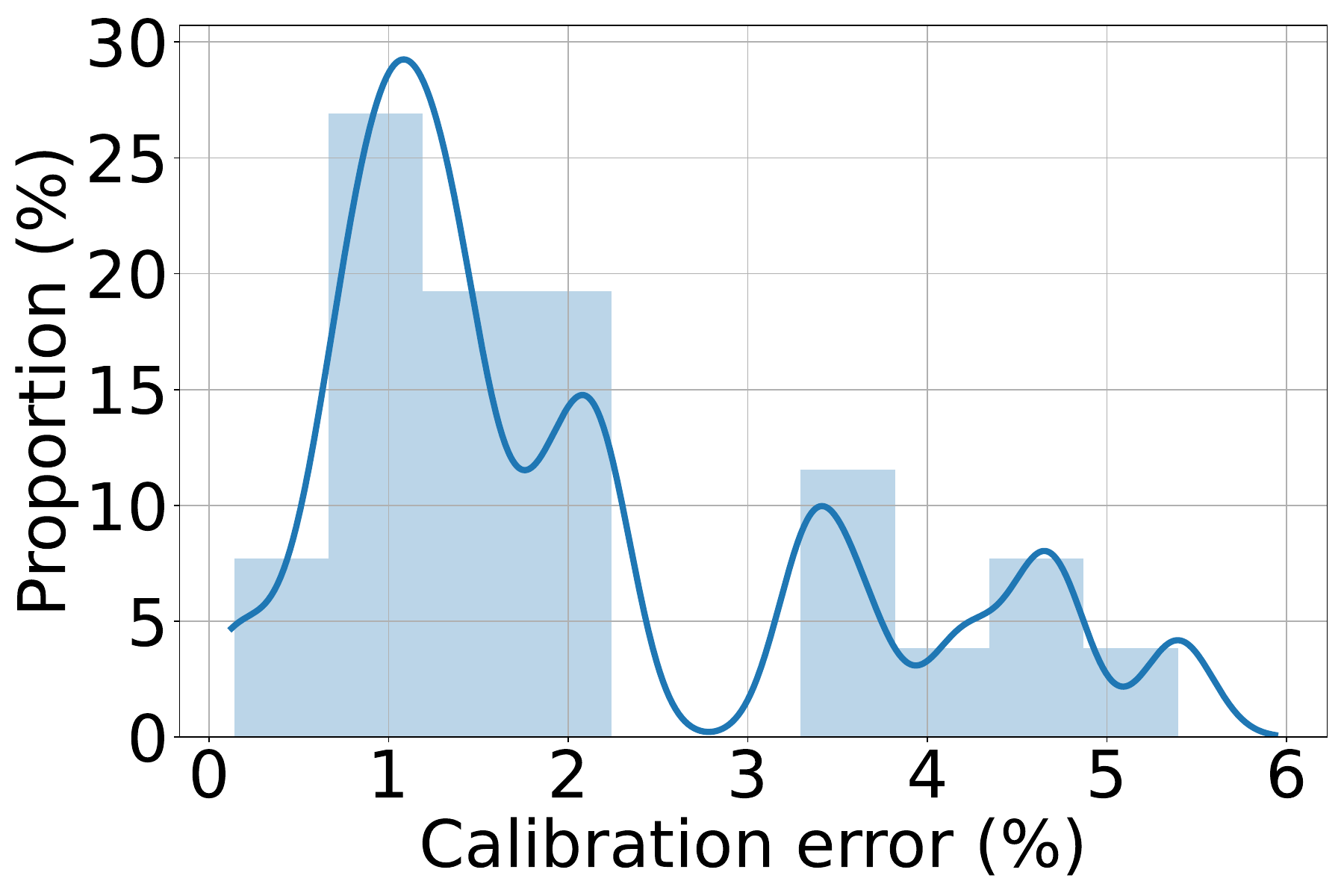}}
    \caption{Calibration error of the OVM.}
    \label{fig:OVM error distribution}
\end{figure}

We run generic algorithms to calibrate the OVM car-following model~\Cref{eq:OVM}. For a vehicle $i$, the  objective function is to minimize the  calibration error for the gap and speed:
\begin{align}
   \sum_j \left( \hat{s}_i(t_j) - s_i(t_j) \right)^2  + \alpha \left(\hat{v}_i(t_j) - v_i(t_j) \right)^2,
\end{align}
where $j$ denotes the time step, and $\alpha = (s_i^*/v_i^*)^2$ is the  coefficient with $s_i^*$ and $v_i^*$ being the average gap and speed of the trajectory. 
We set the range for the OVM parameters as: $\tau \in[0,20]$, $\beta\in[0,5]$, $s^0\in[1,20]$, $T\in[0,5]$, $V^{\mathrm{f}} \in [0,50]$. We run the generic algorithm for 500 iterations. We evaluate the calibration accuracy via:
\begin{align}
    E_s = \frac{\sum_j (\hat{s}(t_j)-s(t_j))^2}{\sum_j (s(t_j))^2} \times 100\%, \quad E_v = \frac{\sum_j (\hat{v}(t_j)-v(t_j))^2}{\sum_j (v(t_j))^2} \times 100\%.
\end{align}
Note that we don't use the RMSE as $\left(\frac{  \hat{v}(t_j)-v(t_j))^2}{v(t_j)} \right)^2 $ since   the vehicle is at stop, i.e., $v(t)=0$, during some time periods.   We present the calibration error  in the four datasets in~\Cref{fig:OVM error distribution}. 
The results show that for the three AV datasets, the calibrated OVM accurately describes car-following dynamics, with the gap error less than 10~\% and the speed error  less than 1~\%. For the HV Ring dataset, the calibration error is higher, as human drivers exhibit more uncertain driving behaviors that are difficult to formulate accurately. Nevertheless, the OVM model still provides an accurate representation of the trajectory. For the gap, the majority of the error is lower than 10~\%.  And for speed, the error is around 1~\%. 

\subsection*{C. Simulation settings for vehicle motion in~\Cref{sec:macro}}

We simulate the motion of a vehicle platoon on a ring road with a total length of $L = 600$ m.  We adopt the OVM car-following dynamics  as in~\Cref{eq:OVM}. 
We consider three types of vehicles: one type of HV, and two types of autonomous vehicles: AV1 and AV2. 
All vehicles are assumed to have a uniform length of $l = 5$ m.
For the HV, the OVM parameters are set as $\tau = 1.7$, $\beta = 0.9$, $s^0 = 15$ m, $T = 0.5$ s, and $V^{\mathrm{f}} = 40$ m/s.
For AV1, we set the OVM parameters as $\tau = 1.7$, $\beta = 0.9$, $s^0 = 5$ m, $T = 0.7$ s, and $V^{\mathrm{f}} = 50$ m/s. For AV2, the OVM parameters are $\tau = 1$, $\beta = 0.6$, $s^0 = 5$ m, $T = 1.2$ s, and $V^{\mathrm{f}} = 30$ m/s.
\Cref{fig:example FD}(a) and \Cref{fig:example FD}(b) show the results for pure HV traffic and pure AV1 traffic, respectively. 
In~\Cref{fig:example FD}(c), we have a mixed traffic of AV1 and HV, and they are evenly distributed along the road.  In~\Cref{fig:example FD}(d), we have all three types of vehicles: HV, AV1, and AV2. They are placed on the road in the repeating order of HV–AV1–HV–AV2–$\cdots$.
The initial gap for vehicle $i = 1, \dots, N-1$ is set as $s_i(0) = s^* \cdot \left(1 + 0.5\sin\left(\frac{2\pi i}{N}\right)\right)$, where $N$ is the total number of vehicles and $s^* = L/N - l$. For the last vehicle, the initial gap is decided by $s_N(0) = L - \sum_{i=1}^{N-1} s_i(0) - Nl$.  The initial speed of each vehicle is set as $v_i(0) = V(s_i(0))$, using the corresponding OVM parameters for that vehicle type. Based on the simulated position data, we reconstruct the macroscopic traffic density $\rho$, speed $v$, and flow $q$ using Edie's formula, with a  grid size of 30 m $\times$  30 s.

\end{document}